\documentclass[a4paper,12pt]{article}
\usepackage[english]{babel}
\usepackage{authblk}
\usepackage[utf8]{inputenc}
\emergencystretch=\maxdimen
\usepackage{hyperref}
\usepackage{fullpage}
\usepackage[pdftex]{graphicx}
\usepackage{csquotes}
\usepackage{cite}
\usepackage{amsmath}
\usepackage{amssymb}
\usepackage{amsfonts}
\usepackage{graphicx}
\usepackage{epstopdf}
\usepackage{multirow}
\usepackage{multimedia}
\usepackage{mathptmx}
\usepackage{listings}
\usepackage{color}
\usepackage{courier}
\usepackage{wrapfig}
\usepackage{tikz}
\usepackage{framed}
\usepackage{xcolor}
\usepackage{fix-cm}
\usepackage{times}
\usepackage{verbatim}
\usepackage{caption}
\usepackage[mathscr]{euscript}
\usepackage{mathtools,xparse}
\usepackage{subfigure}
\usepackage{fixltx2e}
\usetikzlibrary{arrows,shapes}
\usepackage{booktabs}
\usepackage[font=scriptsize,labelfont=bf]{caption}
\providecommand{\keywords}[1]
{
	\small	
	\textbf{\textit{Keywords:}} #1
}
\numberwithin{equation}{section}

\newcommand{\eps}{\varepsilon}
\newcommand{\R}{\mathbb R}
\newcommand{\Sbb}{\mathbb S^{n-1}}
\newcommand{\om}{\omega}
\newcommand{\Om}{\Omega }
\newcommand{\pom}{\partial \Omega}
\newcommand{\xbf}{{\bf x}}
\newcommand{\Xbf}{{\bf X}}
\newcommand{\vbf}{{\bf v}}
\newcommand{\ubf}{{\bf u}}
\newcommand{\thetabf}{\mbox{\boldmath$\theta$}}
\newcommand{\nubf}{\mbox{\boldmath$\nu$}}
\newcommand{\xibf}{\mbox{\boldmath$\xi$}}

\newcommand{\proof}{{\sc Proof.} \quad}

\newcounter{expcounter}			

\newtheorem{theorem}{Theorem}[section]
\newtheorem{assumpt}{Assumption}[section]
\newtheorem{lemma}{Lemma}[section]
\newtheorem{remark}{Remark}[section]

\newenvironment{experiment}
{	\refstepcounter{expcounter}
	{~\\\textbf{Experiment \theexpcounter~---~}}
}

\usepackage{titlesec}
\titleformat{\section}
{\normalfont\fontfamily{phv}\fontsize{14}{17}\bfseries}{\thesection}{0.75em}{}
\titleformat{\subsection}
{\normalfont\fontfamily{phv}\fontsize{12}{17}\selectfont}{\thesubsection}{0.75em}{}
\titleformat{\subsubsection}
{\normalfont\fontfamily{phv}\fontsize{13}{17}\selectfont}{\thesubsubsection}{0.75em}{}
\title{
Multiscale modeling of glioma pseudopalisades: contributions from the tumor microenvironment}
\author{Pawan Kumar$^{1}$, Jing Li$^{2}$, and Christina Surulescu$^{1}$\\
	{\small $^{1}$ TU Kaiserslautern, Felix-Klein-Zentrum f\"{u}r Mathematik,} \\
	{\small Paul-Ehrlich-Str. 31, 67663 Kaiserslautern, Germany}\\	
	{\small $^{2}$	College of Science, Minzu University of China,}\\
	{\small Beijing, 100081, P.R. China}\\
	{\small (kumar, surulescu@mathematik.uni-kl.de, 2010023@muc.edu.cn)}\\
}	

\begin{document}

\maketitle

\begin{abstract}
	
\noindent
Gliomas are primary brain tumors with a high invasive potential and infiltrative spread. Among them, glioblastoma multiforme (GBM) exhibits microvascular hyperplasia and pronounced necrosis triggered by hypoxia. Histological samples showing garland-like hypercellular structures (so-called pseudopalisades) centered around the occlusion site of a capillary are typical for GBM and hint on poor prognosis of patient survival. We propose a multiscale modeling approach in the kinetic theory of active particles framework and deduce by an upscaling process a reaction-diffusion model with repellent pH-taxis. We prove existence of a unique global bounded classical solution for a version of the obtained macroscopic system and investigate the asymptotic behavior of the solution. Moreover, we study two different types of scaling and compare the behavior of the obtained macroscopic PDEs by way of simulations. These show that patterns\footnote{not necessarily of Turing type} (including pseudopalisades) can be formed for some parameter ranges, in  accordance with the tumor grade. This is true when the PDEs are obtained via parabolic scaling (undirected tissue), while no such patterns are observed for the PDEs arising by a hyperbolic limit (directed tissue). This suggests that brain tissue might be undirected - at least as far as glioma migration is concerned.  We also investigate two different ways of including cell level descriptions of response to hypoxia and the way they are related.
\end{abstract}

\medskip
\section{Introduction}\label{sec:intro}

\noindent
Classified as grade IV astrocytoma by WHO \cite{louis20072007}, glioblastoma multiforme (GBM) is considered to be the most aggressive type of glioma, with a median overall survival time of 60 weeks, in spite of state-of-the-art treatment \cite{rong2006pseudopalisading,wrensch02}. It is characterized by fast, infiltrative spread and unchecked cell proliferation which triggers hypoxia and upregulation of glycolysis, usually accompanied locally by exuberant angiogenesis \cite{Brat2002,Fischer2006}; one of the typical features of GBM is the development of a necrotic core \cite{louis20072007}.
Increased extracellular pressure from edema and expression of procoagulant factors putatively lead to vasoocclusion and thrombosis \cite{Brat2004}, hence impairing oxygen supply at the affected site, which becomes hypoxic and induces tissue necrotization. As a consequence, glioma cells actively and radially migrate away from the acidic area \cite{brat2004pseudopalisades}, forming palisade-like structures exhibiting arrangements of elongated nuclei stacked in rows at the periphery of the hypocellular region around the occlusion site \cite{wippold2006neuropathology}. Such histopathological patterns are typically observed in GBM and are used as an indicator of tumor aggressiveness \cite{brat2003malignant,kleih}. Pseudopalisades can be narrow, with a width less than $100\ \mu m$ and a fibrillar interior structure, medium-sized ($200-400\ \mu m$ wide) with central necrosis and vacuolization, but with a fibrillar zone in the immediate interior proximity of the hypercellular garland-like formation. Finally, the largest ones exceed $500\ \mu m$ in width and are surrounding extensive necrotic areas, most often containing central vessels \cite{brat2004pseudopalisades}.\\[-2ex]

\noindent
Mathematical modeling has become a useful means for supporting the investigation of glioma dynamics in interaction with the tumor  microenvironment. Over the years several modeling approaches have been proposed. While discrete and hybrid models (see e.g. \cite{boettger,khain,sander}) use computing power to assess the rather detailed interplay between glioma cells and their surroundings, the continuum settings enable less expensive simulations and mathematical analysis of the resulting systems of differential equations. Since the structure of brain tissue with its patient-specific anisotropy is (among other factors) essential for the irregular  spread of glioma, the mathematical models should be able to include in an appropriate way such information, which is available from diffusion tensor imaging (DTI) data. The macroscopic evolution of a tumor is actually determined by processes taking places on lower scales, thus it is important to deduce the corresponding population dynamics from descriptions of cell behavior on mesoscopic or even subcellular levels, thereby taking into account the interactions with the underlying anisotropic tissue and possibly further biochemical and/or biophysical traits of the extracellular space. This has been done e.g. in \cite{Corbin2018,engwer2015glioma,engwer2015effective,EKS,Hunt2016,PH13} upon starting on the mesoscale from the kinetic theory of active particles (KTAP) framework \cite{Bell-KTAP} and obtaining with an adequate upscaling the macroscopic PDEs of reaction-(myopic) diffusion(-taxis) type for the tumor dynamics. Depending on whether the modeling process included subcellular events, these PDEs contain in their coefficients information from that modeling level, thus receiving a multiscale character. It is this approach  that we plan to follow here, however with the aim of obtaining cell population descriptions for the pseudopalisade formation rather than for the behavior of the whole tumor.\\[-2ex]

\noindent
Mathematical models addressing glioma pseudopalisade formation are scarce; we refer to \cite{cai2016mathematical,caiazzo} for agent-based approaches and to \cite{Alfonso,martinez2012hypoxic} for continuous settings. Of the latter, \cite{Alfonso} investigated the impact of blood vessel collapse on glioma invasion and the phenotypic switch in the migration/proliferation dichotomy. It involves a system of PDEs coupling the nonnlinear dynamics of glioma population with that of nutrient concentration and vasculature, thus not explicitly including acidity. The PDE for the evolution of tumor cell density was obtained upon starting from a PDE/ODE system for migrating/proliferating glioma densities and performing transformations relying on several assumptions. The work in \cite{martinez2012hypoxic} describes interactions between normoxic/hypoxic glioma, necrotic tissue, and oxygen concentration. The model confirms the histological pattern behavior and shows by simulations a traveling wave concentrically moving away from the highly hypoxic site toward less acidic areas. The PDE system therein was set up in a heuristic manner directly on the macroscale; it features reaction-diffusion equations without taxis or other drift. \\[-2ex]

\noindent
In the present work we are interested in deducing effective equations for the space-time evolution of glioma cell density in interaction with extracellular acidity (concentration of protons), thereby accounting for the multiscality of the involved processes and for the anisotropy of brain tissue. The deduced model should be able to reproduce pseudopalisade-like patterns and to investigate the influence of acidity and tissue on their behavior. The rest of this paper is organized as follows: in Section \ref{sec:model1} we formulate our model upon starting from descriptions of cell dynamics on the microscopic and mesoscopic scales. Section \ref{sec:upscaling} is concerned with obtaining the macroscopic limits of that setting; we will investigate parabolic as well as hyperbolic upscalings, corresondingly leading to diffusion and drift-dominated evolution, respectively, and depending on tissue properties (directed/undirected). An alternative modeling approach and its parabolic limit will be addressed as well. In Section \ref{sec:numerics} we provide an assessment of parameters and functions involved in the deduced macroscopic PDE systems and perform numerical simulations, also providing a comparison between the studied modeling approaches. Section \ref{sec:analysis} is dedicated to establishing the existence and uniqueness of a  global bounded classical solution to the macroscopic system obtained by parabolic scaling. A result concerning the asymptotic behavior of such solution is proved as well. Finally, Section \ref{sec:discussion} contains a discussion of the obtained results and an outlook on further problems of interest related to GBM pseudopalisades.

\section{Model set up on subcellular and mesoscopic scales}\label{sec:model1}

\noindent
The approach in \cite{engwer2015glioma,engwer2015effective} led to (hapto)taxis of glioma cell population on the macroscale upon taking into account receptor binding of cells to the surrounding tissue. As such, it was a simplification of subcellular dynamics as considered in \cite{KeSu2011,kelkel2012multiscale,Lorenz2014}, where the cancer cells were supposed to interact with the tissue and with a soluble ligand acting as a chemoattractant. The latter works, however, were concerned with the micro-meso-macro formulation of cancer cell evolution in dynamic interaction with the tissue (as a mesoscopic quantity) and the ligand (obeying a nonlocal macroscopic PDE), along with the analysis therewith, whereas here we intend to obtain a system of effective macroscopic PDEs for glioma population density in interaction with space-time dependent acidity. Here the macroscopic scale is smaller than in the mentioned previous works: it is not the scale of the whole tumor, but that of a subpopulation, localized around one or several vasoocclusion sites in a comparatively small area of the tumor - corresponding to a histological sample. Since (see end of first paragraph in Section \ref{sec:intro}) the size of such samples is too small to allow a reliable assessment of underlying tissue distribution via DTI\footnote{the typical size of a voxel is ca. $1\ mm^3$}, we do not describe a detailed cell-tissue interaction via cell activity variables as in the mentioned works. However, tissue anisotropy might be relevant even on such lower scale, therefore we consider, instead, an artificial structure by way of some given water diffusion tensor, in order to be able to test such influences on the glioma pattern formation. Therefore, on the subcellular level we only account explicitly for interactions between extracellular acidity and glioma transmembrane units mediating them. The latter can be ion channels and membrane transporters ensuring proton exchange, or even proton-sensing receptors \cite{Holzer2009}. \\[-2ex]

\noindent
We denote by $y(t)$ the amount of transmembrane units occupied with protons (in the following we will call this the activity variable, in line with the KTAP framework in \cite{Bell-KTAP}) and by $R_0$ the total amount of such units (ion channels, receptors, etc), which for simplicity we assume to be constant. Let $S$ denote the concentration of (extracellular) protons and $S_{max}$ be a threshold value, which, when exceeded, leads to cancer cell death. The corresponding binding/occupying kinetics are written
\begin{center}
	$\mathrm{\frac{S}{S_{max}}+(R_0 -y)} \xrightleftharpoons[k^{-}]{k^{+}} \mathrm{y},$
\end{center}
so that we can write for the corresponding subcellular dynamics (upon rescaling $y\leadsto y/R_0$)
\begin{equation}\label{eq_ode_0}
\dot{y} = G(y,S):= k^{+} \frac{S}{S_{max}}  (1 - y) -k^{-} y,
\end{equation}
where $k^+$ and $k^{-}$ represent the reaction rates.
We denote by $y^*$ the steady-state of the above ODE, thus we have
\begin{equation}\label{eq_stedy_state}
y^{*}  = \frac{k^{+}S/S_{max}}{k^{+}S/S_{max}+k^{-}}=\frac{S/S_{max}}{S/S_{max}+k_D}, \qquad k_D:=\frac{k_-}{k_+}.
\end{equation}
As in \cite{engwer2015glioma,engwer2015effective} we will consider deviations from the equilibrium of subcellular dynamics:
$$z:=y^{*}-y.$$
Since the events on this scale are much faster than those on the mesoscopic and especially macroscopic levels, the equilibrium is supposed to be quickly attained, so $z$ is very small. We will use this assumption in the subsequent calculations; as in \cite{engwer2015glioma,engwer2015effective} it will allow us to get rid of higher order moments during the upscaling process, thus to close the system of moments leading to the macroscopic formulation. This assumption also allows us to ignore on this microscopic scale the time dependency of $S$.\footnote{In fact, keeping this dependency and accounting for the correct scales during the macroscopic limit leads to omission of the arising supplementary term, so that the outcome is the same whether we do such assumption at this step or later on.}
Next, we consider the path of a single cell starting at position  $\mathbf{x}_0$ and moving with velocity $\mathbf{v}$ in the acidic environment. Since the glioma cells are supposed to move away from the highly hypoxic site, we take:
$$\mathbf{x} := \mathbf{x_0} - \mathbf{v}t,$$
which leads to
\begin{equation}\label{eq_solution_z}
\dot{z} = -k^{+}( \frac{S}{S_{max}} + k_D)z - \frac{k_D/S_{max}}{(S/S_{max}  + k_D)^2} \mathbf{ v} \cdot \nabla S.
\end{equation}

\noindent
We denote by $p(t, \mathbf{ x, v},y)$ the density function of glioma cells at time $t$ , position $\mathbf{x}  \in \mathbb{R}^n$, velocity $\mathbf{v} \in V \subset  \mathbb{R}^n $, and with activity variable $y \in Y=(0,1)$. We assume as in \cite{Corbin2018,engwer2015glioma,engwer2015effective,EKS,Hunt2016,PH13} that the cells have a constant (average) speed $s>0$, so that $V=s\mathbb S^{n-1}$, i.e. only the cell orientation is varying on the unit sphere. In terms of deviations $z\in Z\subset [y^*-1,y^*]$ from the steady-state (we also call $z$ activity variable) we consider for the evolution of $p$ the kinetic transport equation (KTE)
\begin{equation}\label{eq_meso2}
\partial_t p + \mathbf{ v} \cdot \nabla p - \partial_z \left( \left(\left(k^+S/S_{max} + k^-  \right)z +f^{\prime}(S) \mathbf{ v} \cdot \nabla S      \right)p   \right) = \mathscr{L}[\lambda(z)]p+\mathcal P(S,M)p,
\end{equation}
where
$\mathscr{L}[\lambda(z)] p := -\lambda(z)p + \lambda(z) \int_{V} K(\mathbf{x,v,v^\prime}p(\mathbf{v}^\prime))d\mathbf{v}^\prime$ denotes the turning operator modeling cell velocity innovations due to tissue contact guidance and acidity sensing, with $\lambda(z)$ denoting the turning rate of cells. Thereby, $K(\mathbf{ x, v, v^\prime})$ is a turning kernel giving the likelihood of a cell with velocity $\mathbf{v}^\prime$ to change its velocity regime into $\mathbf{v}$. We adopt the choice proposed in \cite{HillenM5}, i.e. $K(\mathbf{x,v,v^{\prime}})=\frac{q(\mathbf{ x},\hat{\mathbf{v}})}{\om}$
where $q(\mathbf{ x},\hat{\mathbf{ v}})$ is the (stationary) orientation distribution of tissue fibers with $\om = \int_{V} q(\hat{\mathbf{v}})d\mathbf{v}= s^{n-1}$ and $ \hat{\mathbf{v}} = \frac{\mathbf{ v}}{|\mathbf{ v}|} \in \mathbb{S}^{n-1}$.
We take the turning rate as in \cite{engwer2015glioma}
\begin{equation}\label{eq:lambda}
\lambda(z)=\lambda_0 -\lambda_1 z  \geq 0,
\end{equation}
where $\lambda_0$ and  $\lambda_1$ are positive constants. The choice means that the turning rate is increasing with the amount of proton-occupied transmembrane units. The turning operator in \eqref{eq_meso2} thus becomes
\begin{equation}
\mathscr{L}[\lambda (z)]p=\mathscr{L}[\lambda_0]p - \mathscr{L}[\lambda_1]zp,
\end{equation}
with \begin{align}
\mathscr{L}[\lambda_i]p(t,\xbf,\vbf ,y) &= -\lambda_i p(t,\xbf,\vbf ,y) +\lambda_i \frac{q}{\om } \int_V p(t,\xbf,\vbf ,y) d\mathbf{ v}   \quad  \text{for} \quad i = 0,1.
\end{align}
We also employ the notation $f(S)=y^*$ to emphasize that the steady-state of subcellular dynamics depends on the proton concentration $S$. The last term in
\eqref{eq_meso2} represents growth or depletion, according to the acidity level in the tumor microenvironment. Similarly to \cite{engwer2015effective}, but now accounting for the effect of acidity, we consider a source term of the form
\begin{equation}\label{eq_proliferation}
\mathscr{P}(S,M):= \mu (M) \int_{Z} \chi (z, z^{\prime}) \left(1 - \frac{S}{S_{max}}\right)  p(t,\mathbf{x},\mathbf{v},z^{\prime}) dz^{\prime},
\end{equation}
where $\chi(z,z')$ represents the likelihood of cells having activity state $z'$ to go into activity state $z$ upon interacting with acidity $S(t,\xbf)$. In particular, $\chi $ is a kernel with respect to $z$, i.e. $\int _Z\chi(z,z')dz=1$. The acidity is reported again to the threshold value $S_{max}$. The growth rate $\mu (M)$ depends on the total amount $M(t,\xbf)=\int _V\int _Z p(t,\xbf ,\vbf,z)dzd\vbf $ of glioma cells, irrespective of their orientation or activity state and takes into account limitations by overcrowding. We will provide a concrete choice later in Subsection \ref{subsec:params}. Hence, the presence of tissue is supporting proliferation, which is maintained until the environment becomes too acidic even for tumor cells.\\[-2ex]

\noindent
The above micro-meso formulation for glioma dynamics is supplemented with the evolution of acidity, described by the macroscopic PDE
\begin{equation}\label{eq_S}
S_t = D_s \Delta S + \beta M - \alpha S,
\end{equation}
where $D_s$ is the diffusion coefficient of protons, $\beta$ is the proton production rate by tumor cells, and $\alpha$ denotes the rate of acidity decay.

\noindent
The high dimensionality of the above setting makes the numerics too expensive, thus we aim to deduce macroscopic equations which can be solved more efficiently and, moreover, facilitate the observation of the glioma cell population and its patterning behavior. In order to investigate the possible effects of the tissue being directed or not\footnote{by 'undirected' we mean as in \cite{HillenM5} that the fibers are symmetric along their axis and both fiber directions are identical, while 'directed' means unsymmetric fibers; since the common medical imaging techniques are not providing the necessary resolution, it is actually not known whether brain tissue is directed, but such feature might play a role in the formation of glioma patterns}, we will perform two kinds of macroscopic limit: the parabolic one, for the diffusion-dominated case of undirected tissue, and the hyperbolic limit for directed tissue, which should be drift-dominated. Both types of limits are performed in a formal way, as the rigorous processes would require analytical challenges which go beyond the aims of this note.

\section{Macroscopic limits}\label{sec:upscaling}
We consider the following moments with respect to $\mathbf{v}$ and $z$:
\begin{equation*}
\begin{aligned}
m(t,\mathbf{ x,v}) &= \int_{Z} p(t,\mathbf{ x,v},z)dz \qquad
& M(t,\mathbf{x}) = \iint_{V \times Z}  p(t,\mathbf{ x,v},z)dzd \mathbf{v} \\
m^z(t,\mathbf{ x,v}) &= \int_{Z} zp(t,\mathbf{ x,v},z)dz \qquad
& M^z(t,\mathbf{x}) = \iint_{V \times Z}  zp(t,\mathbf{ x,v},z)dzd \mathbf{v}
\end{aligned}
\end{equation*}
and neglect higher order moments w.r.t. $z$ due to the assumption of the steady-state of subcellular dynamics being rapidly reached. Moreover, we assume $p$ to be compactly supported in the phase space $\R^n\times V\times Z$.\\[-2ex]

\noindent
Integrating (\ref{eq_meso2}) w.r.t $z$, we get:
\begin{equation}\label{eq_int_meso_dz2}
\partial_t m + \nabla_{\mathbf{x}} \cdot (\mathbf{v}m) = -\lambda_0m + \lambda_1m^z + \lambda_0 \frac{q}{\om }M - \lambda_1 \frac{q}{\om}M^z + \mu(M) \left(1-\frac{S}{S_{max}}\right) m
\end{equation}

\noindent
Multiplying (\ref{eq_meso2}) by $z$ and integrating w.r.t. $z$ we get:
\begin{align}\label{eq_z_int_meso_dz}
\partial_t m^z + \nabla_{\mathbf{x}}\cdot(\mathbf{v}m^z) = -(k^+ S/S_{max} &+ k^-)m^z - f^{\prime}(S)\mathbf{v} \cdot \nabla S\ m - \lambda_0 m^z + \lambda_0 \frac{q}{\om } M^z \notag \\
&+ \mu(M)\left(1-\frac{S}{S_{max}}\right)\int_Z \int_Z z\chi(z,z^{\prime})p(z^{\prime})dz^{\prime}dz.
\end{align}
\noindent
In the following we denote as e.g., in \cite{HillenM5,engwer2015glioma} by
\begin{align*}
&\mathbb{E}_q(\xbf ):= \int_{\Sbb} \thetabf q(\xbf ,\thetabf)d\thetabf \\
&\mathbb V_q(\xbf ):=\int_{\Sbb}(\thetabf-\mathbb E_q) \otimes (\thetabf-\mathbb E_q)q(\xbf ,\thetabf)d\thetabf
\end{align*}
the mean fiber orientation and the variance-covariance matrix for the orientation distribution of tissue fibers, respectively.

\subsection{Parabolic limit}\label{subsec:parab-limit}

\noindent
In this subsection we consider the tissue to be undirected, which translates into the directional distribution function for tissue fibers being symmetric, i.e. $\int _{\Sbb}q(\xbf, \thetabf)d\thetabf=\int _{\Sbb}q(\xbf, -\thetabf)d\thetabf$.
 We rescale the time and space variables by $\tilde{t}:= \epsilon^2 t$, $\tilde {\xbf }:= \epsilon \xbf $. Since proliferation is much slower than migration, we also rescale  with $\epsilon^2$ the corresponding term, as in \cite{engwer2015effective}. For notation simplification we will drop in the following the $\tilde{}$ symbol from the scaled variables $t$ and $\xbf $.\\[-2ex]

\noindent
Thus, from (\ref{eq_int_meso_dz2}) and (\ref{eq_z_int_meso_dz}) we get:
\begin{align}
\epsilon^2 \partial_t m + \epsilon \nabla_{\mathbf{ x}} \cdot (\mathbf{ v}m) &= -\lambda_0 m + \lambda_1 m^z + \lambda_0 \frac{q}{\om }M - \lambda_1 \frac{q}{\om } M^z +\epsilon^2 \mu(M) \left(1-\frac{S}{S_{max}}\right) m\\
\epsilon^2 \partial_t m^z + \epsilon \nabla_{\mathbf{x}} \cdot (\mathbf{ v} m^z) &= -(k^+ S/S_{max} +k^- + \lambda_0)m^z - \epsilon f^{\prime}(S) \mathbf{ v} \cdot \nabla S \ m + \lambda_0\frac{q}{\om }M^z \nonumber \\&+ \epsilon^2 \mu(M)\left(1-\frac{S}{S_{max}}\right) \int_Z \int_Z z \chi(z,z^{\prime})p(z^{\prime})dz^{\prime}dz.
\end{align}
Now, using Hilbert expansions for the moments:
\begin{align*}
m &= m_0 + \epsilon m_1 + \epsilon^2 m_2 + ...\\
m^z &= m_0^z + \epsilon m_1^z + \epsilon^2 m_2^z + ...\\
M &= M_0 + \epsilon M_1 + \epsilon^2 M_2 + ...\\
M^z &= M_0^z + \epsilon M_1^z + \epsilon^2 M_2^z + ...
\end{align*}
and identifying the equal powers of $\epsilon$, we get\\
$\mathbf{\epsilon^0}$:
\begin{align}
0 &= -\lambda_0 m_0 + \lambda_1 m_0^z + \lambda_0 \frac{q}{\om } M_0 - \lambda_1 \frac{q}{\om } M_0^z \label{eq_eps_0_1}\\
0 &= -(k^+ S/S_{max} + k^-)m_0^z -\lambda_0 m_0^z + \lambda_0 \frac{q}{\om } M_0^z  \label{eq_eps_0_2}
\end{align}
$\mathbf{\epsilon^1}$:
\begin{align}
\nabla \cdot (\mathbf{ v} m_0) &= -\lambda_0 m_1 + \lambda_1 m_1^z + \lambda_0 \frac{q}{\om }M_1 - \lambda_1 \frac{q}{\om } M_1^z \label{eq_eps_1_1}\\
\nabla \cdot (\mathbf{ v} m_0^z) &= -(k^+ S/S_{max}+k^-)m_1^z - f^{\prime}(S) \mathbf{ v} \cdot \nabla S m_0 -\lambda_0 m_1^z + \lambda_0 \frac{q}{\om } M_1^z \label{eq_eps_1_2}
\end{align}
$\mathbf{\epsilon^2}$:
\begin{equation}\label{eq_eps^2}
\partial_t m_0 + \nabla \cdot (\mathbf{ v} m_1) = -\lambda_0 m_2 + \lambda_1 m_2^z + \lambda_0 \frac{q}{\om } M_2 -\lambda_1 M_2^z + \mu(M)\left(1-\frac{S}{S_{max}}\right)m_0
\end{equation}
If we also expand $\mu$ around $M_0$, \eqref{eq_eps^2} leads to
\begin{equation}\label{eq_eps^2_2}
\partial_t m_0 + \nabla \cdot (\mathbf{ v} m_1) = -\lambda_0 m_2 + \lambda_1 m_2^z + \lambda_0 \frac{q}{\om } M_2 -\lambda_1 M_2^z + \mu(M_0)\left(1-\frac{S}{S_{max}}\right)m_0.
\end{equation}
Integrating (\ref{eq_eps_0_2}) w.r.t. $\mathbf{ v}$ we get
\begin{align*}
0 &=-(k^+ S/S_{max} + k^-)M_0^z - \lambda_0M_0^z + \lambda_0 M_0^z \\
\implies \qquad  M_0^z &= 0 \quad \text{and} \quad m_0^z = 0.
\end{align*}
Then from (\ref{eq_eps_0_1}) we obtain $m_0 = \frac{q}{\om }M_0$. Integrating (\ref{eq_eps_1_2}) w.r.t. $\mathbf{ v}$ gives
\begin{equation*}
0 = -(k^+S/S_{max} + k^-)M_1^z -f^{\prime}(S) \nabla S \cdot  \int_{V} \mathbf{ v} \frac{q}{\om }d \mathbf{ v} M_0.
\end{equation*}
The assumption of undirected tissue gives $\mathbb E_q={\bf 0}$, thus from the above equation we obtain $M_1^z = 0$, which in virtue of (\ref{eq_eps_1_2}) implies
\begin{equation*}
m_1^z = \frac{-f^{\prime}(S) \mathbf{ v} \cdot \nabla S \ m_0}{(k^+ S/S_{max} + k^- + \lambda_0)}.
\end{equation*}
The compact Hilbert–Schmidt operator $\mathscr{L}[\lambda_0]m_1 = -\lambda_0 m_1 + \frac{q}{\om } M_1$ considered as in \cite{HillenM5} on the weighted space $L^2_{\frac{q}{\om }}(V)$ with measure  $\frac{d\vbf }{\frac{q(\hat{\mathbf{v}})}{\om }}$ has kernel $\langle \frac{q}{\om } \rangle$, thus its pseudo-inverse can be taken on $\langle \frac{q}{\om } \rangle^{\perp} $, to deduce from  (\ref{eq_eps_1_1})
\begin{equation*}
m_1 = -\frac{1}{\lambda_0} \left( \nabla \cdot (\mathbf{ v} m_0) -\lambda_1 m_1^z    \right).
\end{equation*}
We summarize our hitherto information about the moments:
\begin{align}
m_0 &= \frac{q}{\om }M_0 \label{eq_m_0}\\
m_0^z &= M_0^z = M_1 = M_1^z = 0\label{eq:moMo^z}\\
m_1 &= -\frac{1}{\lambda_0} \left( \nabla \cdot (\mathbf{ v} \frac{q}{\om }M_0) -\lambda_1 m_1^z    \right)\\
m_1^z &= \frac{-f^{\prime}(S) \mathbf{ v} \cdot \nabla S\ m_0}{(k^+ S/S_{max} + k^- + \lambda_0)}. \label{eq_m_1^z}
\end{align}
Now integrating (\ref{eq_eps^2_2}) w.r.t. $\mathbf{ v}$ we obtain
\begin{equation*}
\int_V \left( \partial_t \left( \frac{q}{\om }M_0 \right) + \nabla \cdot (\mathbf{ v} m_1)  \right) d\mathbf{ v} =  \mu(M_0) \left(1 - \frac{S}{S_{max}}\right) \int_V m_0 d\mathbf{ v}
\end{equation*}
\color{black}
Using (\ref{eq_m_0})-(\ref{eq_m_1^z}), the previous equation becomes
\begin{equation}\label{macro_eq}
\partial_t M_0 = \nabla \nabla : (\mathbb{D}_T(\mathbf{x})M_0)+\nabla \cdot (g(S)\mathbb{D}_T(\mathbf{x}) \nabla S\ M_0) +\mu(M_0)\left(1-\frac{S}{S_{max}}\right)M_0,
\end{equation}
where:
\begin{equation*}
\begin{aligned}
g(S) &= \lambda_1 (k^{+}S/S_{max} +k^{-}+\lambda_0 )^{-1} f^{\prime}(S),\\
f(S) & = \frac{S/S_{max}}{S/S_{max}+k_D},\\
\mathbf{ u(\xbf )} & = \frac{1}{\lambda_0 \om } \int_{V} \mathbf{v} \otimes \mathbf{ v} \nabla q (\xbf ,\hat \vbf )d\mathbf{v}=\nabla \cdot  \mathbb{D}_{T}(\mathbf{x}),\\
\mathbb{D}_{T}(\mathbf{x}) &=  \frac{1}{\lambda_0 \om  } \int_{V}  q(\xbf ,\hat \vbf ) \mathbf{v} \otimes \mathbf{ v}  d\mathbf{v}.
\end{aligned}
\end{equation*}
This macroscopic PDE forms together with \eqref{eq_S} the system characterizing glioma evolution under the influence of acidity. It involves a term describing repellent pH-taxis (the glioma cells move away from large acidity gradients), in which the tactic sensitivity function contains the tumor diffusion tensor $\mathbb{D}_{T}$ encoding information about the anisotropy of underlying tissue and the function $g(S)$ which relates to the subcellular dynamics of proton sensing and transfer accros cell membranes. The myopic diffusion
$$\nabla \nabla : (\mathbb{D}_T(\mathbf{x})M_0)=\nabla \cdot \left(\mathbb{D}_T(\mathbf{x}) \nabla M_0 +\ubf (\xbf )M_0\right )$$
is common to this and previous models \cite{engwer2015glioma,engwer2015effective,Hunt2016,PH13} obtained by parabolic scaling from the KTAP framework.

\subsubsection*{An alternative approach to including acidity effects}\label{subsec:alternative}

\noindent
In the previous derivation the term with repellent pH-taxis was obtained as a consequence of including subcellular level dynamics in the mesoscopic KTE \eqref{eq_meso2} by way of the transport term w.r.t. the activity variable $z$ and by letting the turning rate depend on it. In the context of bacteria motion an alternative approach was proposed in \cite{othmer2002diffusion} and re-employed in \cite{loy2019kinetic} also for eukaryotes having a more complex motility behavior. It does not explicitly include subcellular dynamics (thus no activity variables and corresponding transport terms are considered), but lets instead the cell turning rate depend on the pathwise gradient of some  chemoattractant concentration which is supposed to bias the cell motion. The relationship between the two mesoscopic modeling approaches was studied for bacteria dispersal in
\cite{PTV16}, where it was rigorously shown that the alternative approach follows from the former one, under certain assumptions made on the receptor binding dynamics on the subcellular level (along with fast relaxation towards equilibrium of external signal transduction and stiff response of the activity variables), on the turning rate, and on the initial data. \\[-2ex]

\noindent
Here we intend to investigate two such approaches for the problem at hand (glioma cells moving away from acidity) from a less rigorous perspective, namely looking into their (formal) macroscopic limits and comparing the numerical results obtained therewith. Concretely, we want to compare our KTE \eqref{eq_meso2} and its parabolic limit with the following simpler KTE for the cell density function $\rho(t,\xbf, \vbf)$:
\begin{align}
\partial_t \rho +\nabla_{\mathbf{x}} \cdot (\mathbf{v}\rho)  &= \mathscr{L} [\lambda(\vbf ,S)]\rho  +\mathscr{P}(M,S)\rho \nonumber\\
&= - \int_V \lambda(\vbf ,S) \frac{q(\hat{\mathbf{ v}}^{\prime})}{\om } \rho (t,\mathbf{ x, v})d\mathbf{v}^{\prime} + \int_V \lambda(\vbf ^{\prime}, S) \frac{q(\hat{\mathbf{ v}})}{\om } \rho(t,\mathbf{ x,v}^\prime)d\mathbf{v}^{\prime} +\mathscr{P}(M,S)\rho \nonumber\\
&= -\lambda(\vbf ,S)\rho (t,\mathbf{ x,v})+\frac{q(\hat{\mathbf{ v}})}{\om }\int_V \lambda(\vbf ',S)\rho (t,\xbf , \vbf^{\prime}) d\mathbf{v}^{\prime} +\mathscr{P}(M,S)\rho  \label{eq_meso4}
\end{align}
and its parabolic limit.
Here the proliferation operator is defined with the same $\mu (M,S)$ as previously used in this section and takes the form
\begin{equation}
\mathscr{P}(M,S) = \mu(M,S)\rho,
\end{equation}
while for the turning rate we set\footnote{notice the difference of sign in the exponent: the cells are supposed to follow the decreasing gradient of signal $S$}
\begin{align}\label{eq:alt-turning_rate}
\lambda(\vbf ,S) &:= \lambda_0 \exp \left( h(S) D_tS\right)\\
&\simeq \lambda_0(1+ h(S)D_tS),
\end{align}
where $D_tS:=S_t+\vbf \cdot \nabla _\xbf S$ is the pathwise gradient of $S$. The coefficient function $h(S)$ is to be chosen later. \\[-2ex]

\noindent
We use again a parabolic scaling $\tilde{t}:= \epsilon^2 t$, $\tilde{x}:= \epsilon x$ and rescale as before the proliferation term by $\epsilon^2$, thus \begin{align}
\epsilon^2 \partial_t \rho + \epsilon \nabla \cdot (\vbf \rho) &= -\lambda_0 (1+h(S)(\eps ^2\partial_tS+\eps \vbf \cdot \nabla S))\rho + \lambda _0\frac{q}{\om } \Big (M+h(S)(\eps^2\partial_tS\ M+\eps \int _V\vbf '\rho (\vbf ')d\vbf '\cdot \nabla S)\Big )\nonumber\\
&+ \epsilon^2 \mathscr{P}(M,S)\rho. \label{eq_meso5}
\end{align}
Performing a Hilbert expansion $\rho =\rho_0+\eps \rho_1+\eps^2 \rho _2+\dots$ and equating the powers of $\epsilon$ yields\\[-2ex]

\noindent
$\mathbf{\epsilon^0}$:
\begin{align*}
0 &= -\lambda_0\rho _0 + \lambda_0  \frac{q(\xbf ,\hat \vbf )}{\om }  M_0, 
\end{align*}
thus
\begin{align}
\rho_0 &= \frac{q(\hat{\mathbf{ v}})}{\om} M_0. \label{eq_rho0}
\end{align}
\noindent
$\mathbf{\epsilon^1}$:
\begin{align*}\label{eq_withoutz_eps1}
\nabla \cdot (\mathbf{ v} \rho_0) &= -\lambda_0\rho_1-\lambda_0h(S)\rho_0\vbf \cdot \nabla S+\lambda_0\frac{q}{\om }\left (M_1+h(S)\int _V\vbf '\rho_0(\vbf ')d\vbf '\cdot \nabla S\right),\notag
\end{align*}
thus by \eqref{eq_rho0} and the assumption of undirected tissue
\begin{align}
\nabla \cdot (\mathbf{ v}\frac{q(\hat{\mathbf{ v}})}{\om} M_0)=-\lambda_0\rho_1-\lambda_0h(S)\frac{q}{\om }M_0\vbf \cdot \nabla S+\lambda_0\frac{q}{\om }M_1.
\end{align}
\noindent
This can be rewritten as
\begin{align}
\mathscr{L}[\lambda_0]\rho_1=\lambda_0h(S)\frac{q}{\om }M_0\vbf \cdot \nabla S+\nabla \cdot (\mathbf{ v}\frac{q(\hat{\mathbf{ v}})}{\om} M_0).
\end{align}
Since the integral of the right hand side w.r.t. $\vbf $ vanishes we can pseudo-invert $\mathscr{L}[\lambda_0]$ as before, to get
\begin{align}
\rho_1 = -h(S)\frac{q}{\om }M_0\vbf \cdot \nabla S-\frac{1}{\lambda_0}\nabla \cdot (\mathbf{ v}\frac{q(\hat{\mathbf{ v}})}{\om} M_0). \label{eq_withoutz_m1}
\end{align}
\noindent
$\mathbf{\epsilon^2}$:
\begin{align}\label{eq_withoutz_eps2}
\partial_t \rho_0 + \nabla \cdot (\mathbf{ v} \rho_1) 
&=\lambda_0(\frac{q}{\om }M_2-\rho_2)+\lambda_0 h(S)\partial_tS\ (\frac{q}{\om }M_0-\rho_0)+\lambda_0h(S)\left(\frac{q}{\om}\int _V\vbf '\rho_1(\vbf ')d\vbf '-\vbf \right)\cdot \nabla S\notag \\
&+\mu (M_0,S)\rho_0.
\end{align}
Integrating (\ref{eq_withoutz_eps2}) w.r.t $\mathbf{ v}$ gives
\begin{align*}
\partial_t M_0 + \nabla \cdot \int_V \mathbf{ v} \rho_1 d\mathbf{v} = \mu(M_0,S)M_0.
\end{align*}
Using \eqref{eq_withoutz_m1} leads to
\begin{align}\label{eq:macro-alternative}
\partial_tM_0=\nabla \nabla :(\mathbb D_T M_0)+\nabla \cdot \left (\lambda _0h(S)\mathbb D_TM_0\nabla S\right )+\mu (M_0,S)M_0,
\end{align}
which differs from \eqref{macro_eq} only by the tactic sensitivity coefficient. This suggests to choose
$$h(S)=\frac{g(S)}{\lambda _0}=\frac{\lambda _1}{\lambda_0}\frac{f'(S)}{k^+S/S_{max}+k^-+\lambda _0}.$$

\noindent
Notice that this function corresponds to the rate of change of $f(S)$ (equilibrium of transmembrane interactions between glioma cells and acidity), scaled by a constant $\lambda _1$ to account for changes in the turning rate per unit of change in $dy^*/dt$, and multiplied by $1/(k^+S/S_{max}+k^-+\lambda_0)$. While a factor $\text{const}\cdot f'(S)$ is also encountered in \cite{othmer2002diffusion}, the denominator obtained here for the tactic sensitivity appears due to the specific choice of the turning rate $\lambda (z)$ in \eqref{eq:lambda}, which also facilitated the upscaling. The whole sensitivity function is tightly related to the first order moment w.r.t. the receptor binding ('activity') variable $z$, recall \eqref{eq:moMo^z} and \eqref{eq_m_1^z}.

\subsection{Hyperbolic scaling}\label{subsec:hyper-limit}

\noindent
In this subsection we investigate the macroscopic limit of \eqref{eq_meso2} in the case where the tissue is directed. In particular, this means that the mean fiber orientation $\mathbb E_q$ is nonzero, as the orientation distribution  $q$ is unsymmetric. We will only treat the case explicitly accounting for the dynamics of activity variables. \\[-2ex]

\noindent
Consider on $L^2_{\frac{q}{\om }}(V)=<q/\om >\oplus <q/\om >^\perp$ the Chapman-Enskog expansion of the cell distribution function $p(t,\xbf ,\vbf ,z)$ in the form
\begin{equation}
p(t,\xbf,\vbf,z) = \bar{p}(t,\xbf ,z) \frac{q}{\om } (\xbf ,\hat \vbf )+ \epsilon p^{\perp} (t,\xbf ,\vbf ,z),
\end{equation}
where $\int_V p^{\perp}(t,\xbf ,\vbf ,z)d\vbf=0$ and $\bar{p}(t,\xbf ,z):= \int_V p(t,\xbf ,\mathbf{ v},z)d\mathbf{ v}$. Then for the moments introduced at the beginning of this Section \ref{sec:upscaling} we have
\begin{align}\label{moments_eq}
\begin{aligned}
m(t,\xbf ,\mathbf{v}) = \int_Z p(t,\xbf ,\mathbf{v},z)dz  &= \int_Z \bar{p} (t,\xbf ,z) \frac{q(\xbf ,\hat{\mathbf{v}})}{\om }dz + \epsilon \int_Z p^{\perp} (t,\xbf ,\mathbf{v},z)dz\\
&=: \frac{q(\xbf , \hat{\mathbf{v}})}{\om } M(t,\xbf) + \epsilon m^{\perp}(t,\xbf ,\mathbf{v})\\
m^z (t,\xbf,\mathbf{v}) = \int_Z z p(t,\xbf ,\mathbf{v},z)dz &= \frac{q(x, \hat{\mathbf{v}})}{w} \int_Z z \bar{p}(t,x,z) dz + \epsilon \int_Z z p^{\perp}(t,\xbf ,\mathbf{ v},z)dz\\
&=:\frac{q(\xbf , \hat{\mathbf{v}})}{\om } M^z(t,\xbf) + \epsilon m_\perp^z(t,\xbf,\vbf).
\end{aligned}
\end{align}
Now we rescale the time and space variables by $\tilde t:=\eps t,\ \tilde \xbf :=\eps \xbf$ and drop again the $\tilde {}$ symbol to simplify the notation.    As before, the proliferation term is scaled by $\eps ^2$. With these, the equations \eqref{eq_int_meso_dz2} and \eqref{eq_z_int_meso_dz} become, respectively:

\begin{align}
\epsilon \partial_t m + \epsilon\nabla_{\mathbf{x}} \cdot (\mathbf{v}m) &= -\lambda_0m + \lambda_1m^z + \lambda_0 \frac{q}{\om }M - \lambda_1 \frac{q}{\om }M^z + \epsilon^2 \mu(M)\left(1-\frac{S}{S_{max}}\right)m
\end{align}
and
\begin{align}
\epsilon \partial_t m^z + \epsilon \nabla_{\mathbf{x}}\cdot(\mathbf{v}m^z) = -(k^+ S/S_{max} + k^-)m^z &-  {\epsilon} f^{\prime}(S)\mathbf{v} \cdot \nabla S\ m -\lambda_0 m^z + \lambda_0 \frac{q}{\om } M^z \nonumber\\
& + \epsilon^2 \mu(M)\left(1-\frac{S}{S_{max}}\right)\int_Z \int_Z z\chi(z,z^{\prime})p(z^{\prime})dz^{\prime}dz.
\end{align}
Using (\ref{moments_eq}) we write
\begin{align}
\epsilon \partial_t \left(\frac{q}{\om } M + \epsilon m^{\perp}\right) + \epsilon \nabla \cdot (\mathbf{v}\left(\frac{q}{\om } M + \epsilon m^{\perp}\right))
&= -\lambda_0\left(\frac{q}{\om } M + \epsilon m^{\perp}\right) + \lambda_1 \left(\frac{q}{\om } M^z +\epsilon m_\perp^z \right)  \nonumber\\
&+ \lambda_0 \frac{q}{\om }M - \lambda_1 \frac{q}{\om }M^z + \epsilon^2 \mu(M)\left(1 -\frac{S}{S_{max}}\right)m\\
\epsilon \partial_t \left(\frac{q}{\om } M^z + \epsilon m_\perp^z\right) + \epsilon \nabla\cdot(\mathbf{v} \left(\frac{q}{\om } M^z + \epsilon m_\perp^z\right)) &= -(k^+ S/S_{max} + k^-)\left(\frac{q}{\om } M^z + \epsilon m_\perp^z\right) \nonumber\\
&- {\epsilon} f^{\prime}(S) \mathbf{v} \cdot \nabla S \left(\frac{q}{\om } M + \epsilon m^{\perp}\right) - \lambda_0 \left(\frac{q}{\om } M^z + \epsilon m_\perp^z\right)\nonumber\\
& + \lambda_0 \frac{q}{\om } M^z \notag \\
&+ \epsilon^2 \mu(M)\left(1-\frac{S}{S_{max}}\right)\int_Z \int_Z z\chi(z,z^{\prime})p(z^{\prime})dz^{\prime}dz.
\end{align}
Since $q$ is independent of time, these equations imply
\begin{align}\label{meso_2_eq}
\frac{q}{\om } \partial_tM + \epsilon \partial_t m^{\perp} + \nabla \cdot (\mathbf{v}M\frac{q}{\om }) + \epsilon \nabla \cdot (\mathbf{v} m^{\perp}) &= - \lambda_0 m^{\perp} + \lambda_1 m_\perp^z +\epsilon \mu(M)\left(1-\frac{S}{S_{max}}\right)m\\
\epsilon \frac{q}{\om } \partial_tM^z + \epsilon^2 \partial_t m_\perp^z + \epsilon \nabla \cdot (\mathbf{v} M^z \frac{q}{\om }) + \epsilon^2 \nabla \cdot (\mathbf{v} m_\perp^z) &= -(k^+ S/S_{max}+k^-)\frac{q}{\om }M^z -\epsilon (k^+S/S_{max} +k^-)m_\perp^z \nonumber\\&-\epsilon f^{\prime} (S) \mathbf{v} \cdot \nabla S \frac{q}{\om }M -\epsilon^2 f^{\prime}(S) \mathbf{v} \cdot \nabla S\ m^{\perp} - \epsilon \lambda_0 m_\perp^z \nonumber\\& + \epsilon^2 \mu(M)\left(1-\frac{S}{S_{max}}\right)\int_Z \int_Z z\chi(z,z^{\prime})p(z^{\prime})dz^{\prime}dz. \label{eq_meso_Mz}
\end{align}
Integrating (\ref{meso_2_eq}) w.r.t. $\mathbf{v}$ gives
\begin{equation}\label{macro_1}
\partial_t M + \nabla \cdot (\tilde {\mathbb{E}}_q M) + \epsilon \nabla \cdot \int_V \mathbf{v} m^{\perp} d\mathbf{v} = \epsilon \mu(M)\left(1-\frac{S}{S_{max}}\right) M,
\end{equation}
where we used the notation $\tilde {\mathbb{E}}_q(\xbf):=\int _V\vbf \frac{q}{\om}(\xbf ,\hat \vbf )d\vbf =s\mathbb{E}_q$. \\[-2ex]

\noindent
From (\ref{eq_meso_Mz}) we get at leading order\\[-2ex]
\begin{equation*}
-(k^+ S/S_{max} +k^-)\frac{q}{\om }M^z = 0\qquad \Rightarrow \quad M^z = 0.
\end{equation*}
\noindent
Plugging this in (\ref{eq_meso_Mz}), we obtain (again at leading order)
\begin{equation*}
0 = -(k^+ S/S_{max} + k^-)m_{\perp}^z - f^{\prime}(S) \mathbf{v} \cdot \nabla S \frac{q}{\om } M - \lambda_0 m_{\perp}^z,
\end{equation*}
whence
\begin{equation}\label{eq_m_z_perp}
m_\perp^z = \frac{-f^{\prime}(S) \mathbf{v} \cdot \nabla S}{k^+ S/S_{max} + k^- + \lambda_0} \frac{q}{\om } M.
\end{equation}
From (\ref{macro_1}):
\begin{equation}
\partial_tM = \epsilon \mu(M)\left(1-\frac{S}{S_{max}}\right) M - \nabla \cdot (\tilde {\mathbb{E}}_q M) - \epsilon \nabla \cdot \int_V \mathbf{v} m^{\perp}d\mathbf{v}.
\end{equation}
Plugging this into (\ref{meso_2_eq}) we get (at leading order)
\begin{align}
\mathscr{L}[\lambda_0]m^{\perp}=-\frac{q}{\om }\nabla \cdot \left(\tilde {\mathbb{E}}_q M\right)+\nabla \cdot (\vbf M\frac{q}{\om })-\lambda _1m_\perp^z.
\end{align}
Since the right hand side vanishes when integrated w.r.t. $\vbf $, we can pseudo-invert $\mathscr{L}[\lambda_0]$ and use \eqref{eq_m_z_perp} to get
\begin{align}
m^{\perp} &= \frac{-1}{\lambda_0} \left( \nabla \cdot \left( \mathbf{v} M \frac{q}{\om }\right) - \frac{q}{\om }  \nabla \cdot \left( \tilde{ \mathbb{E}}_q M\right) - \lambda_1  \frac{-f^{\prime}(S) \mathbf{v} \cdot \nabla S}{k^+ S/S_{max} + k^- + \lambda_0} \frac{q}{\om } M  \right), 
\end{align}
hence
\begin{align}
\nabla \cdot \int_V \mathbf{v} m^{\perp} dv = -\nabla \nabla :(\mathbb D_TM)+\nabla \cdot \left (\frac{1}{\lambda_0}\tilde {\mathbb E}_q\nabla \cdot (\tilde {\mathbb E}_qM)\right )-\lambda _1\nabla \cdot \left (\frac{f^{\prime} (S) }{k^+ S/S_{max} + k^-+\lambda _0}\mathbb D_TM\nabla S\right ),
\end{align}
so that (\ref{macro_1}) becomes
\begin{align}\label{eq:macro-hyp}
\partial_tM + \nabla \cdot \left(s\mathbb{E}_q M\right) & =\eps \nabla \nabla :(\mathbb D_TM)-\eps \nabla \cdot \left (\frac{s^2}{\lambda_0}\mathbb E_q\nabla \cdot (\mathbb E_qM)\right )+\eps \nabla \cdot \left (g(S)\mathbb D_TM\nabla S\right )\nonumber \\
& +  \epsilon \mu(M)\left(1-\frac{S}{S_{max}}\right) M.
\end{align}
Comparing this with the parabolic limit obtained in \eqref{macro_eq} we observe that we obtain the same form for the (myopic) diffusion, repellent pH-taxis, and proliferation terms, but here they are $\eps$-corrections of the leading transport terms - together with the new advection which drives cells with velocity $\frac{\eps}{\lambda_0}\mathbb E_q\nabla \cdot \mathbb E_q$ in the direction of the dominating drift.


\section{Numerical simulations}\label{sec:numerics}

\subsection{Parameters and coefficient functions}\label{subsec:params}

We assume a logistic type growth of the tumor cells and choose
\begin{equation}\label{eq_mu}
\mu(\mathbf{ x},M) = \mu_0 \left(1-\frac{M}{M_{max}}\right),
\end{equation}
where $\mu_0$ is the growth rate and $M_{max}$ is the carrying capacity of tumor cells.

\subsection*{Nondimensionalization}
Considering the following nondimensional quantities:
\begin{align*}
&\tilde{M} := \frac{M_0}{M_{max}},\qquad \tilde{S} := \frac{S}{S_{max}},
\qquad \tilde{t} := \frac{\beta M_{max}}{S_{max}}t,\qquad \tilde{x} := x\sqrt{\frac{\beta M_{max}}{D_s S_{max}}}, \\
&\tilde{\mathbb{D}}_T := \frac{\mathbb{D}_T}{D_s}, \qquad \tilde{\alpha} := \frac{\alpha}{\beta} \frac{S_{max}}{M_{max}}, \qquad
\tilde \lambda _1:=\frac{\lambda _1}{k^+}, \qquad \tilde \lambda _0:=\frac{\lambda _0}{k^+},\\
&\tilde{g}(\tilde S) = \frac{\tilde \lambda_1k_D}{(\tilde S+k_D)^2(\tilde S+k_D+\tilde \lambda_0)},\qquad   
\tilde{\mu}_0 = \frac{\mu_0 S_{max}}{\beta M_{max}},
\end{align*}
and dropping the tildes for simplicity of notation, we get the following nondimensionalized system:
\begin{align}
\partial_t M &= \nabla \nabla : (\mathbb{D}_T M) +\nabla \cdot (g(S)M\mathbb{D}_T \nabla S)
+\mu_0 \left(1-M\right) \left(1-S\right)M\label{macro_eq_non}\\
\partial_tS &= \Delta S + M - \alpha S\label{macro_eq_non_S}
\end{align}
\begin{table}[h]
	\centering
\begin{tabular}{p{2cm} p{7cm} p{4cm} p{2.8cm}}
	\toprule[1.5pt]
	Parameter& Meaning & Value & Reference\\[0.75ex]
	\hline \\[-1.5ex]
	$M_{max}$   & glioma carrying capacity    &$10^5-10^8 \text{ cells/mm}^3$&
	\cite{Banerjee2015,HATHOUT2014,Rockne2010}\\  
	$S_{max}$ &  acidity threshold for cancer cell death   & $10^{-6.4} \text{ mol/l}$  & \cite{Webb2011} \\
	$s$  & speed of glioma cells & $2.8\cdot 10^{-6}\ \text{mm/s}$&  
	 estimated, \cite{Prag}\\
	$\lambda_0$&   turning frequency coefficient   & $0.1\ s^{-1}$&  \cite{engwer2015glioma,sidani2007cofilin}\\
	$\lambda_1$& turning frequency coefficient   & $0.01 \ s^{-1}$ &\cite{engwer2015glioma,sidani2007cofilin}\\
	$k^+$& \begin{minipage}{6cm}interaction rate tumor cells-protons\end{minipage}  & $0.004\ s^{-1}$&\cite{Lauffenburger}\\
	$k^-$ & detachment rate  & $0.01\ s^{-1}$  &\cite{Lauffenburger} \\
	$\beta$ & proton production rate & $10^{-9}$ mol\ /(mm$^3$s)  & estimated, \cite{Martin2} \\
	$\alpha$ & proton removal rate  & $10^{-11}$ /s& estimated\\
	$D_s$ & $H^+$ diffusion coefficient & \begin{minipage}{6cm}$5\cdot 10^{-4}-6\cdot 10^{-2}$ mm$^2$/s\end{minipage} & \cite{gatenby1996reaction} \cite{schornack2003contributions} \\
	$\mu_0$ & glioma growth rate & 0.2/day & \cite{stein2007mathematical} \cite{eikenberry2009virtual} \\
	\bottomrule[1.5pt]
\end{tabular}
\caption{Parameters (dimensional quantities)}\label{table} 
\end{table}
\subsection{Description of tissue}

The structure of brain tissue can be assessed by way of biomedical imaging, e.g. diffusion tensor imaging (DTI) which provides for each voxel the water diffusion tensor $\mathbb{D}_w$. The corresponding resolution is, however, too low and does not deliver information about the (orientation) distribution of tissue fibers below the size of a voxel (ca. 1 mm$^3$). For more details we refer e.g. to \cite{engwer2015glioma,PH13} and references therein. As mentioned in Section \ref{sec:intro}, pseudopalisades are comparatively small structures with a medium width of $200-400\ \mu m$. Thus, in order to investigate the possible effect of (local) tissue anisotropy on these patterns we will create a synthetic DTI data set which will allow to compute the tumor diffusion tensor $\mathbb{D}_T$ in the space points of such a narrow region. To this aim we proceed as in \cite{PH13} and consider the water diffusion tensor
\begin{equation}
\mathbb D_{w}(x,y) = \begin{pmatrix}
0.5-d(x,y) & 0 \\
0 & 0.5+d(x,y)
\end{pmatrix}
\end{equation}
where $d(x,y) = 0.25e^{-0.005(x-450)^2} - 0.25e^{-0.005(y-450)^2}$.
For the fiber distribution function, we consider a mixture between uniform and von Mises-Fisher distributions, as follows:
\begin{equation}\label{eq_q}
q(\mathbf{x},\thetabf) = \frac{\delta}{2\pi} + (1-\delta)\left(\frac{1}{2\pi I_0(k(\mathbf{x}))}\right) \frac{e^{k(\mathbf{x})\varphi_1(\mathbf{x})\cdot \thetabf} + e^{-k(\mathbf{x})\varphi_1(\mathbf{x})\cdot \thetabf}}{2}
\end{equation}
Here, $\delta \in \left[0,1\right]$ is a weighting coefficient, $\varphi_1$ is the eigenvector corresponding to the leading eigenvalue of $\mathbb D_w(\mathbf{ x})$ and $I_0$ is the modified Bessel function of first kind of order $0$. Also, $\thetabf = \left(\cos \xi , \sin \xi \right)$ for $\xi \in \left[0,2\pi\right]$,  and $k(\mathbf{ x}$) = $\kappa$FA$(\mathbf{ x})$, where
FA$(\mathbf{ x})$ denotes the fractional anisotropy: in 2D it has the form \cite{PH13} \begin{equation*}
FA(\mathbf{ x}) = \frac{|\lambda_1 -\lambda_2|}{\sqrt{\lambda_1^2 + \lambda_2^2}},
\end{equation*}
with $\lambda_i$ ($i=1,2$) denoting the eigenvalues of $\mathbb D_w(\xbf )$.
The parameter $\kappa \geq 0$ characterizes the sensitivity of cells towards orientation of tissue fibers. For perfectly aligned tissue (i.e.,  maximum anisotropy), $FA(\xbf)=1$ and $k(\mathbf{ x}) = \kappa$. Taking $\kappa =0$ means, however, that the cells are insensitive to even such alignment and the distribution in \ref{eq_q} becomes a uniform one. Taking $\delta=1$ has the same effect. \\[-2ex]

\noindent
For the model deduced by hyperbolic scaling in Subsection \ref{subsec:hyper-limit}, we consider for the orientation distribution of tissue fibers the following combination of two unsymmetric unimodal von Mises distributions:
\begin{equation}\label{eq_q_h}
q_h(\mathbf{ x, \theta}) = \frac{\delta}{2\pi I_0 (k_h(\mathbf{ x})) }e^{k_h(\mathbf{ x}) \gamma \cdot \theta } + \frac{1-\delta}{2\pi I_0(k(\mathbf{ x}))}e^{k(\mathbf{ x}) \varphi_1 \cdot \theta },
\end{equation}
where $k_h(\mathbf{ x}) = 0.05e^{-10^{-6} \left((x-450)^2 + (y-450)^2 \right)}$, $\gamma = \left(1/\sqrt{2}, 1/\sqrt{2} \right) ^T$ and the rest of parameters are the same as in (\ref{eq_q}). The first summand, similar to the choice in \cite{hillen2013transport}, generates an orientation along the diagonal $\gamma$, while the second leads to alignment along the positive $x$ and $y$ directions. Due to $k_h(\mathbf{ x})$, the strength of diagonal orientation of tissues decreases from the chosen center $(450,450)$.\\[-2ex]

\noindent
The macroscopic tissue density $Q$ is obtained in the same way as in \cite{engwer2015effective} by using the free path length from the diffusivity obtained from the data, more precisely from the water diffusion tensor. In that approach the occupied volume is obtained upon computing a characteristic (diffusion) length $l_c=\sqrt{tr (\mathbb D_w) t_c}$, where $t_c$ is the characteristic (diffusion) time. The latter is determined by assuming the underlying stochastic process behind water diffusion tensor measurements to be the Brownian motion and considering the expected exit time from the minimal ball with radius $r$ containing a square with side length $h$ as smallest unit in our grid. Therefore, the tissue density $Q$ (area fraction occupied by tissue) is:
 \begin{align}
 Q = 1 -  \frac{l_c^2}{h^2},
 \end{align}
where
\begin{align*}
l_c = \sqrt{\frac{tr\left( \mathbb{D}_w \right) h^2}{4 l_1}}
\end{align*}
with $l_1$ being the largest eigenvalue of $\mathbb{D}_w$.


\subsection{Numerical experiments}\label{subsec:simulations}
\begin{figure}[!htbp]
	\subfigure[][]{\includegraphics[width=0.45\textwidth,height=5.5cm]{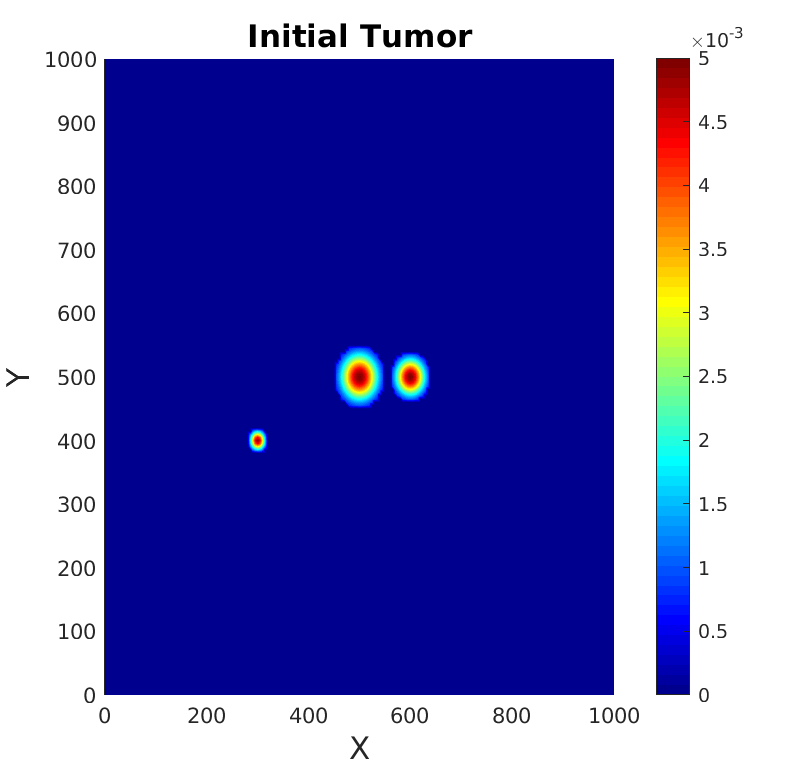}\label{fig:IC-tumor}}\quad \subfigure[][]{\includegraphics[width=0.45\textwidth,height=5.5cm]{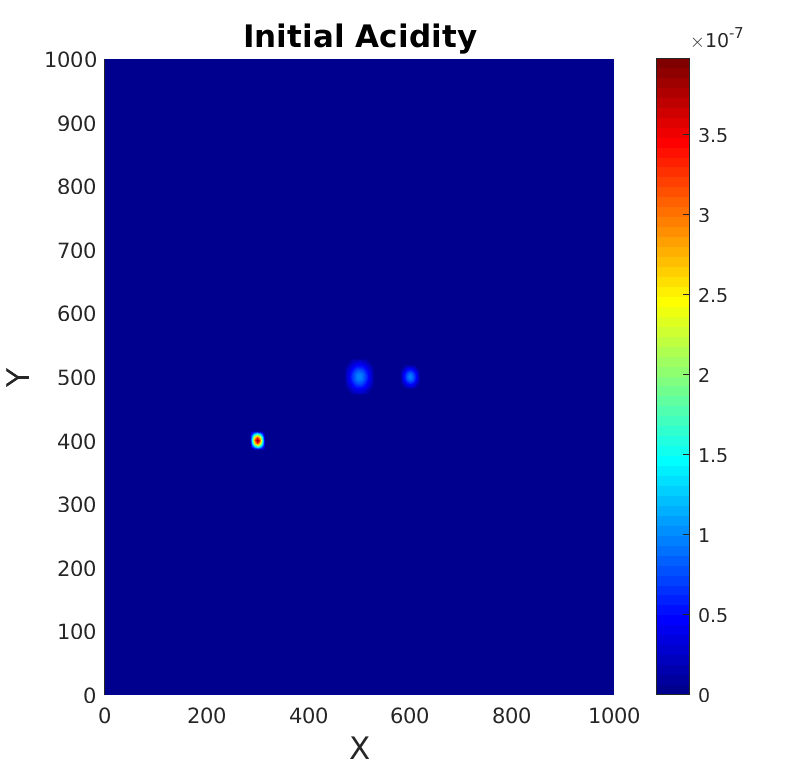}\label{fig:IC-acid}}	\\
	\subfigure[][]{\includegraphics[width=0.45\textwidth]{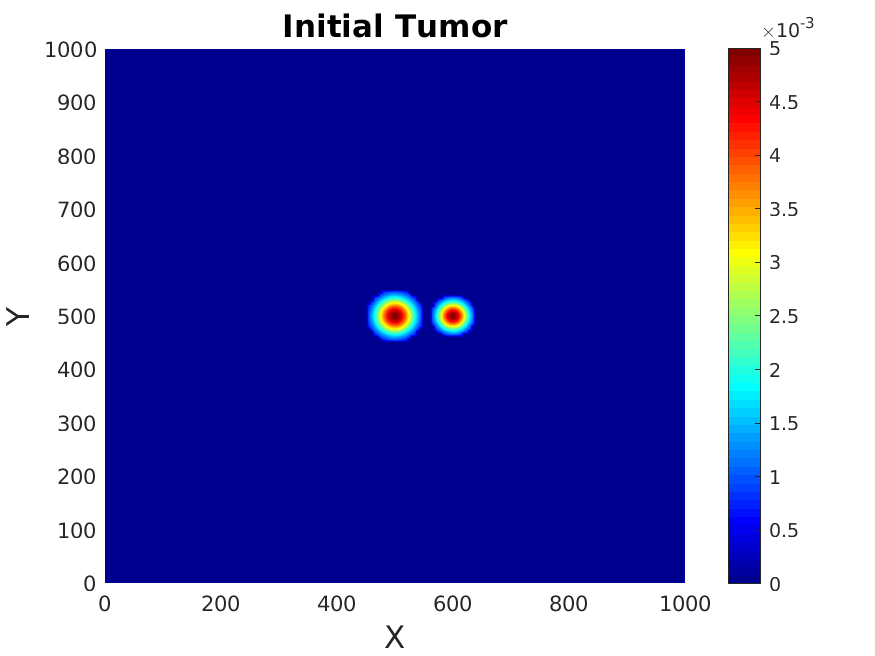}\label{fig:IC-tumor-neu}}\quad \subfigure[][]{\includegraphics[width=0.45\textwidth]{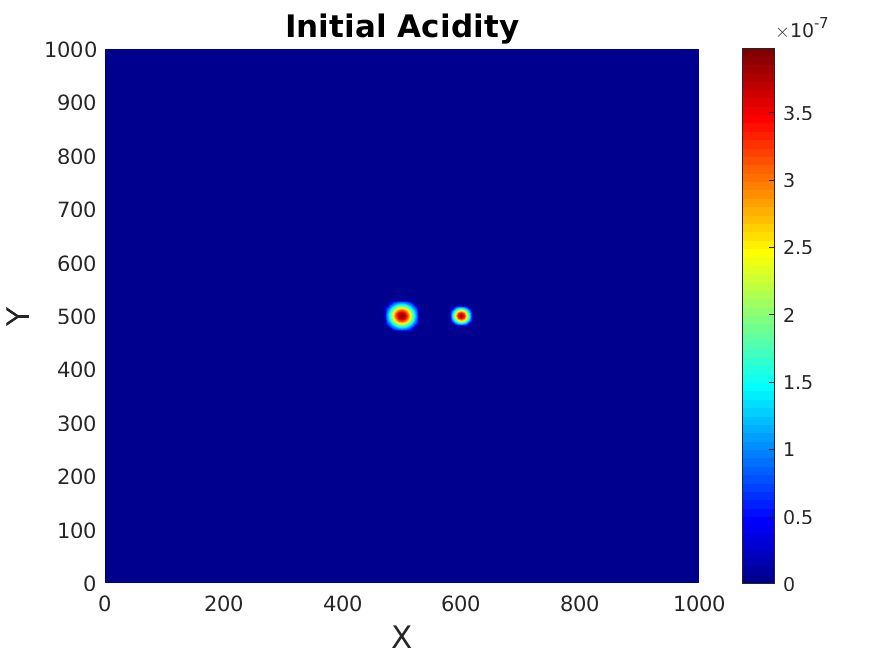}\label{fig:IC-acid-neu}}	
	\caption[]{Initial conditions. Upper row: set \eqref{eq:ICs} for tumor cell density (\ref{fig:IC-tumor}) and acidity distribution (\ref{fig:IC-acid}), lower row: set \eqref{eq:ICs-neu} for tumor cell density (\ref{fig:IC-tumor-neu}) and acidity distribution (\ref{fig:IC-acid-neu})}\label{fig:ICs}
\end{figure}

The system \eqref{macro_eq_non}, \eqref{macro_eq_non_S} is solved numerically on a square domain $[0,1000]\times [0,1000]$ (in $\mu m$) using appropriate finite difference methods for spatial discretization and an IMEX method for time discretization, where the diffusion part is handled implicitly, while the advection and source terms are treated explicitly. We use a standard central difference scheme (5-point stencil) for the acidity diffusion. To avoid numerical instability \cite{mosayebi2010stability} due to negative values in the stencil obtained from discretization of mixed derivative terms in the myopic tumor diffusion, we use the non-negative discretization scheme  proposed in \cite{weickert1998anisotropic} instead of the standard one. Thereby, the derivatives are calculated in newly chosen directions (diagonal directions of the $3 \times 3$-stencil in 2D) in addition to the standard x,y-directions and mixed term derivatives are replaced by directional derivatives. To discretise the advection terms, we use a first order upwind scheme for the parabolic scaling model, while for the system obtained via hyperbolic scaling we employ a second order upwind scheme with Van Leer flux limiter. Implicit and explicit Euler methods are used for IMEX time discretization. The systems are solved with no-flux boundary conditions and the following sets of initial conditions as illustrated in Figures \ref{fig:IC-tumor} and \ref{fig:IC-acid}:
\begin{subequations}\label{eq:ICs}
\begin{align}
M(\mathbf{ x}, 0) &= 0.005 \left( e^{\frac{-(x-500)^2- (y-500)^2}{2(25)^2}}    + e^{\frac{-(x-600)^2- (y-500)^2}{2 (20)^2}} + e^{\frac{-(x-300)^2- (y-400)^2}{2(10)^2}}\right)\label{eq:IC-M}\\
S(\mathbf{ x},0) & = 10^{-7}e^{\frac{-(x-500)^2- (y-500)^2}{2 (15)^2}} + 10^{-7}e^{\frac{-(x-600)^2- (y-500)^2}{2(10)^2}} + 10^{-6.4} e^{\frac{-(x-300)^2- (y-400)^2}{2(7.5)^2}}.\label{eq:IC-S}
\end{align}
\end{subequations}
and \ref{fig:IC-tumor-neu} and \ref{fig:IC-acid-neu}, respectively:
\begin{subequations}\label{eq:ICs-neu}
	\begin{align}
	M(\mathbf{ x}, 0) &= 0.005 \left( e^{\frac{-(x-500)^2 -(y-500)^2}{2(25)^2}}    + e^{\frac{-(x-600)^2 -(y-500)^2}{2 (20)^2}} \right)\label{eq:IC-M-neu}\\
	S(\mathbf{ x},0) & = 10^{-6.4}e^{\frac{-(x-500)^2 -(y-500)^2}{2 (15)^2}} + 10^{-6.4}e^{\frac{-(x-600)^2- (y-500)^2}{2(10)^2}} .\label{eq:IC-S-neu}
	\end{align}
\end{subequations}


\begin{experiment}\textbf{Fully isotropic tissue}\label{exp:FAis0}

\noindent	
We begin by considering a fully isotropic tissue, i.e. taking $\delta =1$ in 	\eqref{eq_q}. The corresponding fractional anisotropy is everywhere $FA=0$, and the macroscopic tissue density $Q$ is shown in Figure \ref{fig:Q-exp_anisotropic}.

\begin{figure}[!htbp]
	\subfigure[][]{\includegraphics[width=0.45\textwidth]{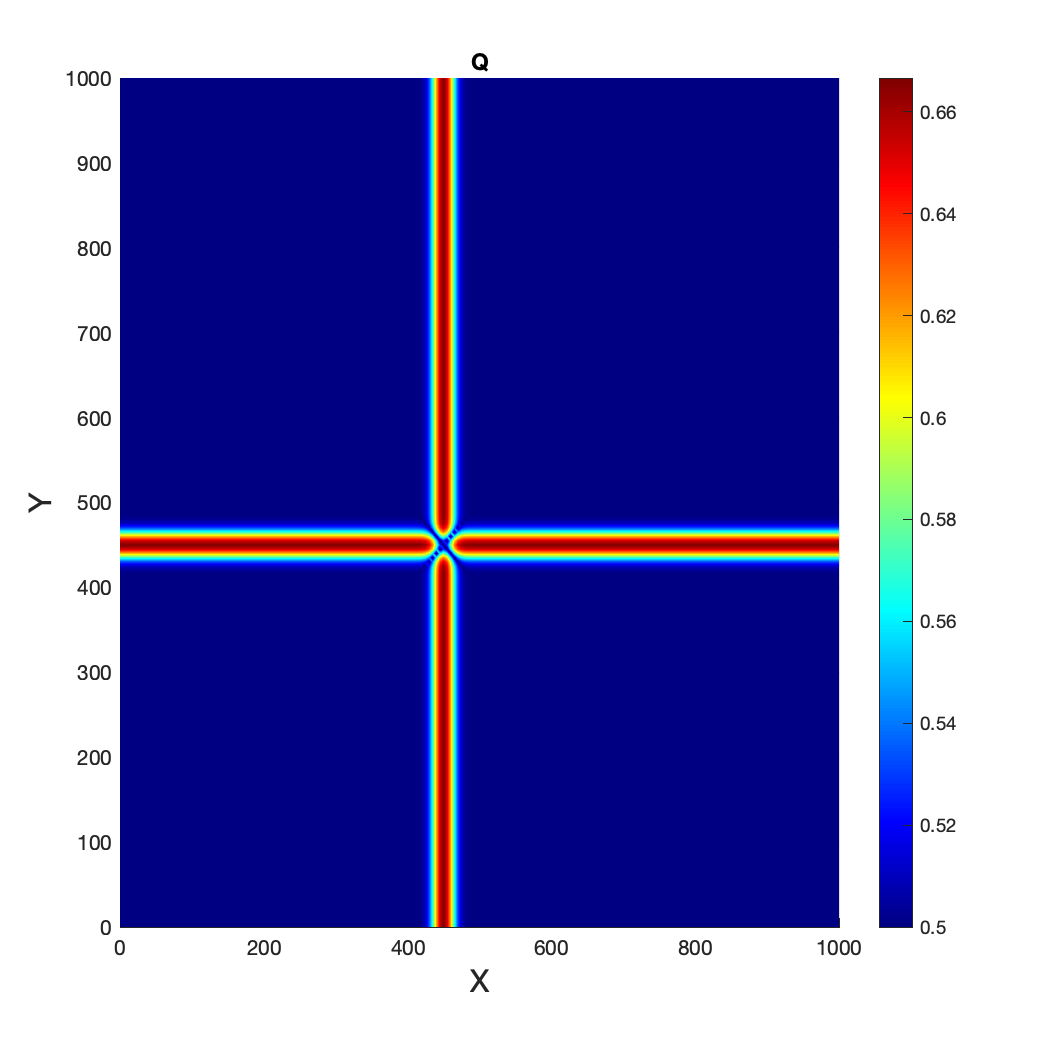}\label{fig:Q-exp_anisotropic}}\qquad 
	\subfigure[][]{\includegraphics[width=0.45\textwidth]{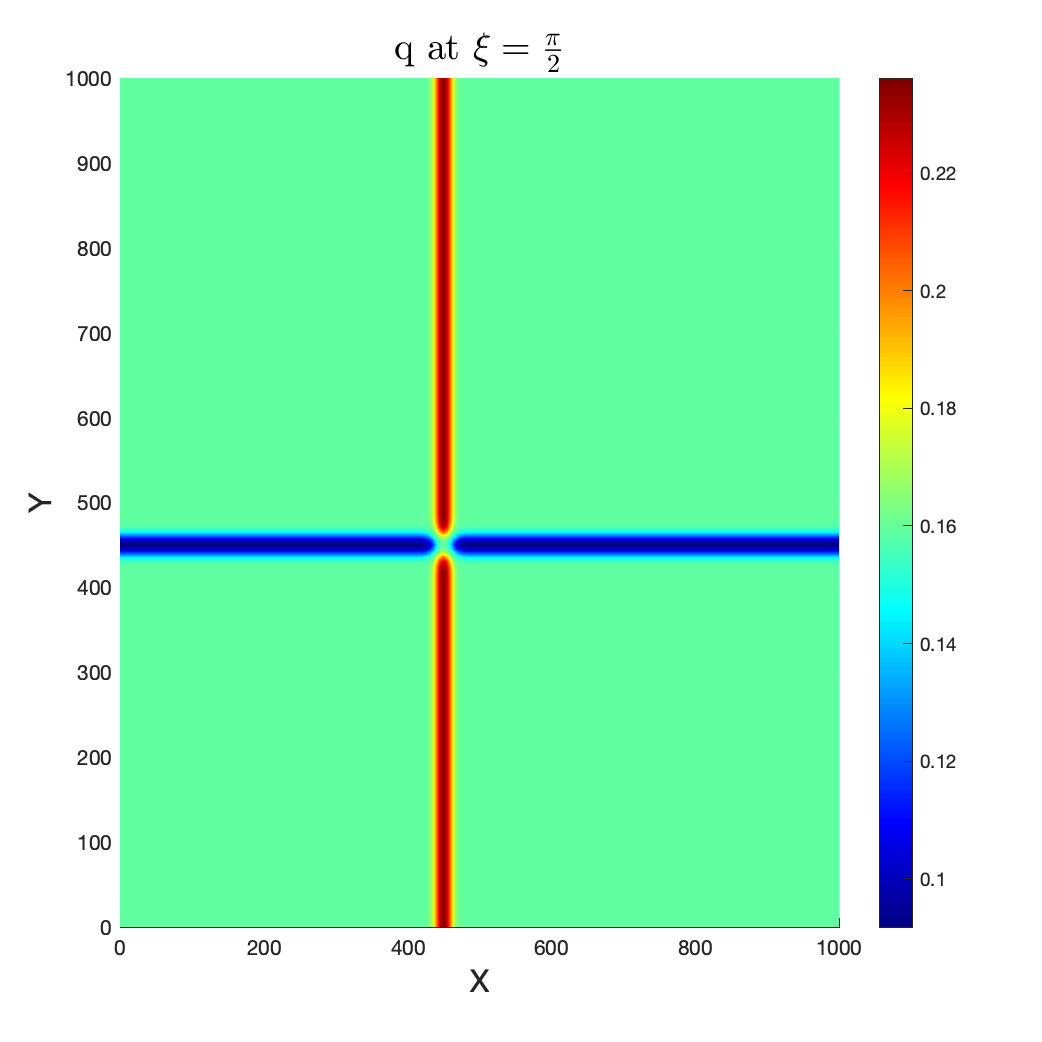}\label{fig:q_set2}}
		\caption[]{\ref{fig:Q-exp_anisotropic}: Macroscopic tissue density (Experiments \ref{exp:FAis0},  \ref{exp:kFAis3}) and \ref{fig:q_set2}: mesoscopic tissue distribution for Experiment \ref{exp:kFAis3}, for a given fiber direction.}
	\label{fig:tissue-exp_anisotropic}	
\end{figure}

\begin{figure}[!htbp]
	\subfigure[][]{\includegraphics[width=0.31\textwidth]{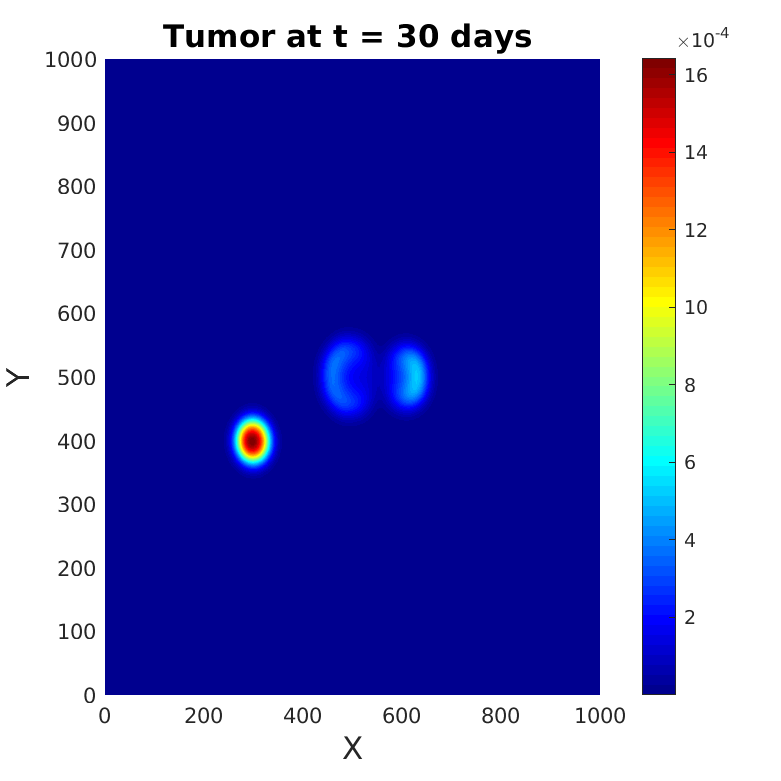}\label{fig:tumor_t30-exp_isotropic}}\quad \subfigure[][]{\includegraphics[width=0.31\textwidth]{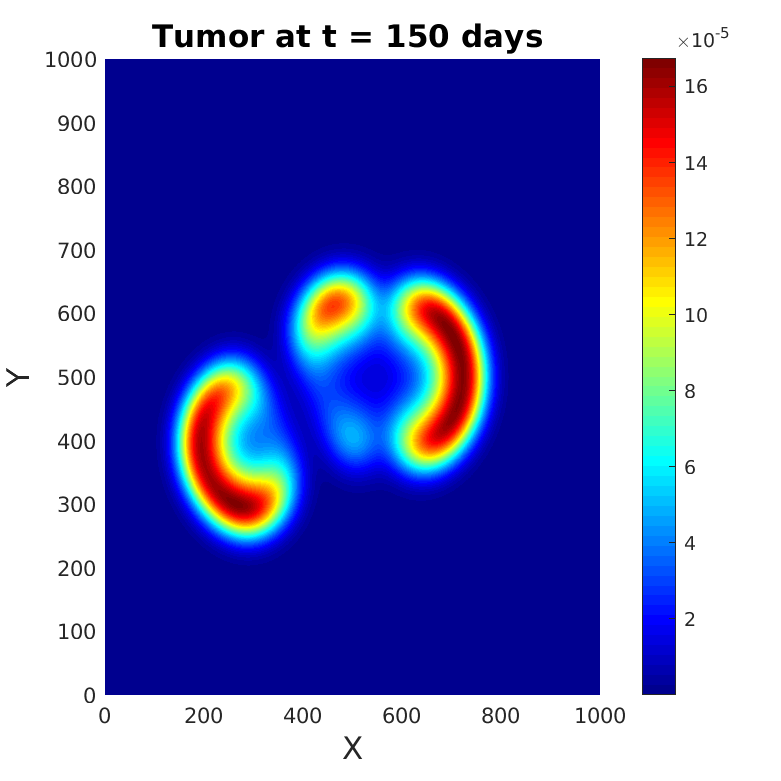}\label{fig:tumor_t150-exp_isotropic}}\quad \subfigure[][]{\includegraphics[width=0.31\textwidth]{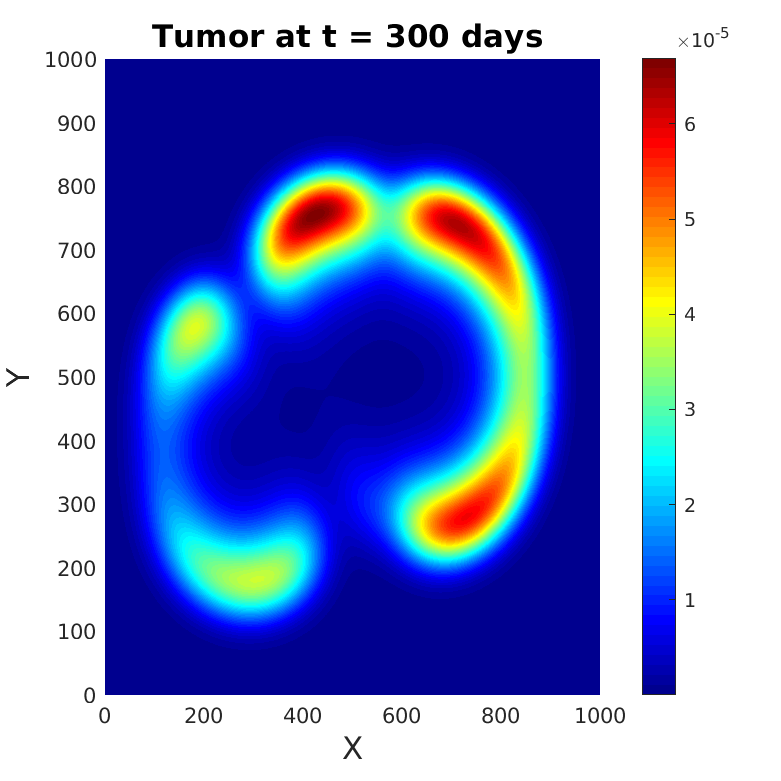}\label{fig:tumor_t300-exp_isotropic}}\\
	\subfigure[][]{\includegraphics[width=0.31\textwidth]{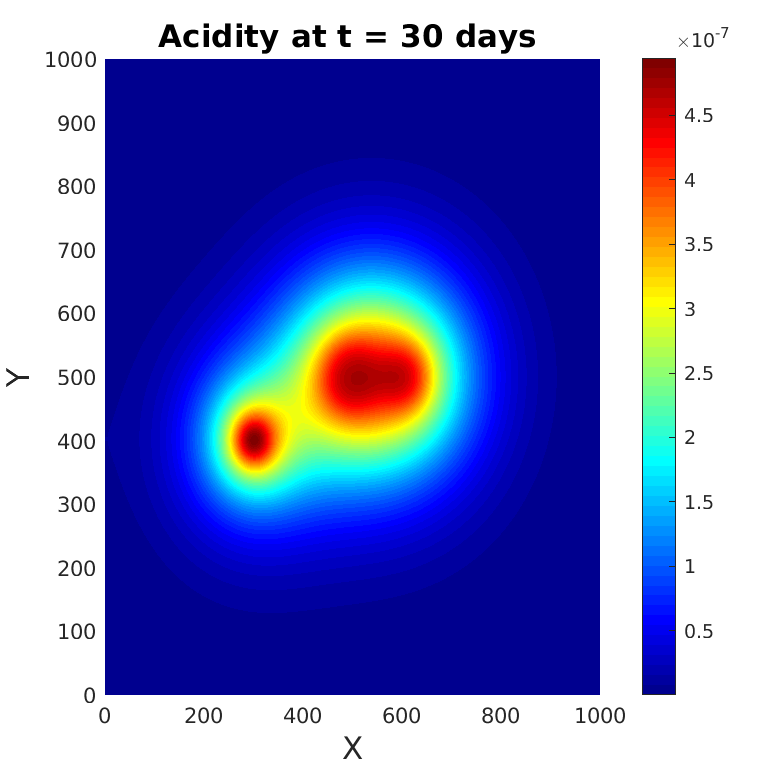}\label{fig:acid_t30-exp_isotropic}}\quad \subfigure[][]{\includegraphics[width=0.31\textwidth]{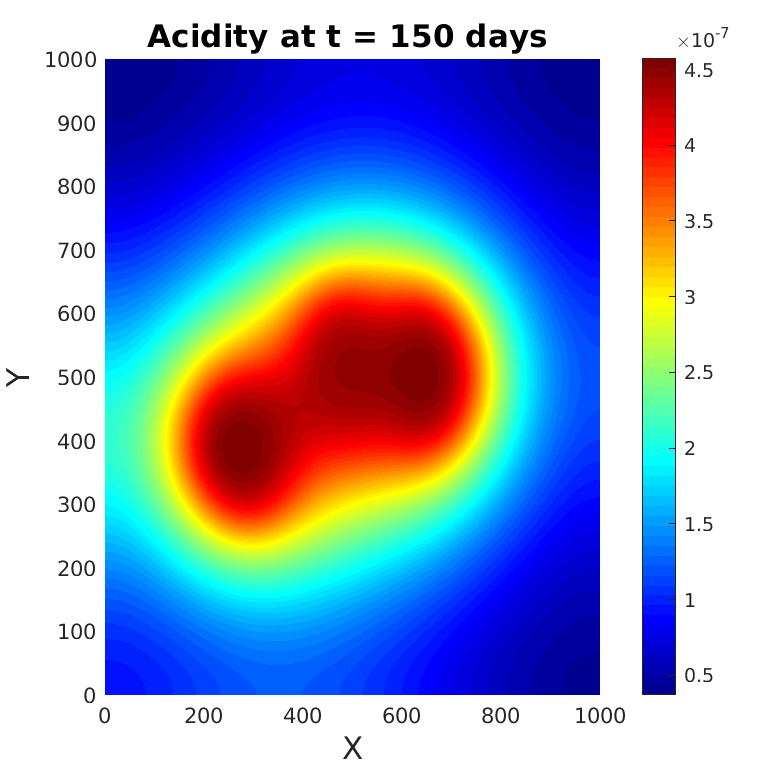}\label{fig:acid_t150-exp_isotropic}}\quad \subfigure[][]{\includegraphics[width=0.31\textwidth]{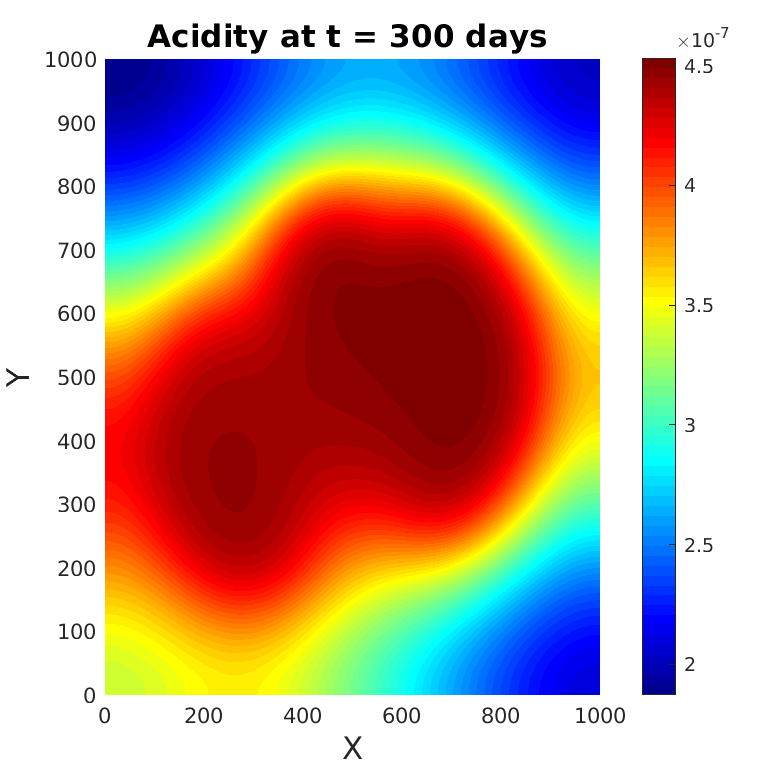}\label{fig:acid_t300-exp_isotropic}}\\	
	\caption[]{Tumor (upper row) and acidity (lower row) at several times for Experiment \ref{exp:FAis0} and initial conditions \eqref{eq:ICs}.}
	\label{fig:tumor-acid-exp_isotropic}
\end{figure}	

\begin{figure}[!htbp]
	\subfigure[][]{\includegraphics[width=0.31\textwidth]{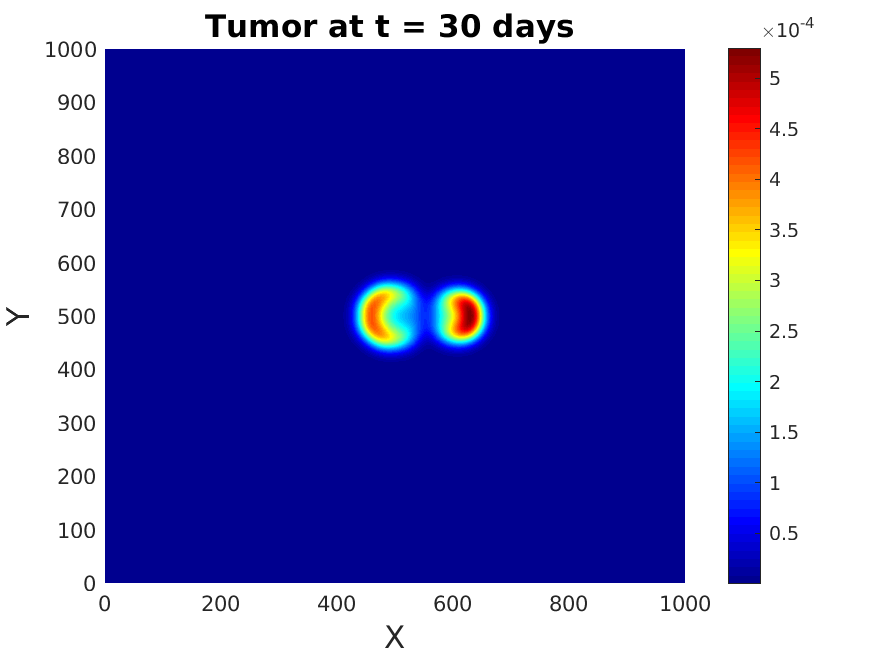}\label{fig:tumor_t30-exp_isotropic-neu}}\quad \subfigure[][]{\includegraphics[width=0.31\textwidth]{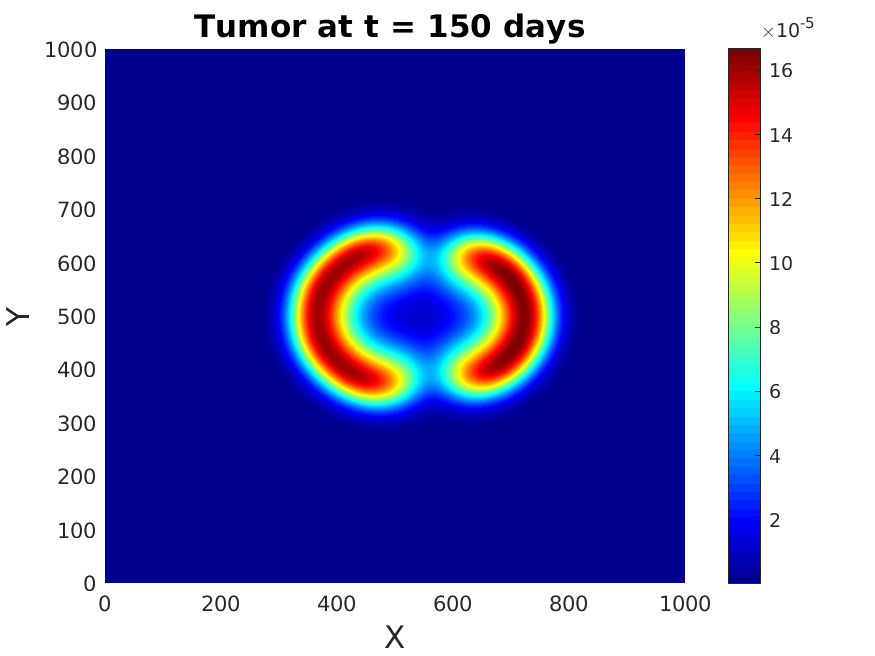}\label{fig:tumor_t150-exp_isotropic-neu}}\quad \subfigure[][]{\includegraphics[width=0.31\textwidth]{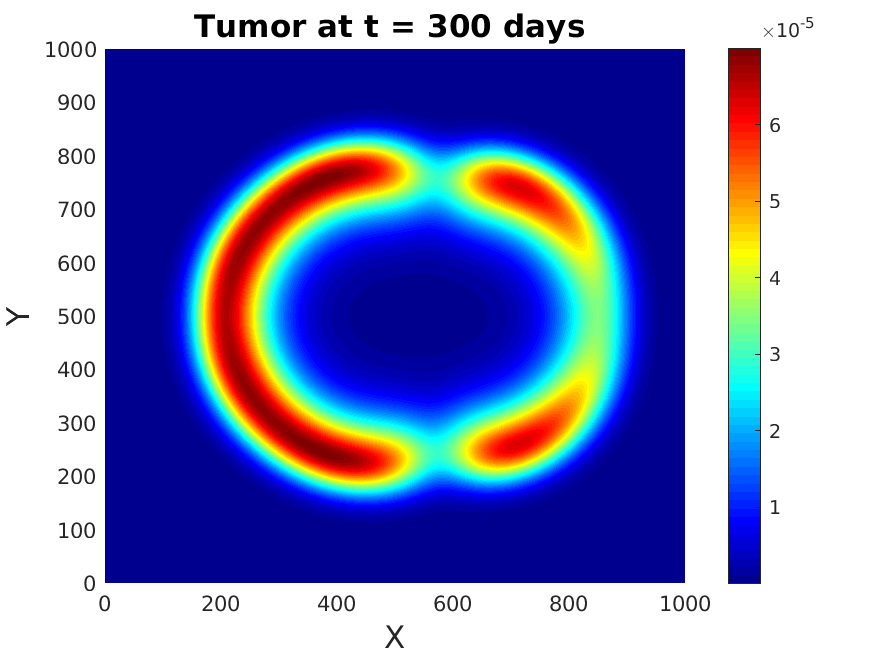}\label{fig:tumor_t300-exp_isotropic-neu}}\\
	\subfigure[][]{\includegraphics[width=0.31\textwidth]{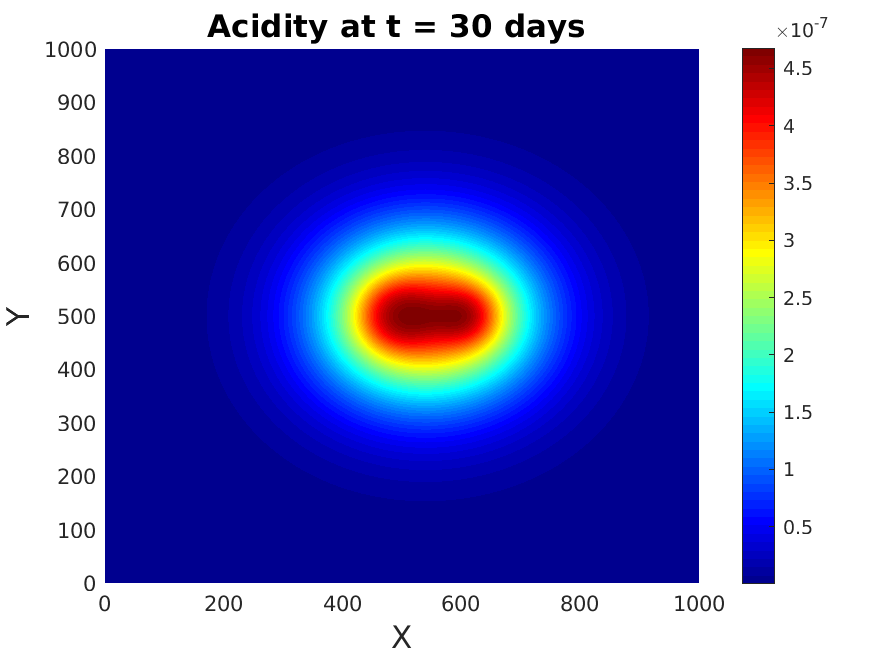}\label{fig:acid_t30-exp_isotropic-neu}}\quad \subfigure[][]{\includegraphics[width=0.31\textwidth]{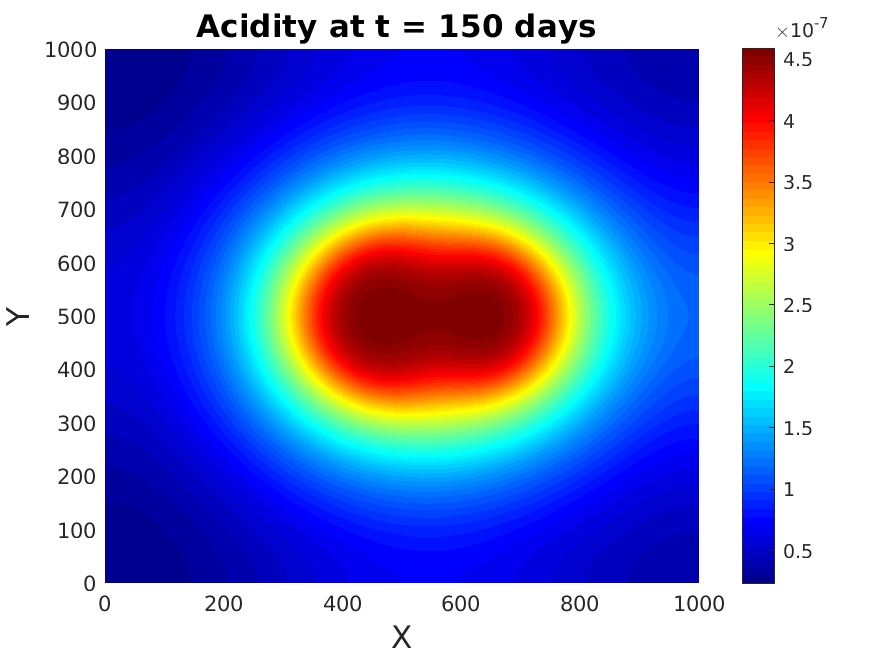}\label{fig:acid_t150-exp_isotropic-neu}}\quad \subfigure[][]{\includegraphics[width=0.31\textwidth]{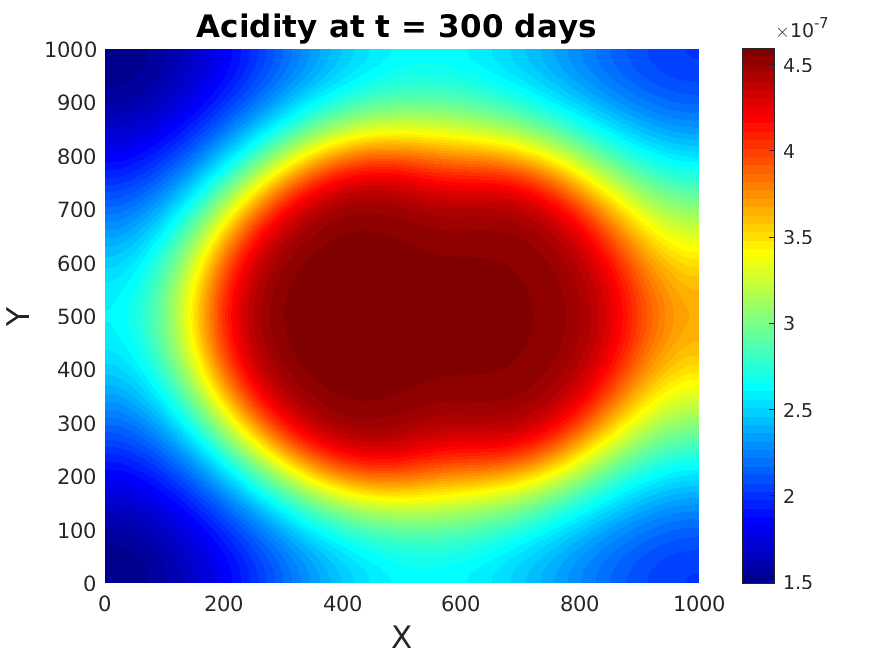}\label{fig:acid_t300-exp_isotropic-neu}}\\	
	\caption[]{Tumor (upper row) and acidity (lower row) at several times for Experiment \ref{exp:FAis0} and initial conditions \eqref{eq:ICs-neu}.}
	\label{fig:tumor-acid-exp_isotropic-neu}
\end{figure}

\noindent	
The simulations show (see Figures \ref{fig:tumor-acid-exp_isotropic} and \ref{fig:tumor-acid-exp_isotropic-neu}) the formation of a pseudopalisade-like pattern, with a very acidic, hypocellular center region surrounded by relatively high glioma cell densities. Thereby, the initial distribution of the tumor cell aggregates and their corresponding pH distribution decisively influence the shape and size of the pattern and the space-time acidity distribution; compare Figures \ref{fig:tumor-acid-exp_isotropic} and \ref{fig:tumor-acid-exp_isotropic-neu}.
	
\end{experiment}

\begin{experiment}\textbf{Anisotropic tissue}\label{exp:kFAis3}

\noindent
With the choice $\delta =0.2$, $\kappa =3$ we describe an underlying tissue with pronounced anisotropy (two crossing fibre bundles). The corresponding mesoscopic fiber distribution $q$
is shown in Figure \ref{fig:q_set2} for a fixed fiber direction, while the macroscopic tissue density $Q$ remains unchanged.

\noindent
The results of this experiment are shown in Figures  \ref{fig:tumor-acid-exp_anisotropic} and \ref{fig:tumor-acid-exp_anisotropic-neu}. The simulated patterns have similar shapes with those in Experiment \ref{exp:FAis0}, but here the tissue anisotropy determines the cells to follow the main orientation of the fiber bundles, which leads to a longer persistence of (small amounts of) cells in the central region with more localized cell aggregates exhibiting higher maxima (see Subfigures \ref{fig:tumor_t150-exp_anisotropic} and \ref{fig:tumor_t150-exp_anisotropic-neu}). The patterns at later times (see Subfigures \ref{fig:tumor_t300-exp_anisotropic} and \ref{fig:tumor_t300-exp_anisotropic-neu}) still bear traits of the degraded tissue; the cells are still forming garland-like structures around the hypoxic centers, with the highest cell density located at one or several peripheral sites with highly aligned tissue, farthest away from the main sources of (initial) acidity. As before, the initial distributions of tumor and acidity influence the shape of the patterns (compare Figures \ref{fig:tumor-acid-exp_anisotropic} and \ref{fig:tumor-acid-exp_anisotropic-neu}). The differences between the acidity distributions in Subfigures \ref{fig:acid_t30-exp_anisotropic}-\ref{fig:acid_t300-exp_anisotropic} and those in Subfigures  \ref{fig:acid_t30-exp_isotropic}-\ref{fig:acid_t300-exp_isotropic} (and correspondingly Subfigures \ref{fig:acid_t30-exp_anisotropic-neu}-\ref{fig:acid_t300-exp_anisotropic-neu} and, respectively, those in  Subfigures  \ref{fig:acid_t30-exp_isotropic-neu}-\ref{fig:acid_t300-exp_isotropic-neu} for the set of initial conditions \eqref{eq:ICs-neu}) are less prominent, since the acidity concentration $h$ obeys in both cases a PDE with linear diffusion, where the tissue anisotropy has minor influence.

\begin{figure}[!htbp]
\subfigure[][]{\includegraphics[width=0.31\textwidth]{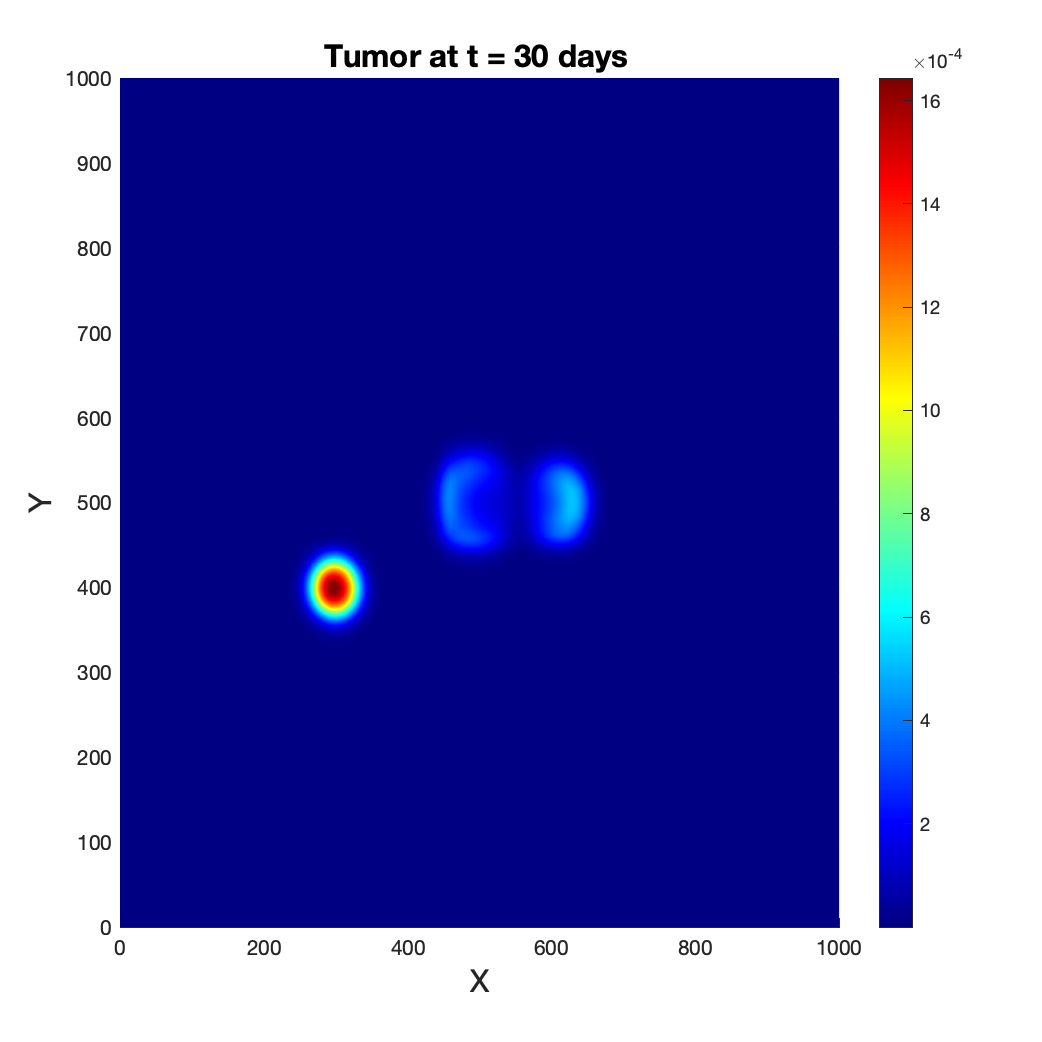}\label{fig:tumor_t30-exp_anisotropic}}\quad \subfigure[][]{\includegraphics[width=0.31\textwidth]{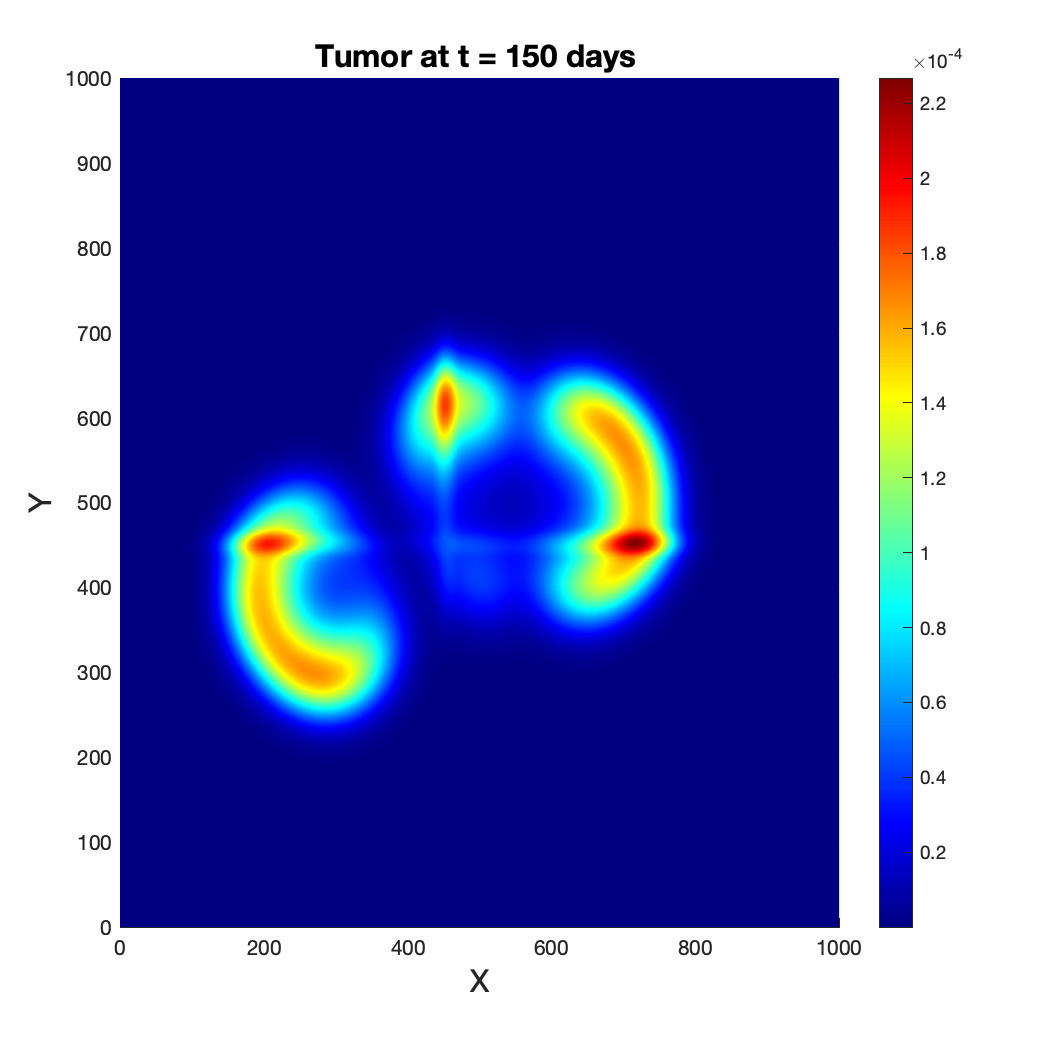}\label{fig:tumor_t150-exp_anisotropic}}\quad \subfigure[][]{\includegraphics[width=0.31\textwidth]{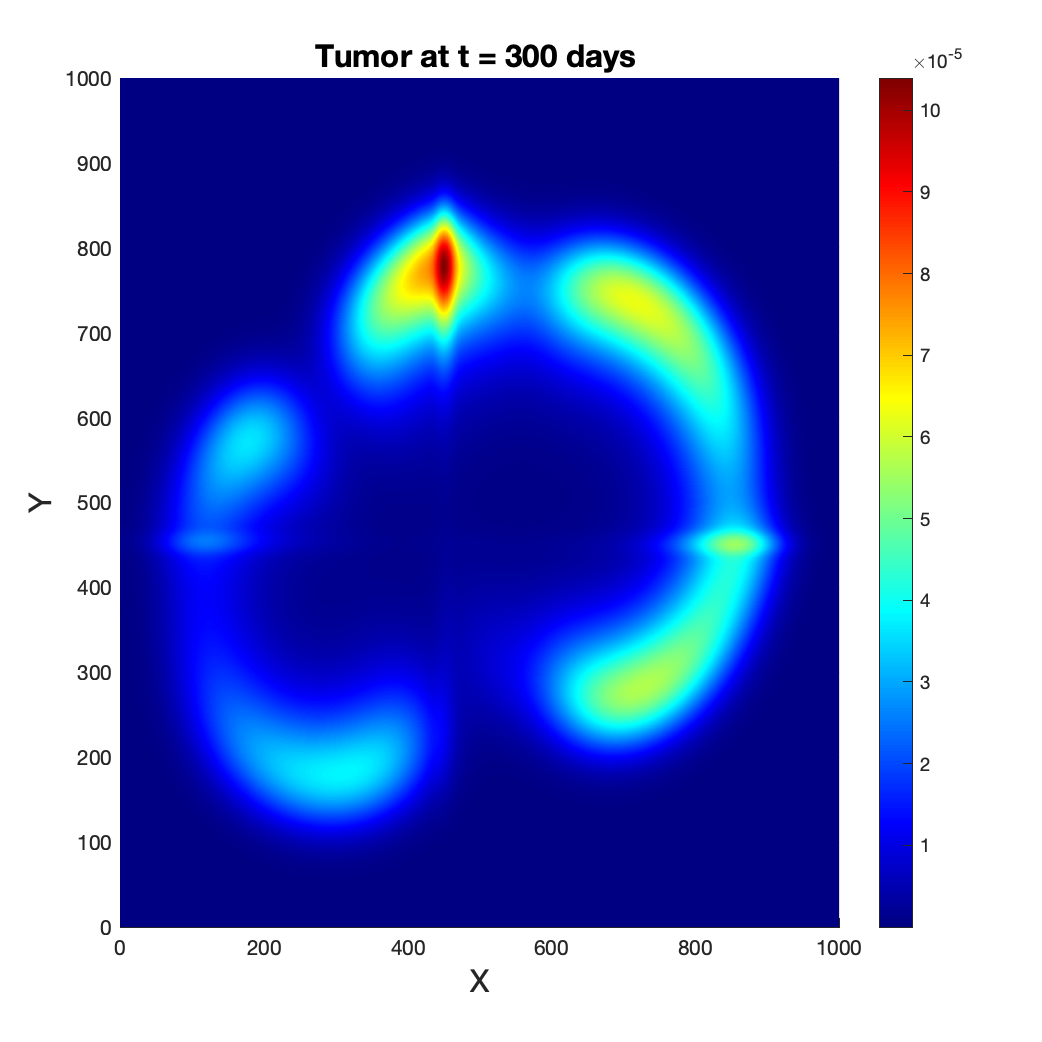}\label{fig:tumor_t300-exp_anisotropic}}\\
\subfigure[][]{\includegraphics[width=0.31\textwidth]{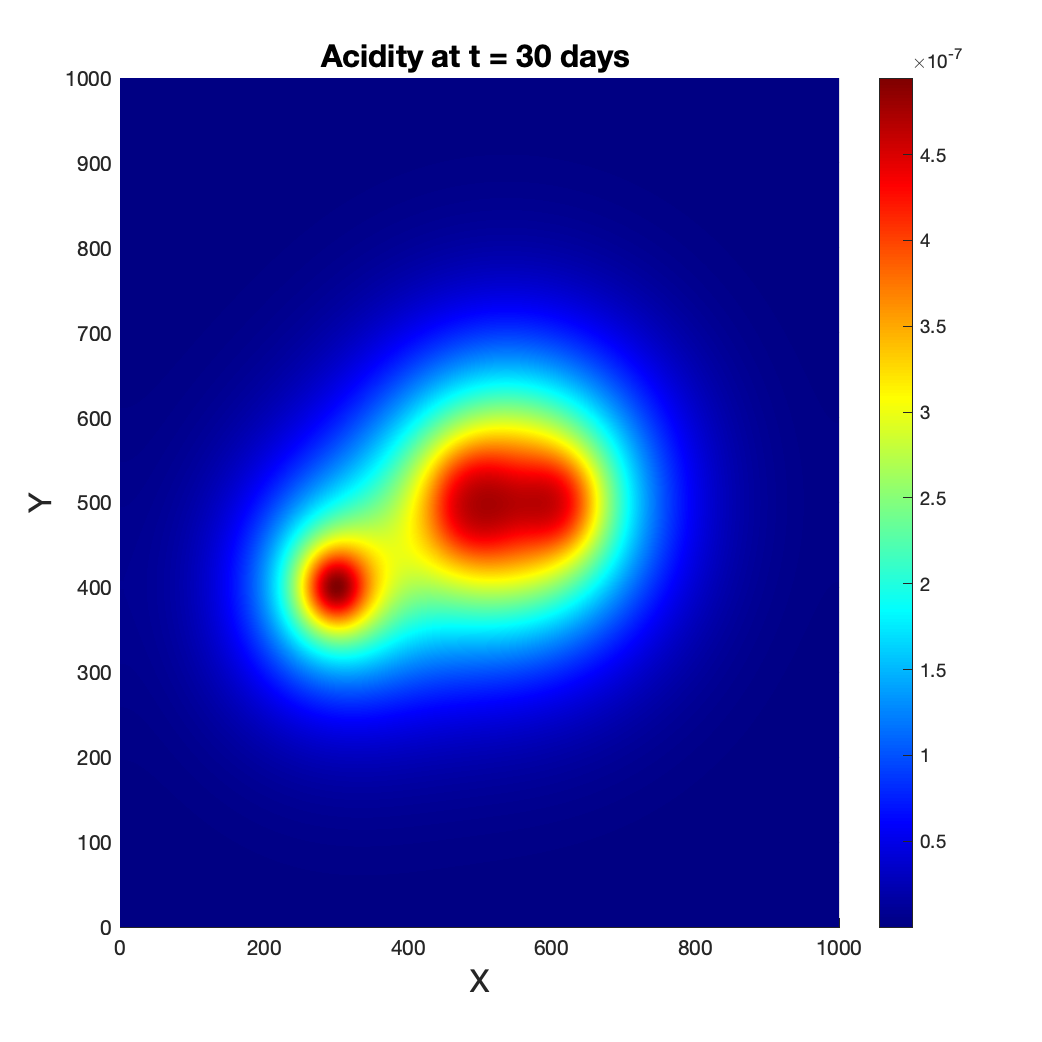}\label{fig:acid_t30-exp_anisotropic}}\quad \subfigure[][]{\includegraphics[width=0.31\textwidth]{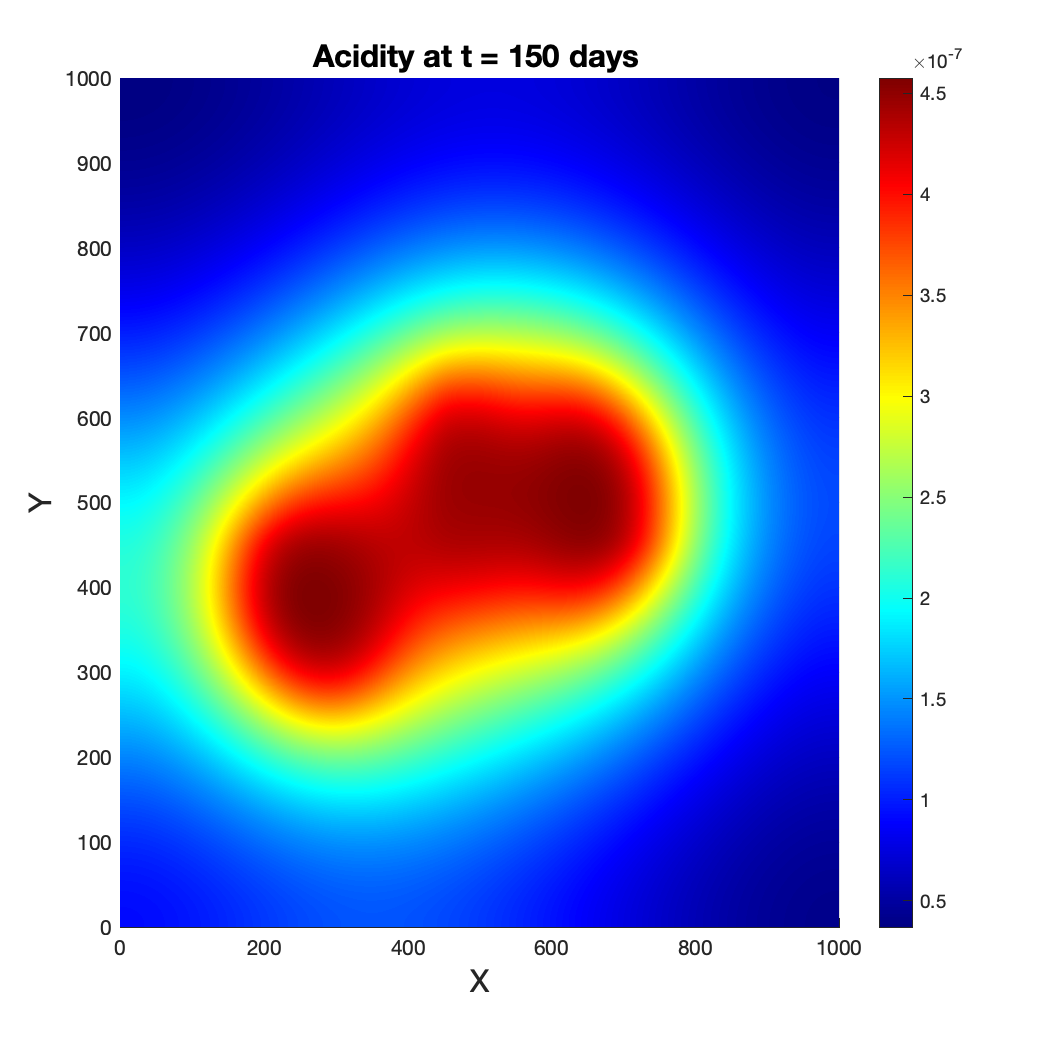}\label{fig:acid_t150-exp_anisotropic}}\quad \subfigure[][]{\includegraphics[width=0.31\textwidth]{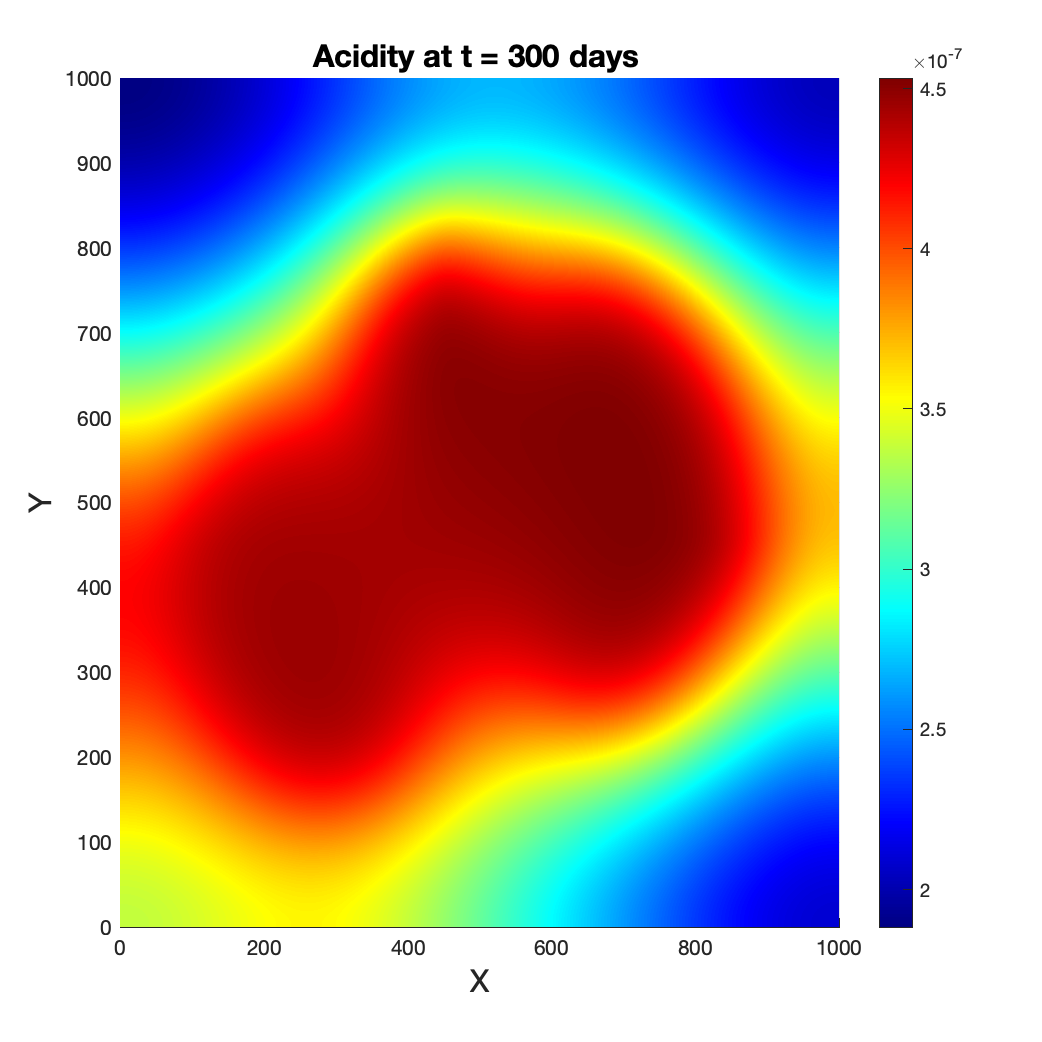}\label{fig:acid_t300-exp_anisotropic}}\\	
\caption[]{Tumor (upper row) and acidity (lower row) at several times for Experiment \ref{exp:kFAis3} and initial conditions \eqref{eq:ICs}}
	\label{fig:tumor-acid-exp_anisotropic}
\end{figure}
	
\begin{figure}[!htbp]
	\subfigure[][]{\includegraphics[width=0.31\textwidth]{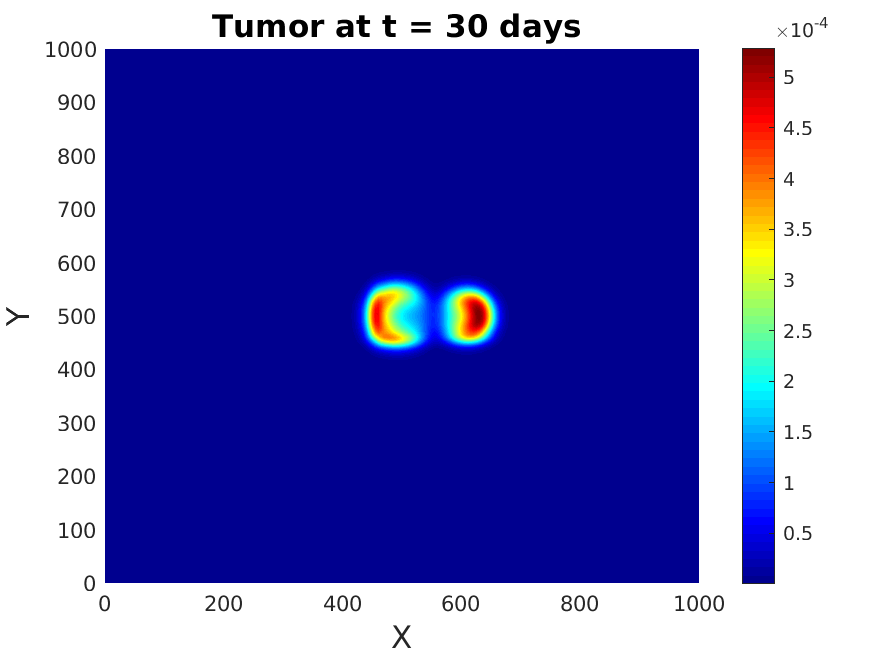}\label{fig:tumor_t30-exp_anisotropic-neu}}\quad \subfigure[][]{\includegraphics[width=0.31\textwidth]{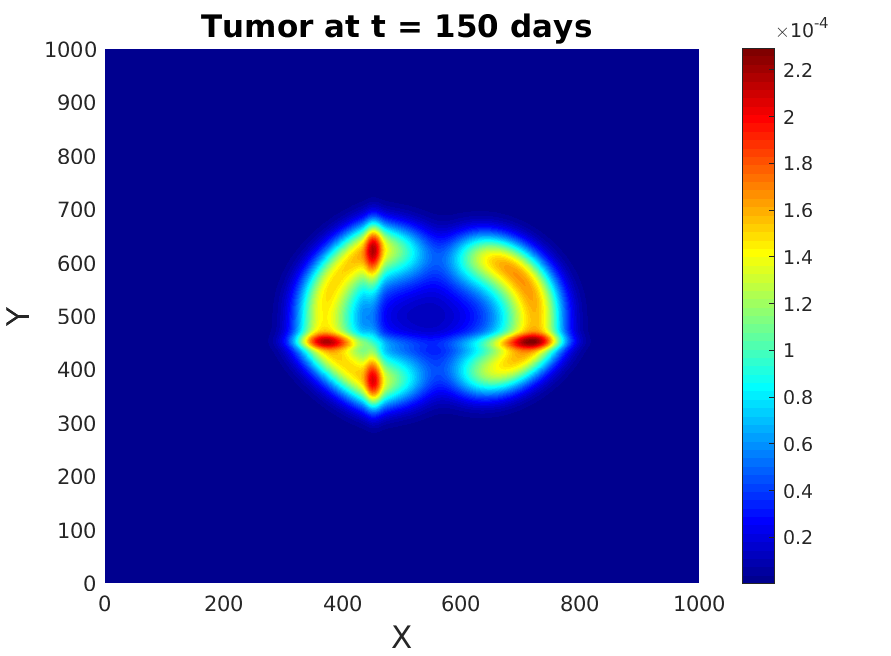}\label{fig:tumor_t150-exp_anisotropic-neu}}\quad \subfigure[][]{\includegraphics[width=0.31\textwidth]{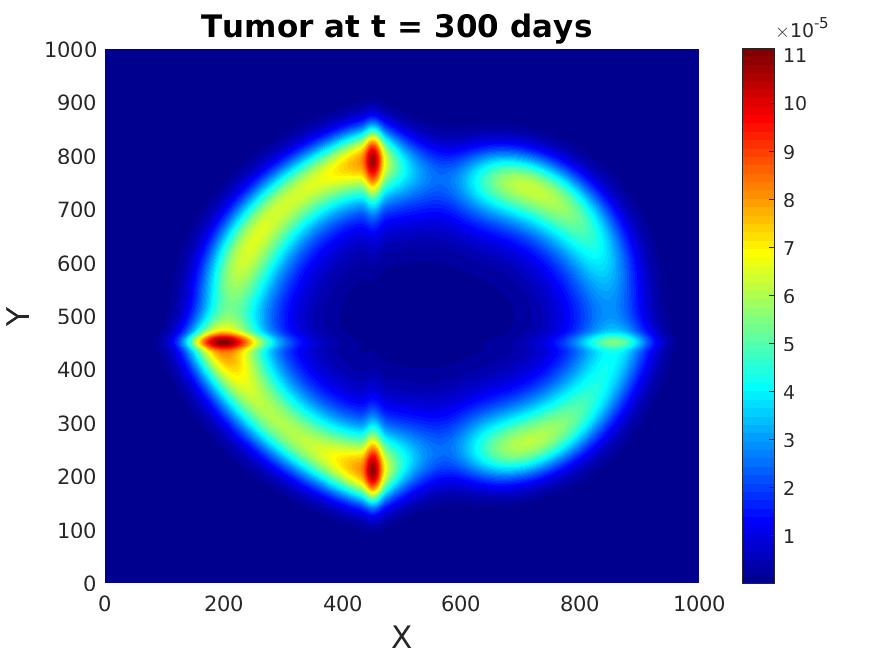}\label{fig:tumor_t300-exp_anisotropic-neu}}\\
	\subfigure[][]{\includegraphics[width=0.31\textwidth]{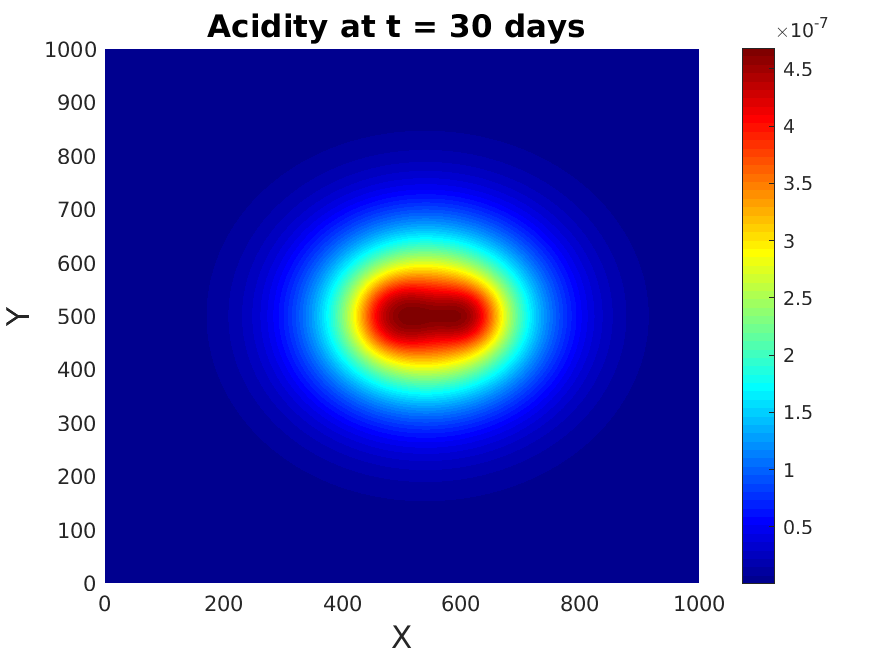}\label{fig:acid_t30-exp_anisotropic-neu}}\quad \subfigure[][]{\includegraphics[width=0.31\textwidth]{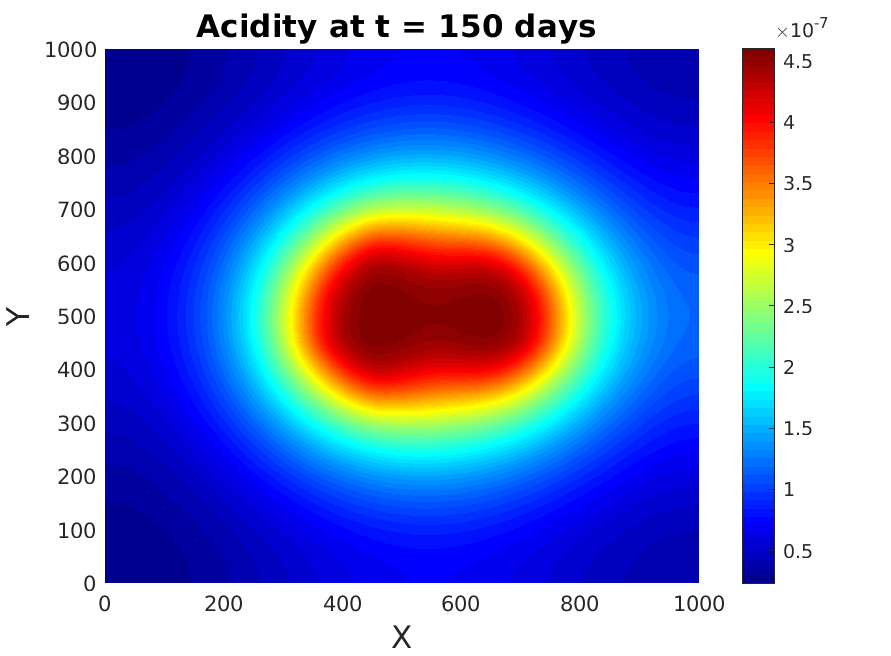}\label{fig:acid_t150-exp_anisotropic-neu}}\quad \subfigure[][]{\includegraphics[width=0.31\textwidth]{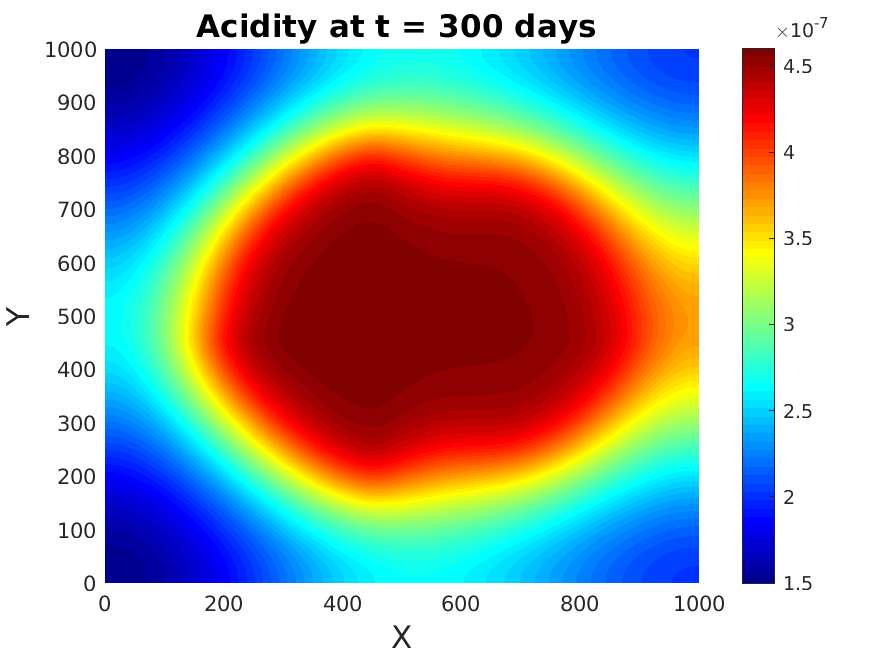}\label{fig:acid_t300-exp_anisotropic-neu}}\\	
	\caption[]{Tumor (upper row) and acidity (lower row) at several times for Experiment \ref{exp:kFAis3} and initial conditions \eqref{eq:ICs-neu}.}
	\label{fig:tumor-acid-exp_anisotropic-neu}
\end{figure}

\end{experiment}

\begin{figure}[!htbp]
	\subfigure[][]{\includegraphics[width=0.23\textwidth]{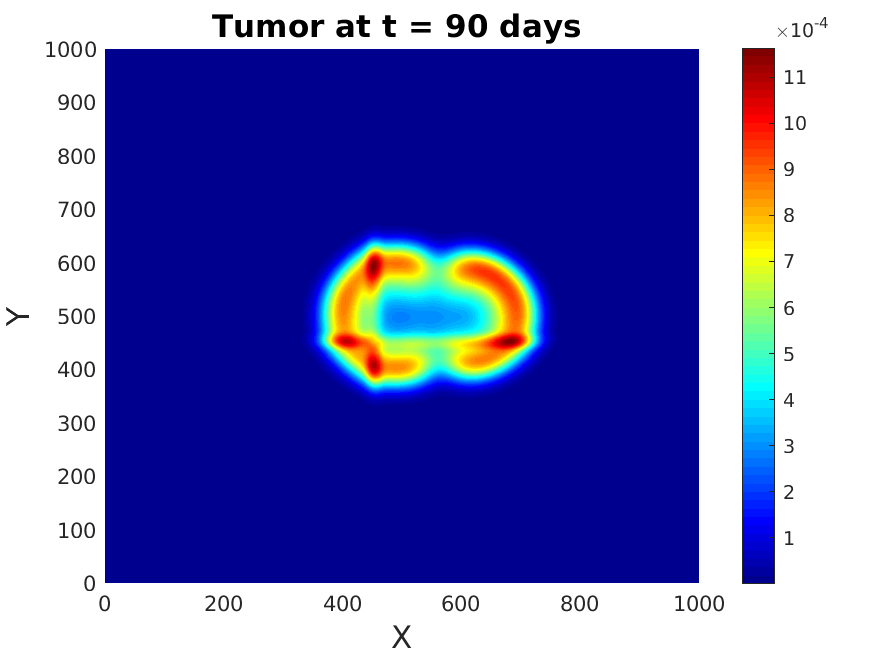}\label{fig:tumor_t90-lowgrade}}\quad \subfigure[][]{\includegraphics[width=0.23\textwidth]{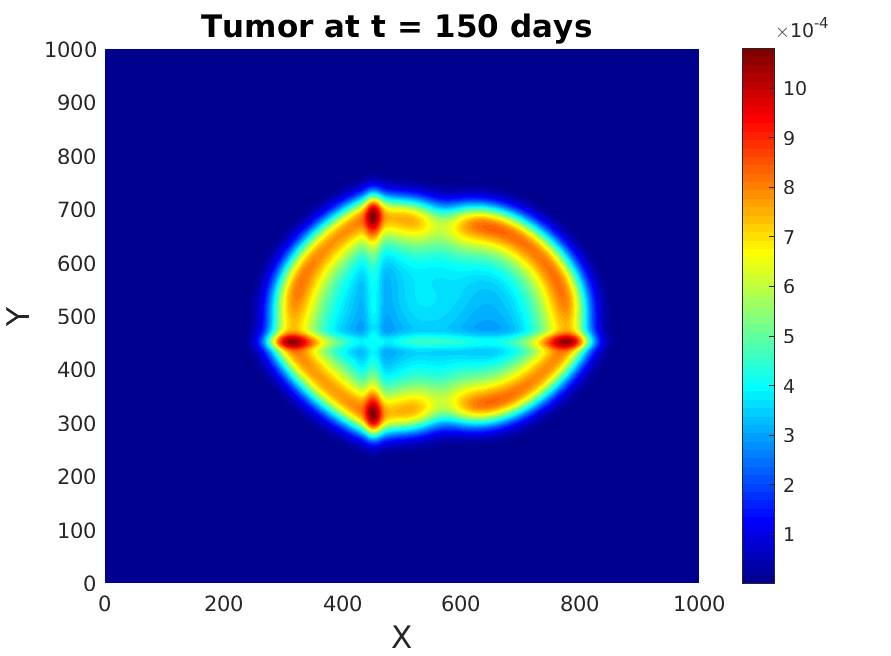}\label{fig:tumor_t150-lowgrade}}\quad \subfigure[][]{\includegraphics[width=0.23\textwidth]{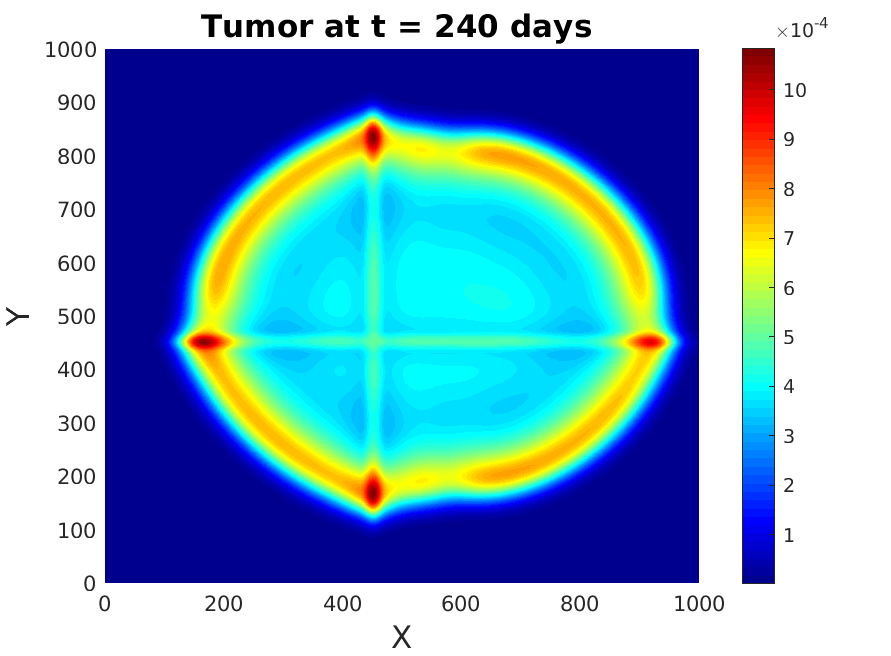}\label{fig:tumor_t240-lowgrade}}\quad \subfigure[][]{\includegraphics[width=0.23\textwidth]{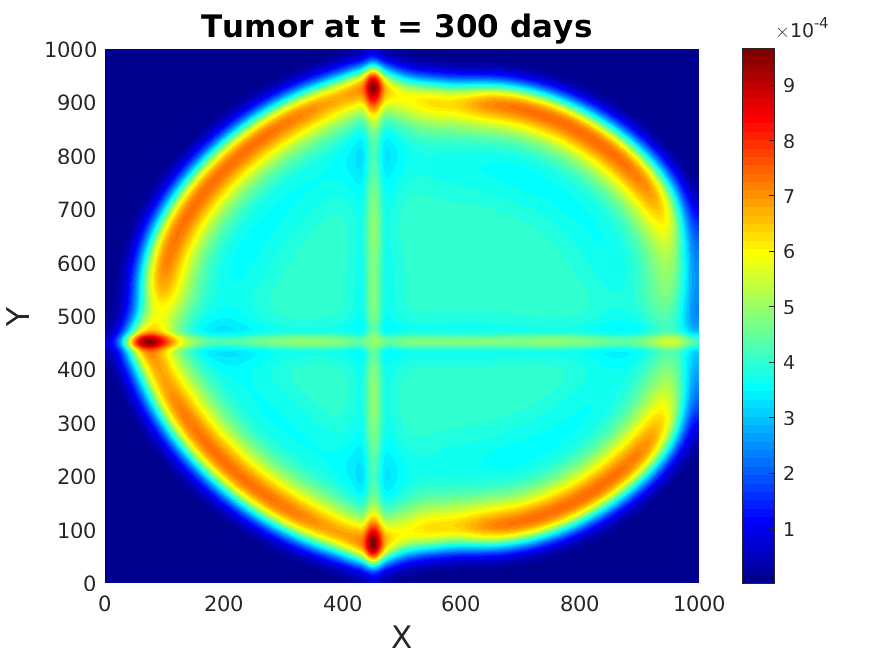}\label{fig:tumor_t300-lowgrade}}\\
	\subfigure[][]{\includegraphics[width=0.23\textwidth]{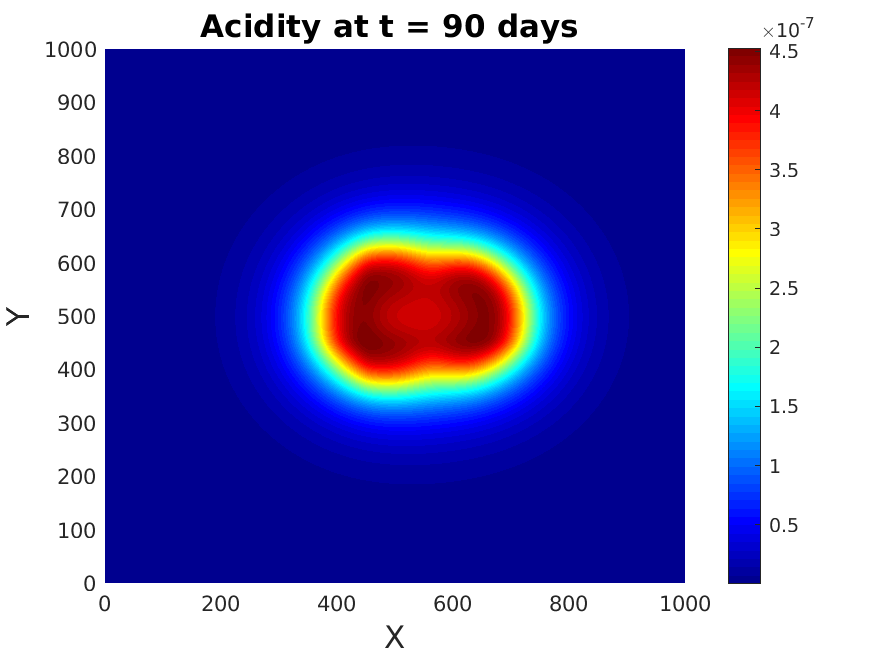}\label{fig:acid_t30-lowgrade}}\quad \subfigure[][]{\includegraphics[width=0.23\textwidth]{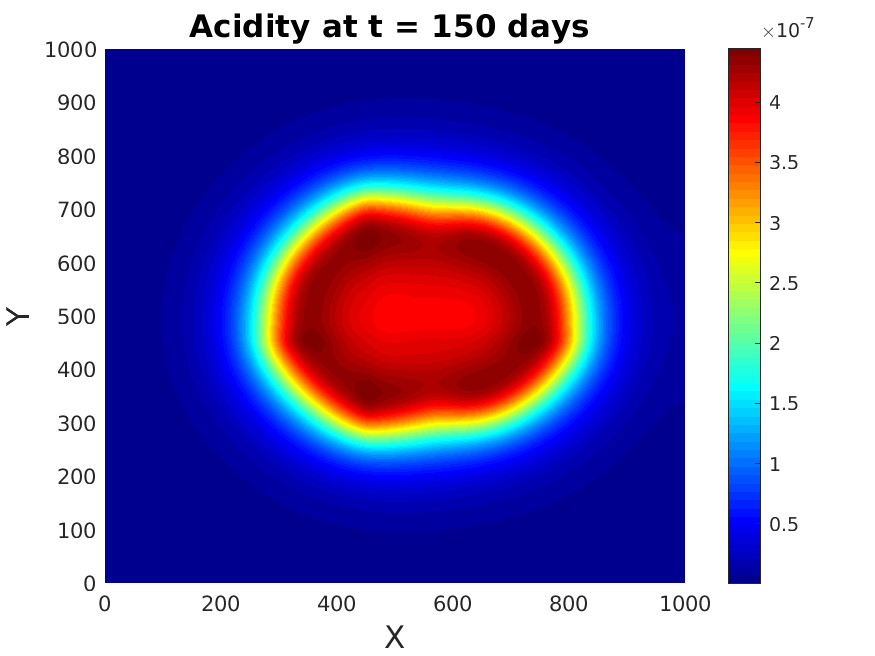}\label{fig:acid_t150-lowgrade}}\quad \subfigure[][]{\includegraphics[width=0.23\textwidth]{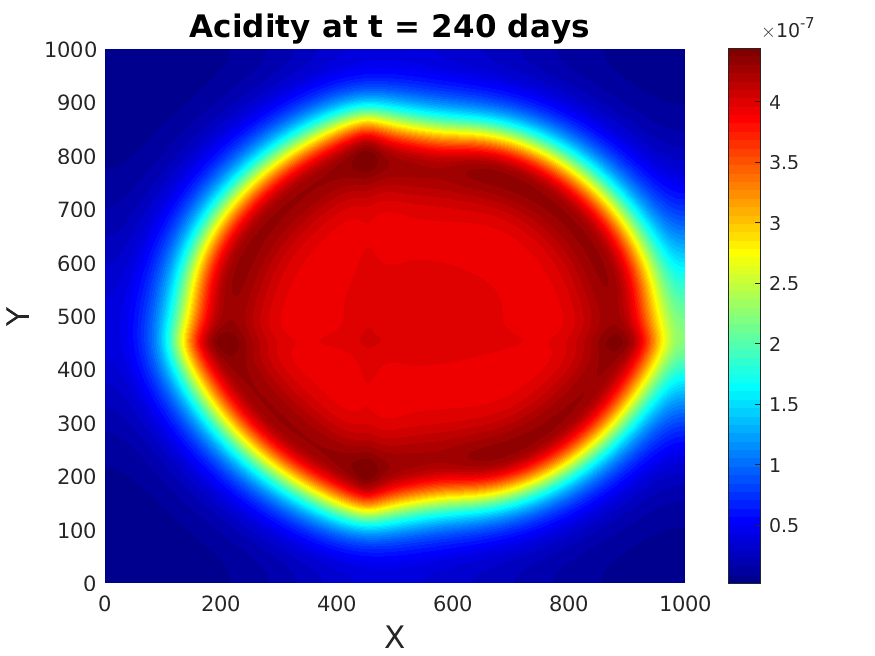}\label{fig:acid_t240-lowgrade}}\quad \subfigure[][]{\includegraphics[width=0.23\textwidth]{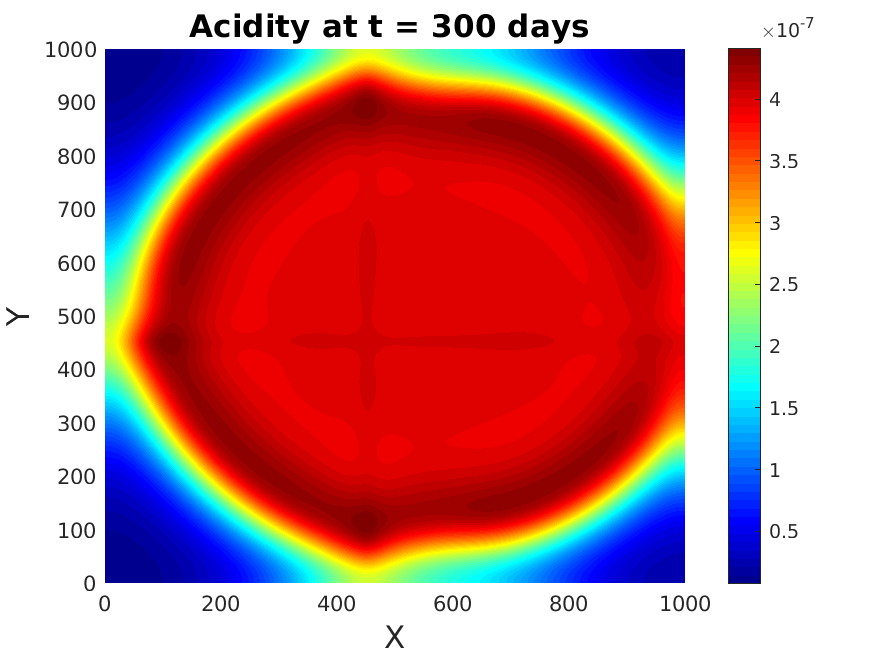}\label{fig:acid_t300-lowgrade}}\\	
	\caption[]{Tumor (upper row) and acidity (lower row) at several times for Experiment \ref{exp:kFAis3}, initial conditions \eqref{eq:ICs-neu}, and stronger proton buffering.}
	\label{fig:tumor-acid-exp_anisotropic-lowgrade}
\end{figure}

\begin{figure}[!htbp]
	\subfigure[][]{\includegraphics[width=0.23\textwidth]{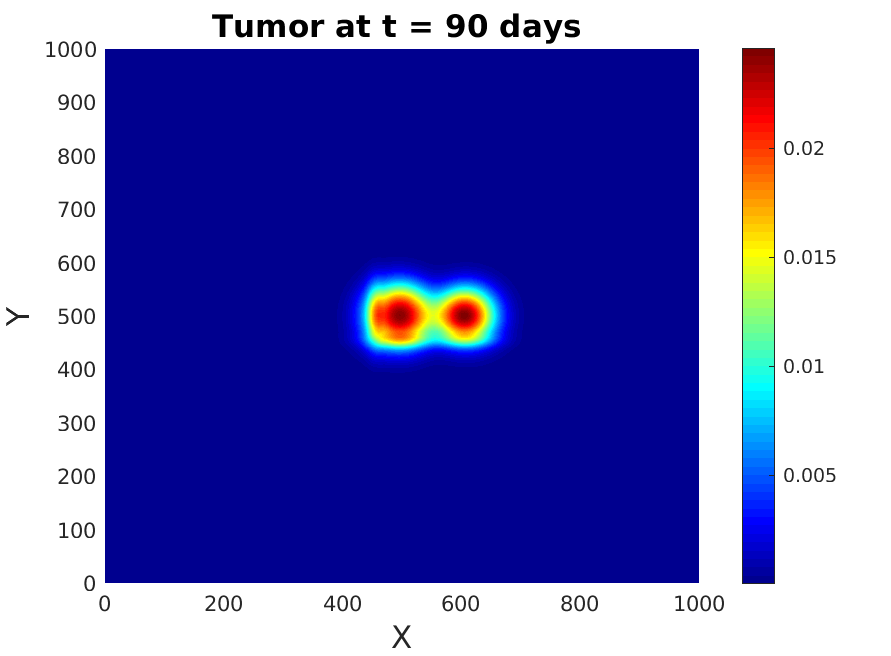}\label{fig:tumor_t90-source_modif}}\quad \subfigure[][]{\includegraphics[width=0.23\textwidth]{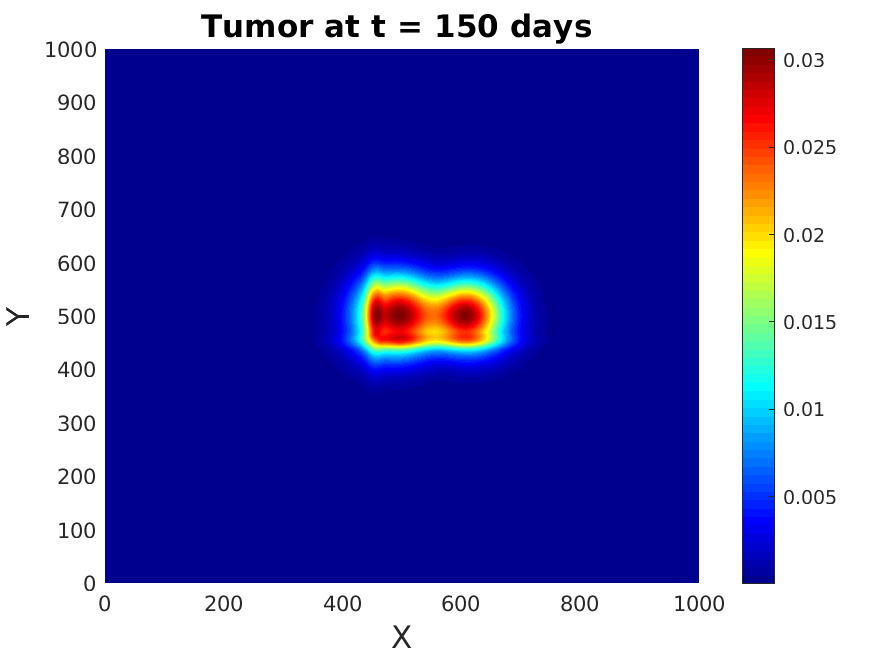}\label{fig:tumor_t150-source_modif}}\quad \subfigure[][]{\includegraphics[width=0.23\textwidth]{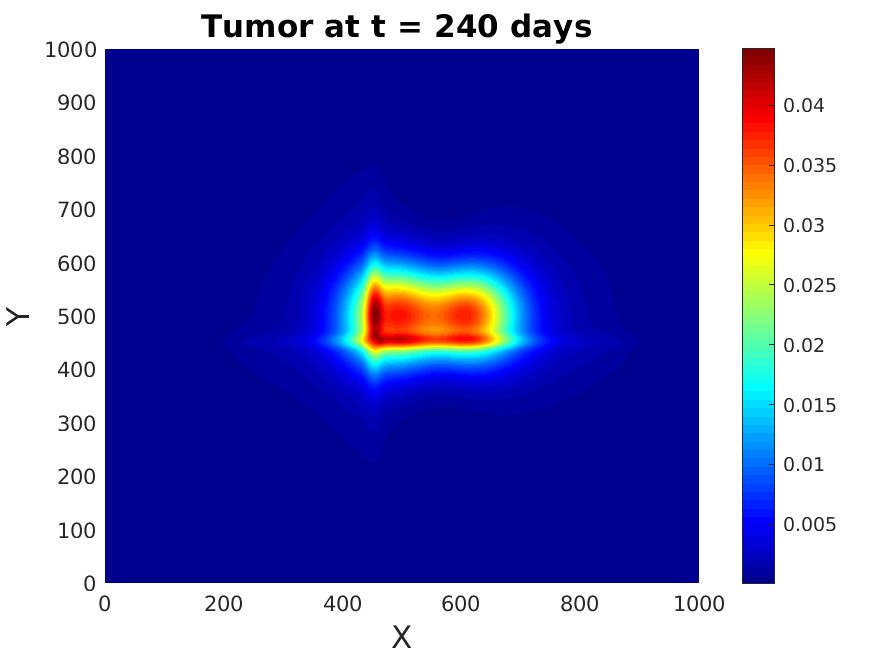}\label{fig:tumor_t240-source_modif}}\quad \subfigure[][]{\includegraphics[width=0.23\textwidth]{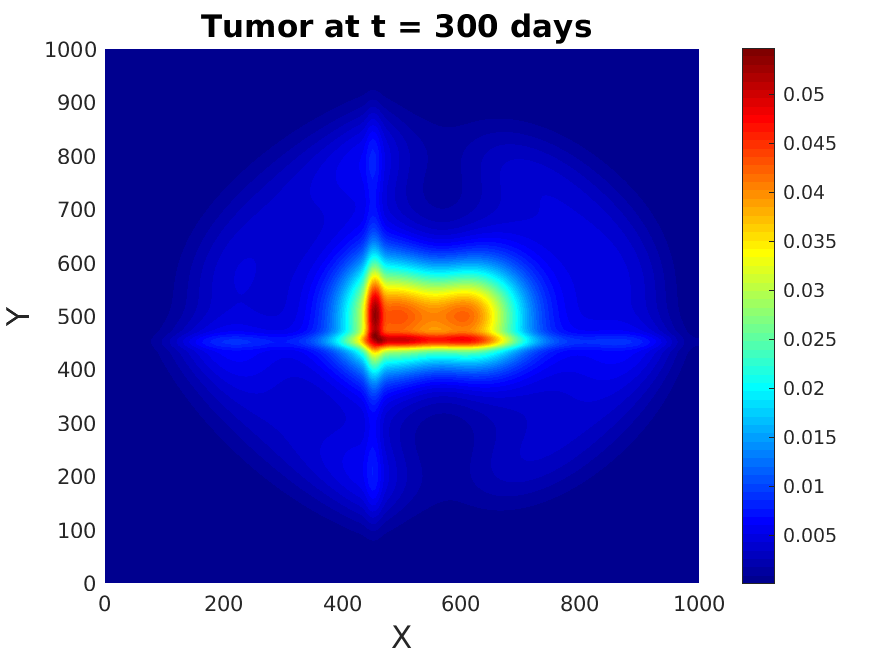}\label{fig:tumor_t300-source_modif}}\\
	\subfigure[][]{\includegraphics[width=0.23\textwidth]{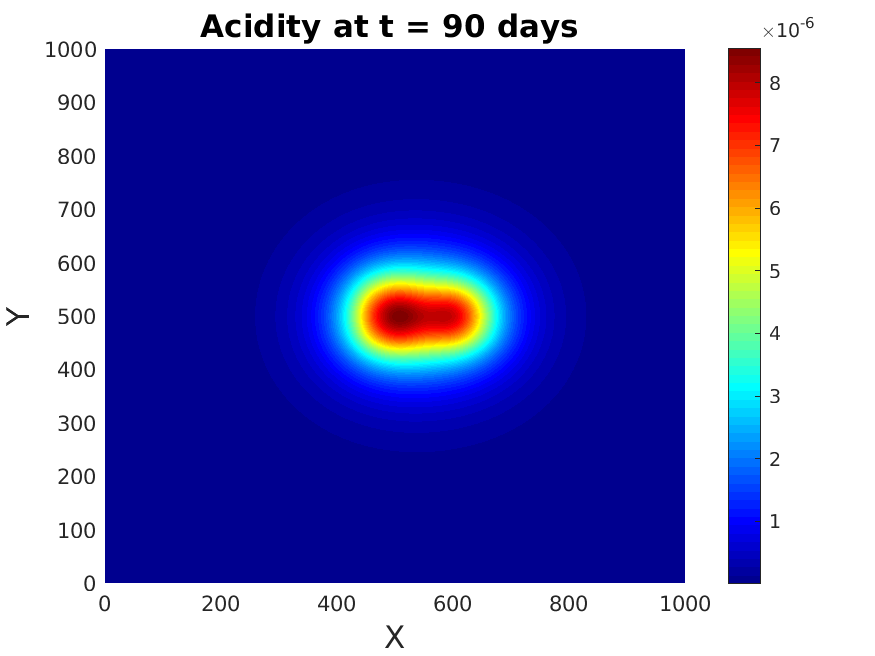}\label{fig:acid_t30-source_modif}}\quad \subfigure[][]{\includegraphics[width=0.23\textwidth]{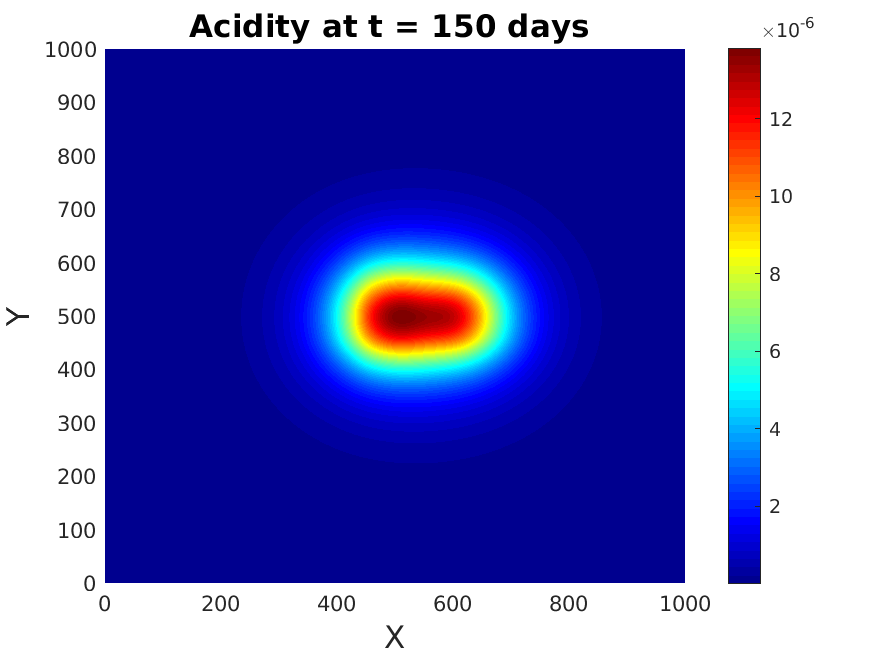}\label{fig:acid_t150-source_modif}}\quad \subfigure[][]{\includegraphics[width=0.23\textwidth]{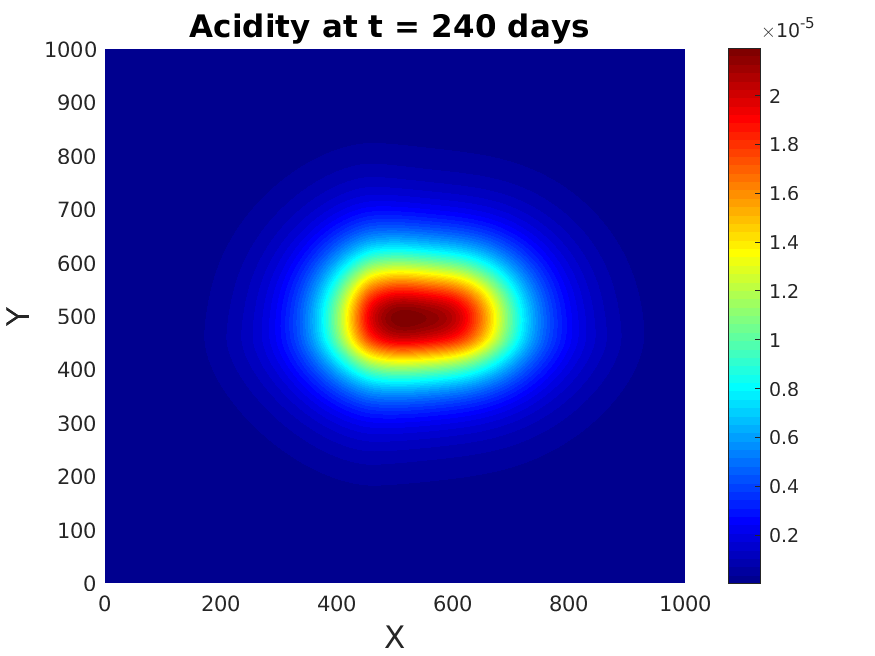}\label{fig:acid_t240-source_modif}}\quad \subfigure[][]{\includegraphics[width=0.23\textwidth]{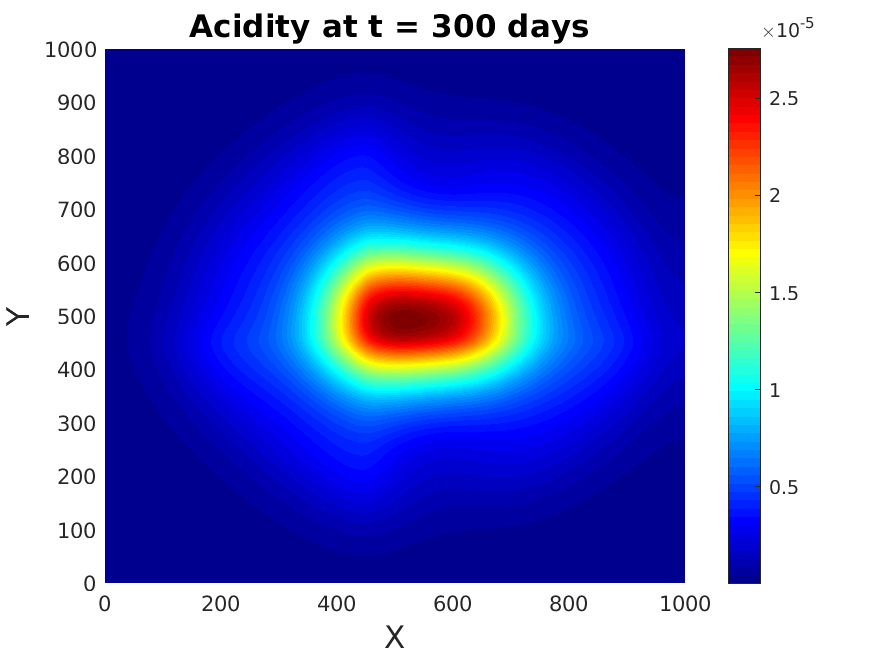}\label{fig:acid_t300-source_modif}}\\	
	\caption[]{Tumor (upper row) and acidity (lower row) at several times for Experiment \ref{exp:kFAis3}, initial conditions \eqref{eq:ICs-neu}, and source term $\mu_0(1-M)\frac{M}{1+S}$ instead of that in \eqref{macro_eq_non}. All parameters as in Table \ref{table}.}
	\label{fig:source-modif}
\end{figure}

\noindent
Running numerical simulations for several different parameter sets led to the following observations:
\begin{itemize}
\item The decisive parameter in this system seems to be $\alpha $, which relates to the proton buffering efficacy (in the nondimensionalized form \eqref{macro_eq_non} it is basically the ratio between the acidity removal and proton production rates). The tumor growth rate $\mu_0$ plays a role, too, but a less prominent one. Concretely, pseudopalisade patterns form for very low values of $\alpha $ (weak buffering). If the system is able to remove protons more efficiently (e.g., because there is a functioning capillary network), then these garland-like patterns typical for GBM do not form in a time span which is relevant for this cancer (less than a year); instead, there are rather homogeneous structures with dense cellular areas and no necrosis - which corresponds to a lower tumor grade, without (local) occlusions of capillaries and corresponding necrotization (anaplastic astrocytoma), some with partially preserving the underlying tissue structure (fibrillar astrocytoma), see \cite{Ramnani} for WHO-grading on the basis of  histopathological samples. Figure \ref{fig:tumor-acid-exp_anisotropic-lowgrade} shows the evolution of glioma and acidity at several times for the system with initial conditions \eqref{eq:ICs-neu} and the same parameters as in the simulations of Experiment \ref{exp:kFAis3}, with the exception of $\alpha$, which is now still very small, but four orders of magnitude higher. The tumor cells are producing acidity (by glycolyis) and the inner region begins to degrade, as in the previous simulations. However, due to the stronger acidity removal ratio, it does not become severely hypoxic, which allows the tumor cells to repopulate it, while the rest of the neoplasm is expanding outwards. The underlying tissue structure is thereby supporting both migraton and growth. Notice the more extensive tumor spread in comparison with Figure \ref{fig:tumor-acid-exp_anisotropic-neu}. In \ref{sec:appendix} we do a short linear stability analysis of system \eqref{macro_eq_non}, \eqref{macro_eq_non_S} with a constant tumor diffusion coefficient; it turns out that no Turing patterns are formed - which does not mean, however, that other types of patterns are not possible.
\item If $\alpha $ exceeds a certain threshold value (in our simulations it was one order of magnitude higher than in the computations for Figure \ref{fig:tumor-acid-exp_anisotropic-lowgrade}) then the solution blows up already in 1D.
\item The shape of the source term in the equation for tumor cell density has itself a substantial influence on the pattern. It should be chosen in such a way that proliferation is reduced for higher acidity levels. This is, however, not enough for pseudopalisade formation: for instance, a source term of the form $\mu_0 (1-M)\frac{M}{1+S}$ instead of that in \eqref{macro_eq_non} does not lead to such patterns, as there is no decay of glioma cells due to hypoxia. Figure \ref{fig:source-modif} shows the behavior of tumor and acidity for this alternative choice of the source term, in the framework of Experiment \ref{exp:kFAis3}.
\end{itemize}

\begin{figure}[!htbp]
	\subfigure[][]{\includegraphics[width=0.23\textwidth]{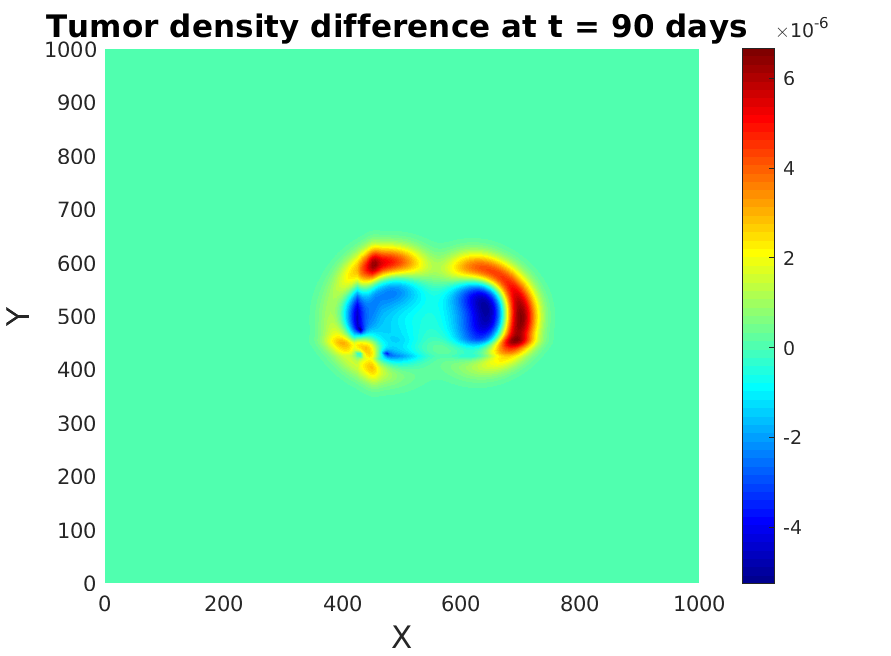}\label{fig:tumor_t90-diff}}\quad \subfigure[][]{\includegraphics[width=0.23\textwidth]{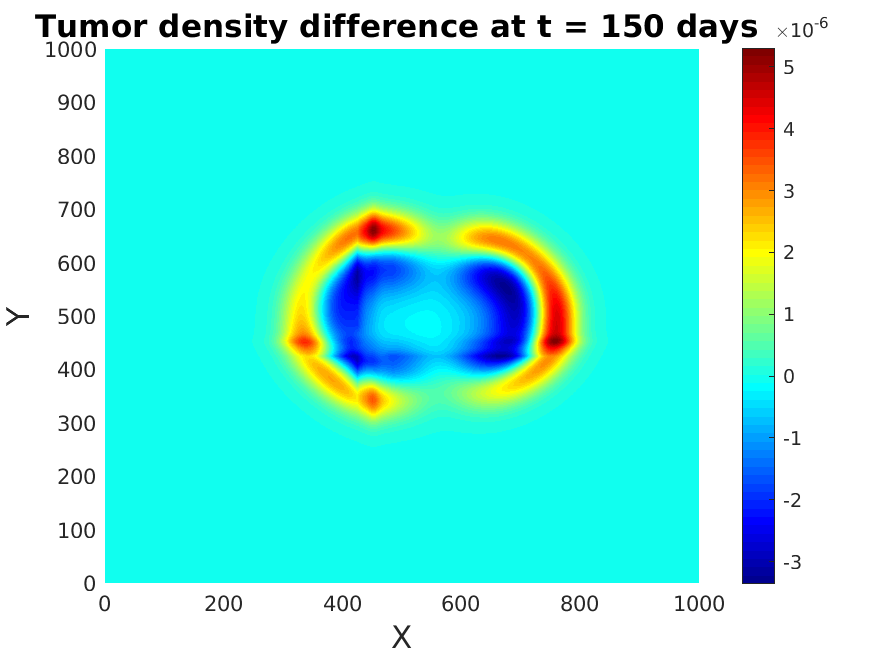}\label{fig:tumor_t150-diff}}\quad \subfigure[][]{\includegraphics[width=0.23\textwidth]{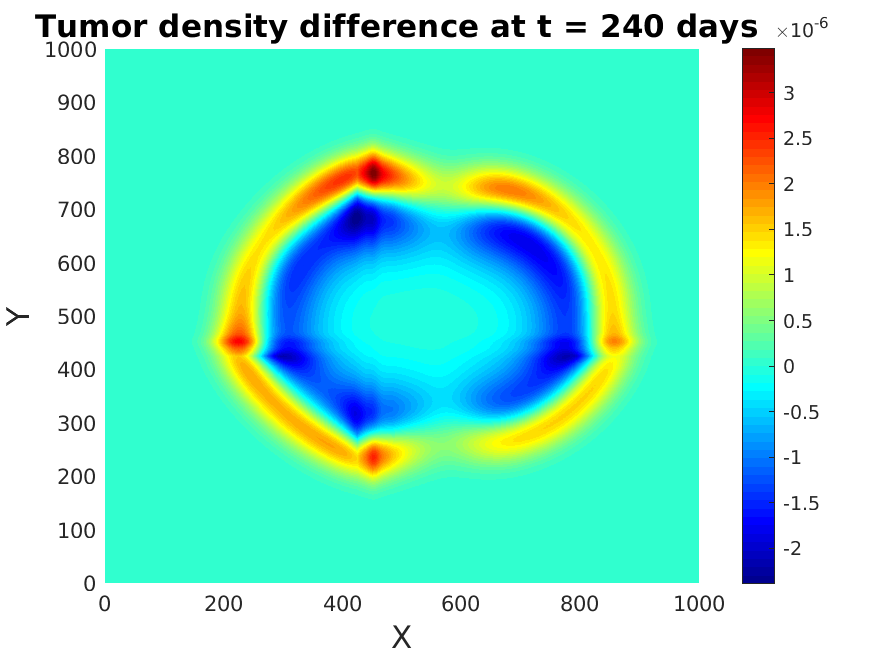}\label{fig:tumor_t240-diff}}\quad \subfigure[][]{\includegraphics[width=0.23\textwidth]{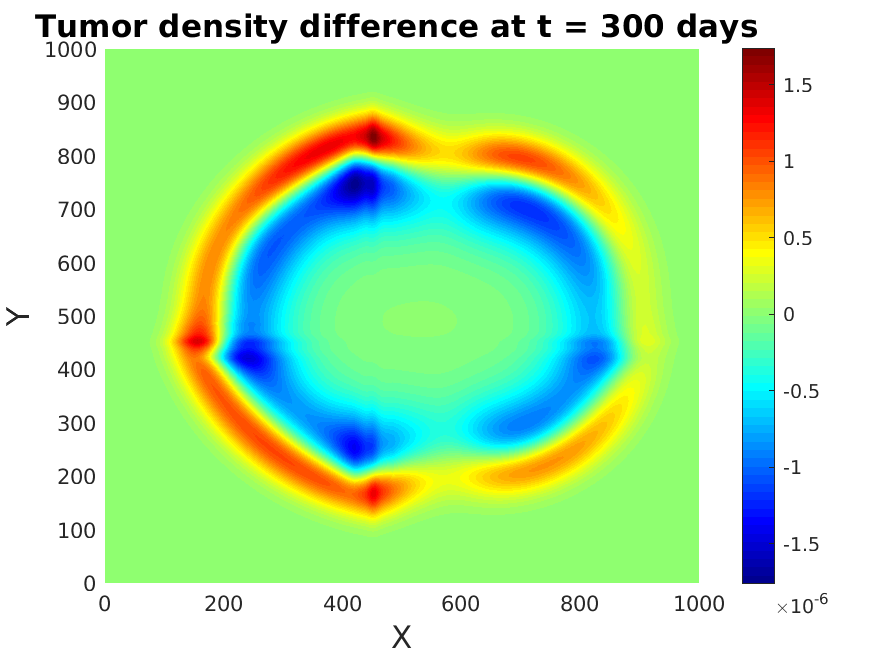}\label{fig:tumor_t300-diff}}\\
	\subfigure[][]{\includegraphics[width=0.23\textwidth]{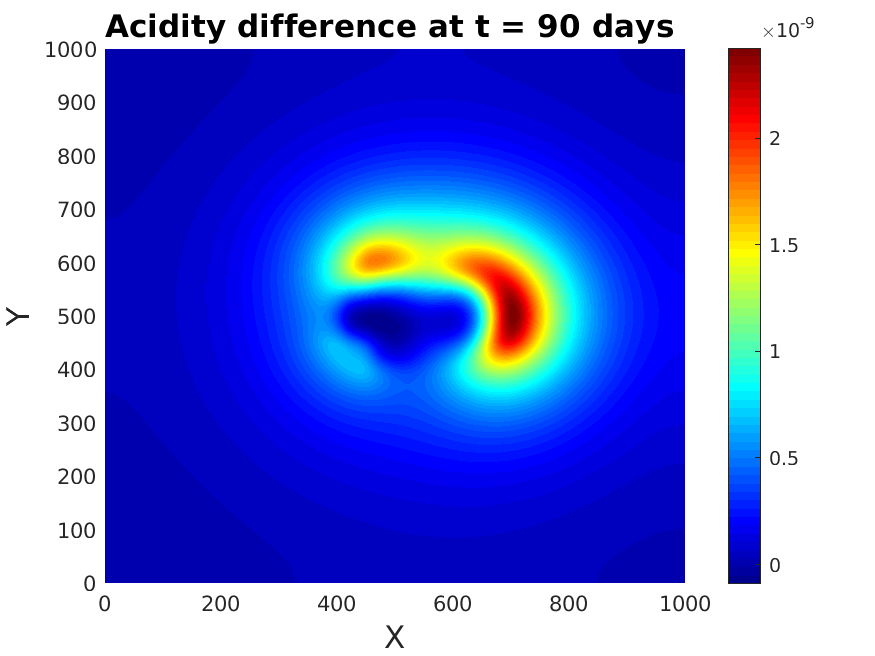}\label{fig:acid_t30-diff}}\quad \subfigure[][]{\includegraphics[width=0.23\textwidth]{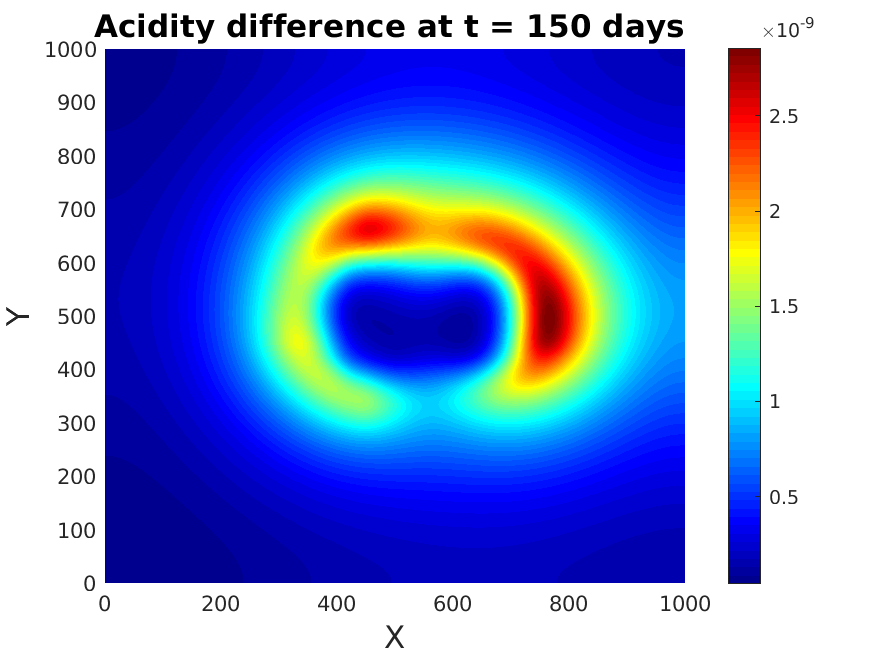}\label{fig:acid_t150-diff}}\quad \subfigure[][]{\includegraphics[width=0.23\textwidth]{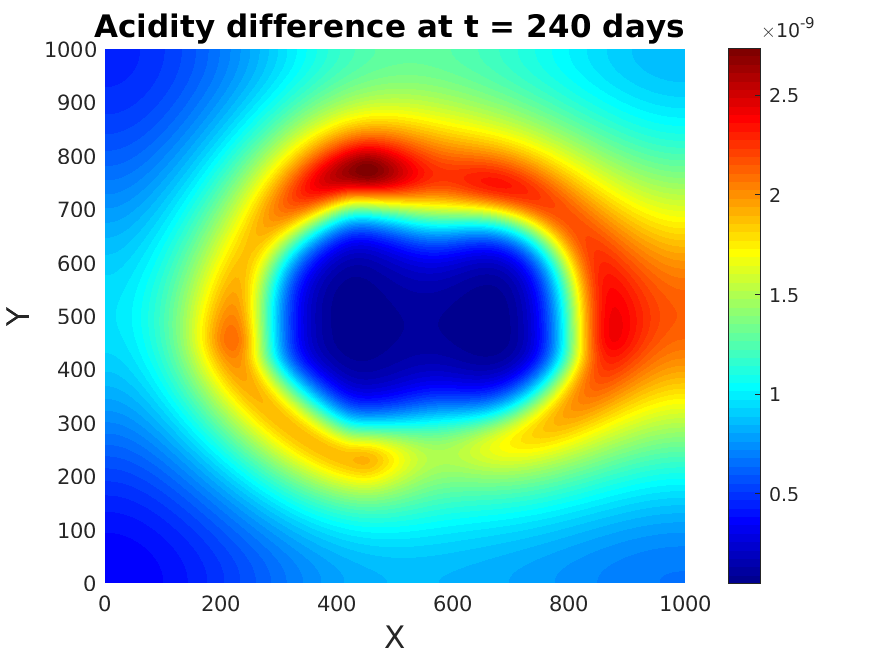}\label{fig:acid_t240-diff}}\quad \subfigure[][]{\includegraphics[width=0.23\textwidth]{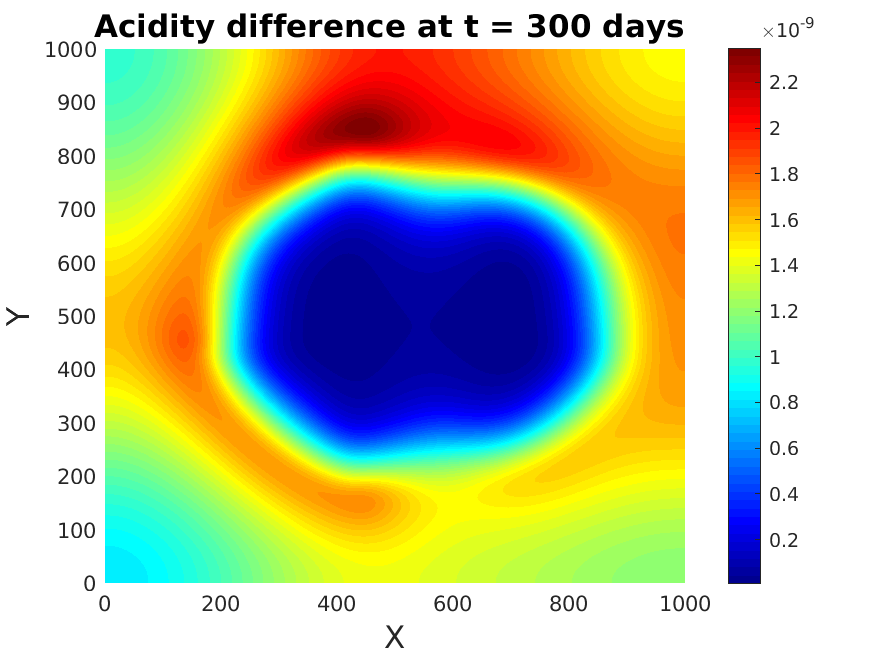}\label{fig:acid_t300-diff}}\\	
	\caption[]{Difference between tumor (upper row) and acidity (lower row) at several times computed for System \eqref{macro_eq_non}, \eqref{macro_eq_non_S} with and without pH-taxis in the framework of Experiment \ref{exp:kFAis3}, initial conditions \eqref{eq:ICs-neu}.}
	\label{fig:diff}
\end{figure}

\noindent
Including repellent pH-taxis leads to the formation of wider pseudopalisades, with thinner cell 'garlands' than in the case where the glioma cells are only performing myopic diffusion and being degraded by excessive acidity.  Figure \ref{fig:diff} illustrates the differences between tumor density and acidity in the two cases, i.e. between solutions of \eqref{macro_eq_non}, \eqref{macro_eq_non_S} and those of the  same system with $g(S)=0$. The differences are more pronounced in earlier stages of pattern formation and become smaller with advancing time. The plots also show that pseudopalisades are formed even if there is no pH-taxis, suggesting that the latter merely enhances the effect of the source/decay term in \eqref{macro_eq_non} who is actually driving the pattern - together with an opportune parameter combination (in particular, adequate proton buffering).\\[-2ex]

\begin{figure}[!htbp]
	\subfigure[][]{\includegraphics[width=0.32\textwidth]{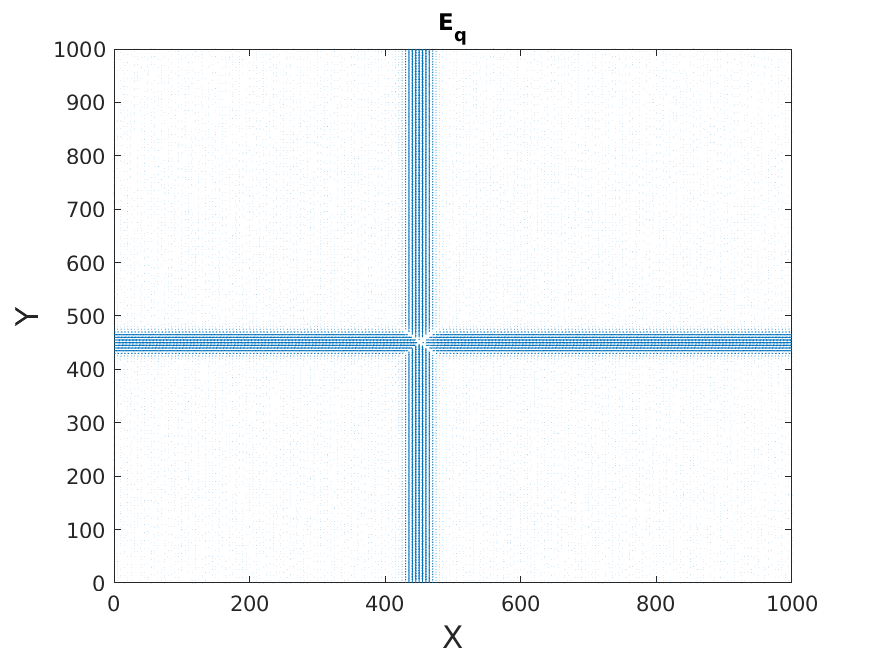}\label{fig:Eq-exp2_delta0p2}}\quad 
	\subfigure[][]{\includegraphics[width=0.32\textwidth]{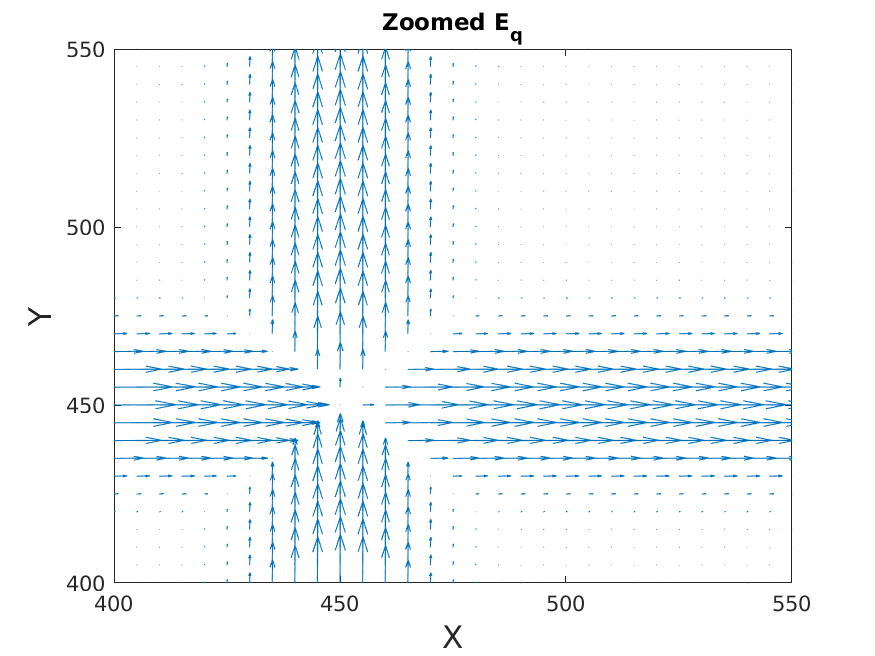}\label{fig:Eq-zoomed-exp2_delta0p2}}\quad
	\subfigure[][]{\includegraphics[width=0.32\textwidth]{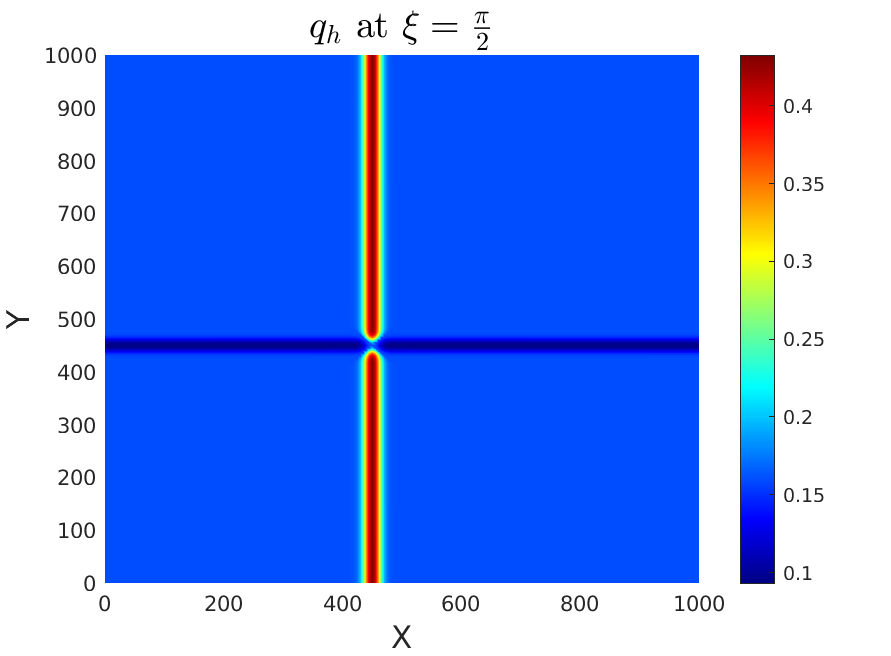}\label{fig:q_3pipe4-exp2_delta0p2}}
	\caption[]{Mean fiber orientation $\mathbb E_q$ (\ref{fig:Eq-exp2_delta0p2}) and zoom near crossing of fiber strands (\ref{fig:Eq-zoomed-exp2_delta0p2}) for $q_h$ as in \eqref{eq_q_h} with $\delta =0.2$. Subfigure \ref{fig:q_3pipe4-exp2_delta0p2}: mesoscopic tissue density $q_h$ for direction $\xi =\pi/2$.}	
	\label{fig:tissue-hyp}
\end{figure}

\noindent
To see the effect of drift dominance we also solve the macroscopic system \eqref{eq:macro-hyp}, \eqref{eq_S} obtained by hyperbolic scaling. Thereby we use (where applicable) the same set of parameters and boundary conditions as before for the parabolic scaling (Table \ref{table}). For the scaling parameter we take $\varepsilon =10^{-5}$. The initial conditions are those in set \eqref{eq:ICs-neu}, as visualized in Subfigures \ref{fig:IC-tumor-neu}, \ref{fig:IC-acid-neu}. Here we  consider an unsymmetric tissue with mesoscopic orientational distribution $q_h$ as in \eqref{eq_q_h}. Figure \ref{fig:tissue-hyp} shows the mean fiber orientation $\mathbb E_q$ corresponding to $q_h$ along with a magnification to observe the directionality in the neighborhood of the crossing fiber strands, and with $q_h$ plotted for $\delta =0.2$ and a specific direction $\xi =\pi/2$. \\[-2ex]

\begin{figure}[!htbp]
	\subfigure[][]{\includegraphics[width=0.23\textwidth]{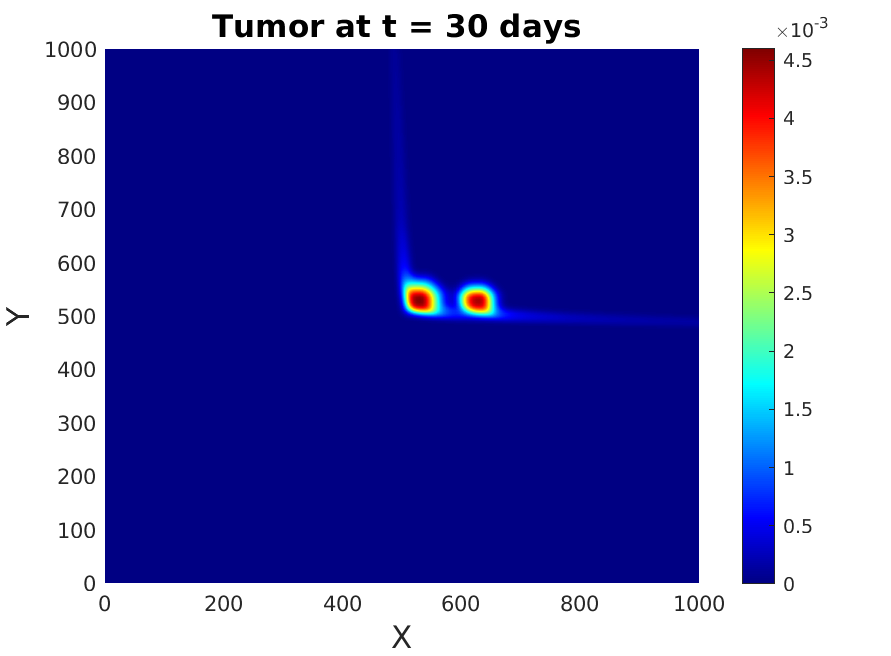}\label{fig:tumor-30days}}\quad \subfigure[][]{\includegraphics[width=0.23\textwidth]{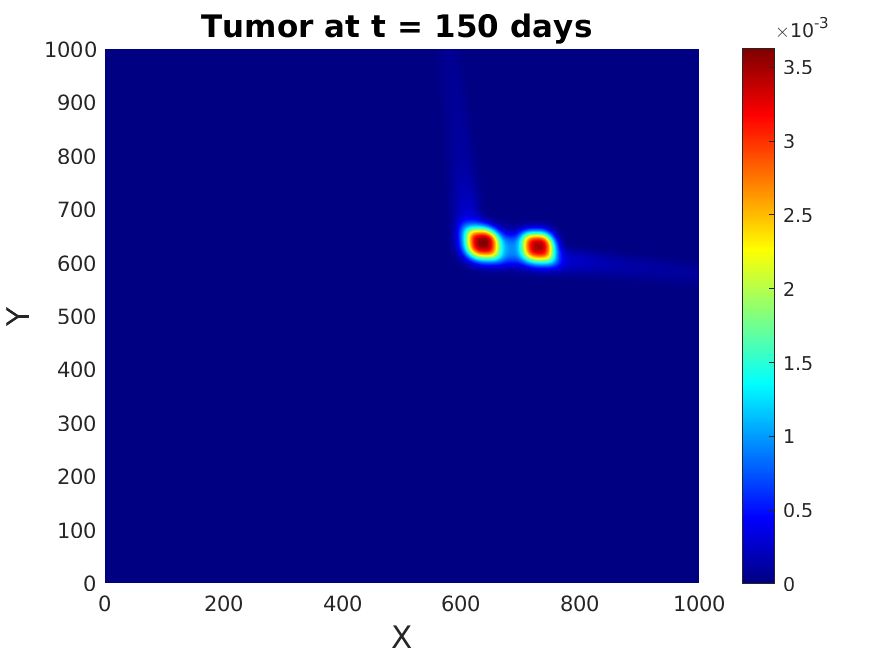}\label{fig:tumor-150days}}\quad \subfigure[][]{\includegraphics[width=0.23\textwidth]{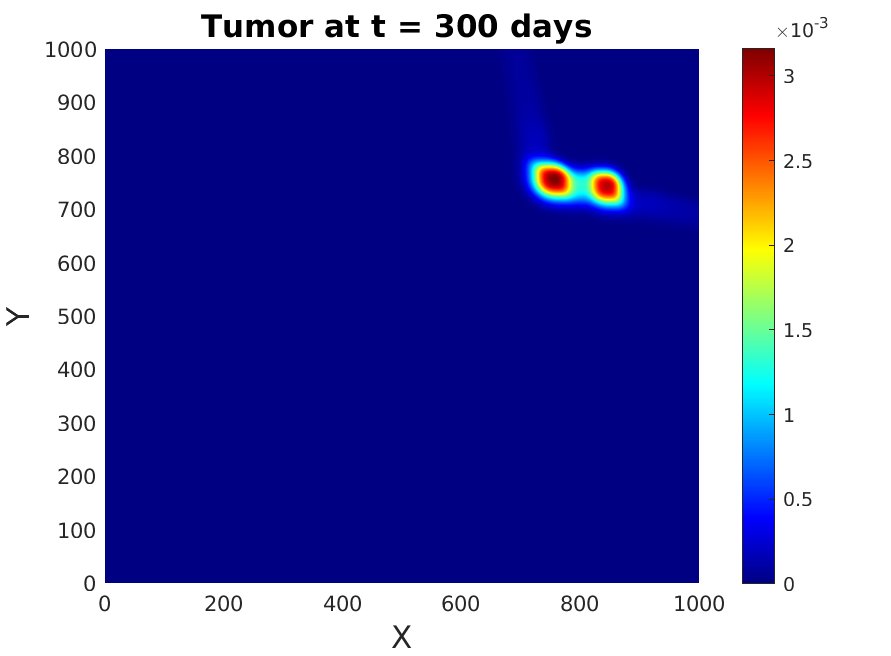}\label{fig:tumor-300days}}\quad \subfigure[][]{\includegraphics[width=0.23\textwidth]{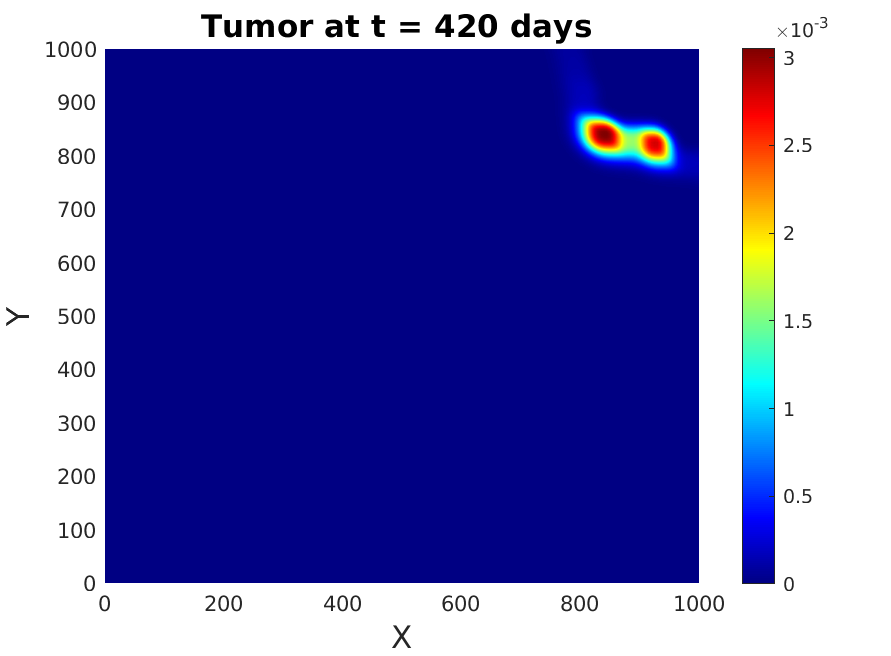}\label{fig:tumor-420days}}\\
\subfigure[][]{\includegraphics[width=0.23\textwidth]{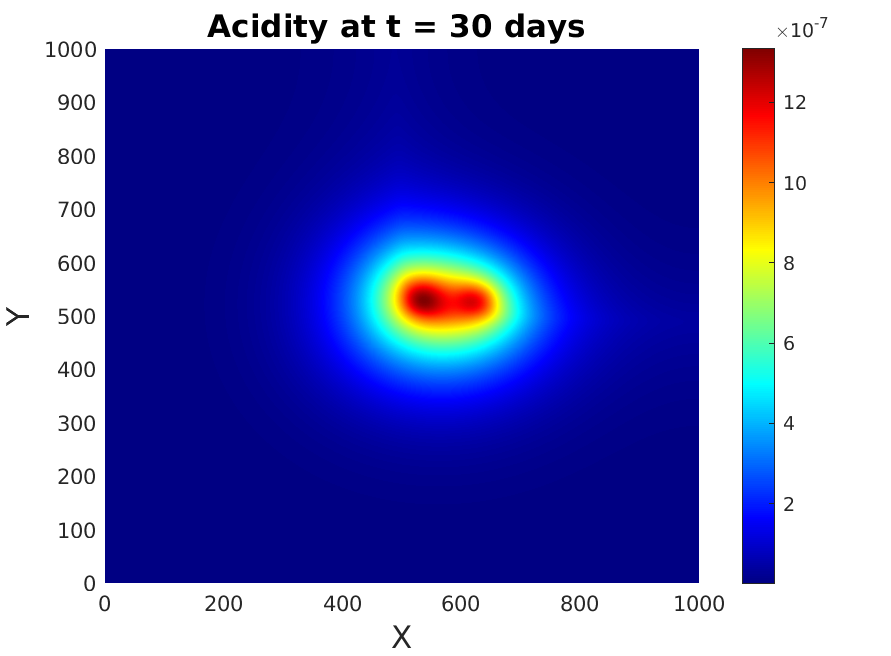}\label{fig:acid-30days}}\quad \subfigure[][]{\includegraphics[width=0.23\textwidth]{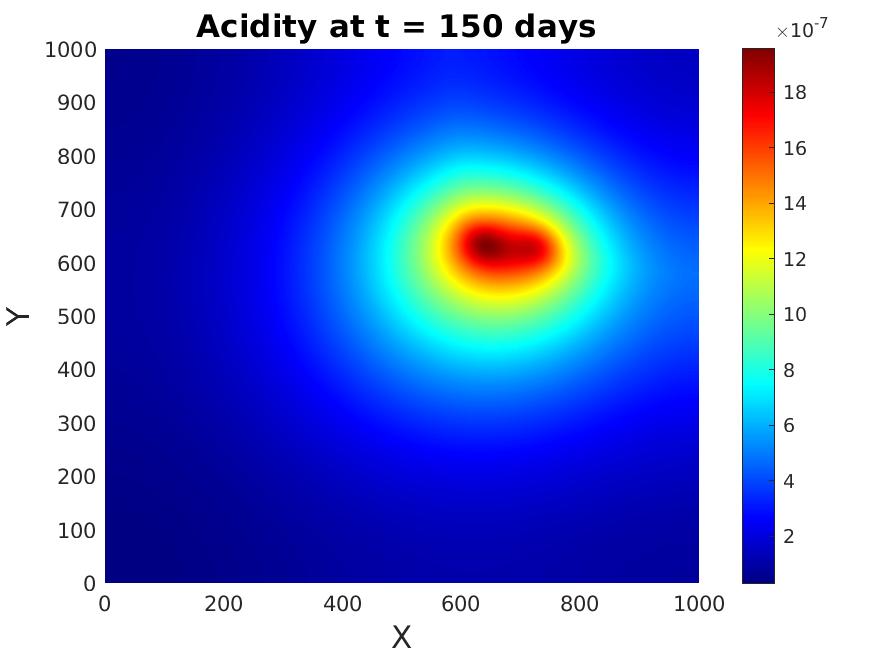}\label{fig:acid-150days}}\quad \subfigure[][]{\includegraphics[width=0.23\textwidth]{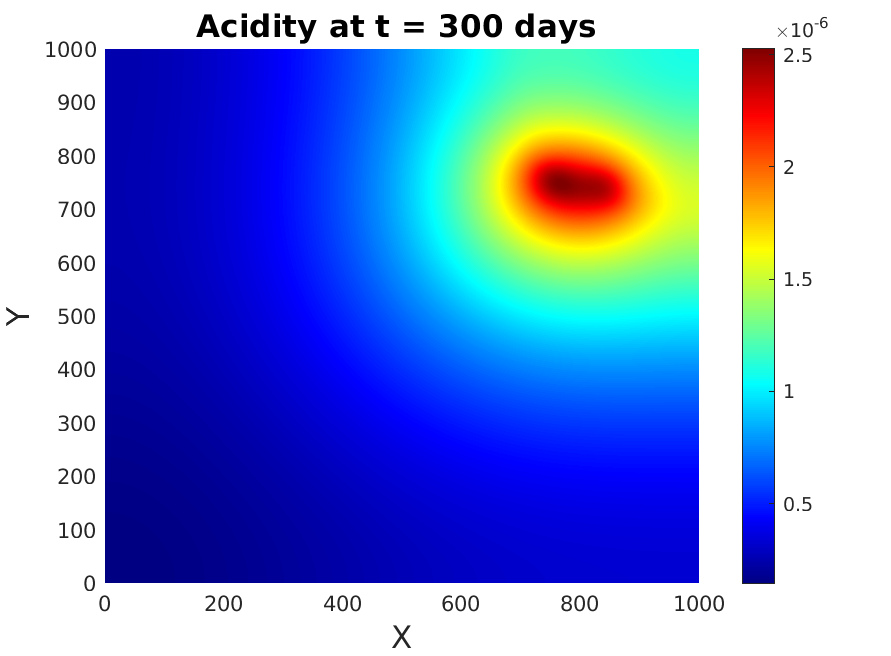}\label{fig:acid-300days}}\quad \subfigure[][]{\includegraphics[width=0.23\textwidth]{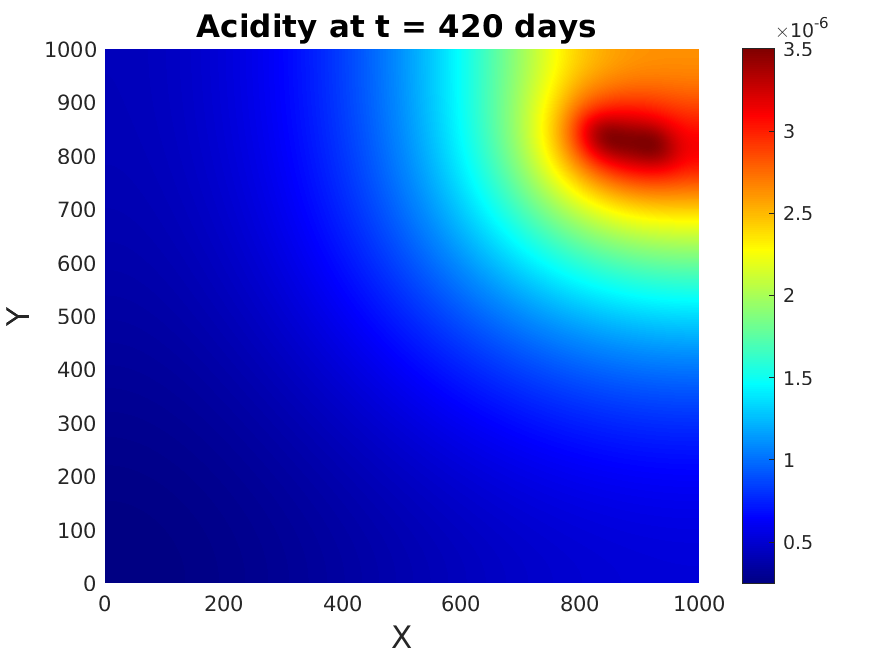}\label{fig:acid-420days}}\\	
\caption[]{Tumor (upper row) and acidity (lower row) at several times for $q_h$ as in \eqref{eq_q_h} with $\delta =0.2$ and initial conditions \eqref{eq:ICs-neu}. Solutions of system \eqref{eq:macro-hyp}, \eqref{eq_S} obtained by hyperbolic scaling.}
\label{fig:hyp-scaling-ex2}
\end{figure}

\noindent
The results obtained by solving the system for the evolution of tumor cells and acidity are shown in Figure \ref{fig:hyp-scaling-ex2}. Although we ran the simulations for a longer time than we did for the system obtained via parabolic scaling no pseudopalisade patterns are formed. Rather, the drift-dominated PDE for glioma cell density drives the cells along the positive $x$ and $y$ directions (as $\delta =0.2$ makes the second term in \eqref{eq_q_h} dominant). The cells 'escaping' that influence move fast along the diagonal $\gamma $ towards the right upper corner and cannot form the pattern in due time. A quick comparison with Figure \ref{fig:tumor-acid-exp_anisotropic-neu} obtained for the parabolic limit and $q$ as in \eqref{eq_q} shows the radically different behavior w.r.t. the two approaches.\\[-2ex]

\begin{figure}[!htbp]
	\subfigure[][]{\includegraphics[width=0.32\textwidth]{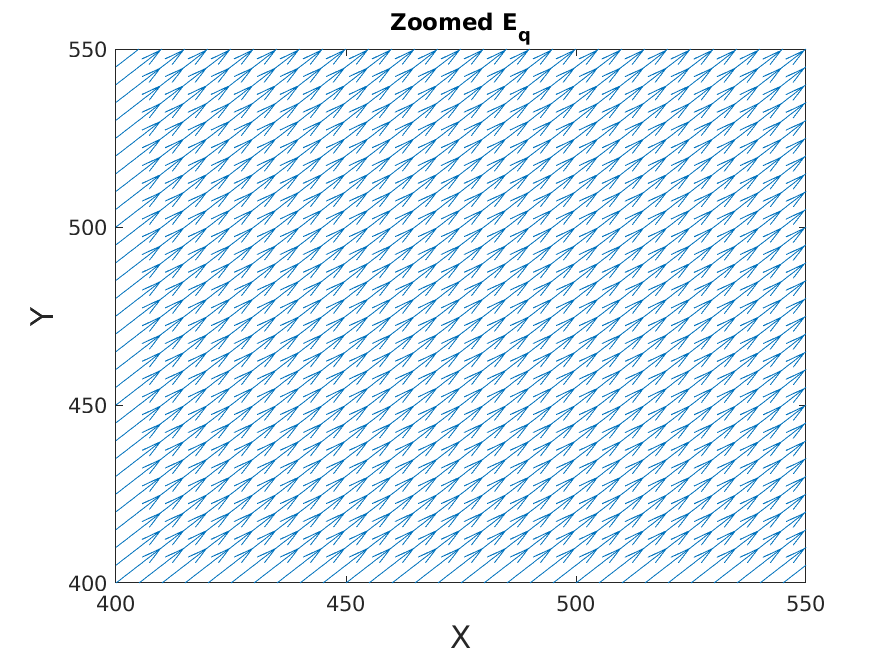}\label{fig:Eq-zoomed_delta1}}\quad 
	\subfigure[][]{\includegraphics[width=0.32\textwidth]{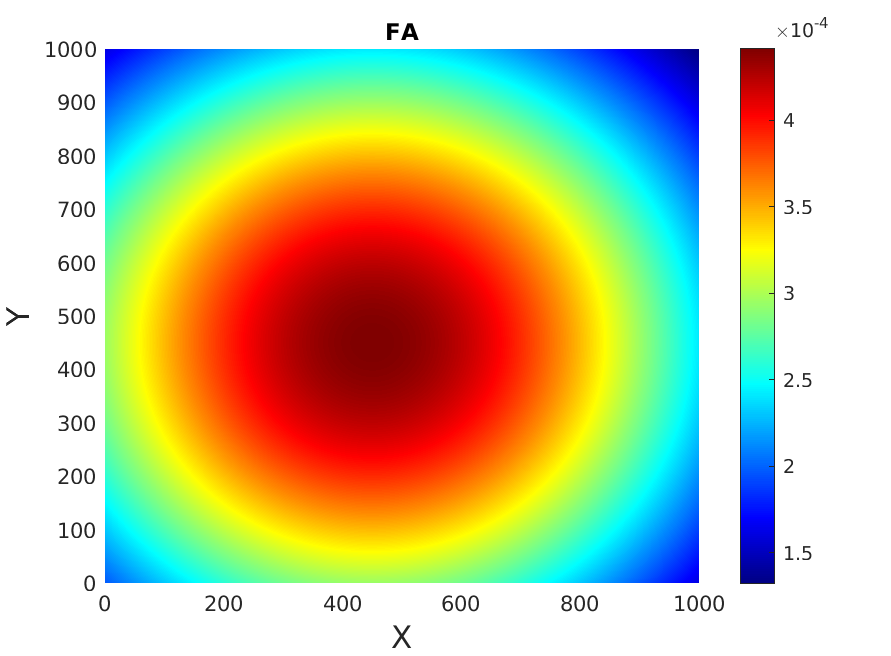}\label{fig:FA_delta1}}\quad
	\subfigure[][]{\includegraphics[width=0.32\textwidth]{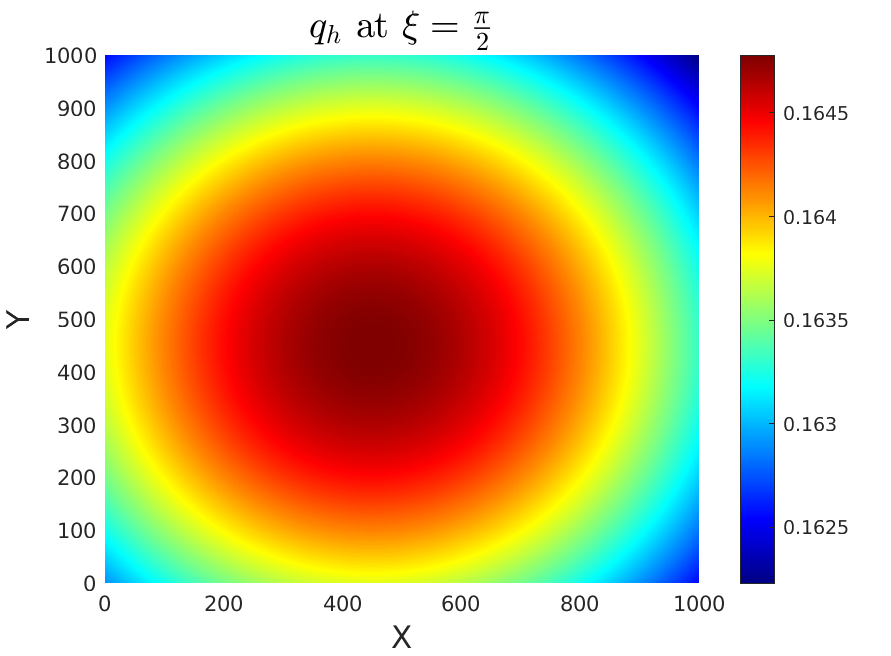}\label{fig:q_pipe2_delta1}}
	\caption[]{Zoomed mean fiber orientation $\mathbb E_q$  (\ref{fig:Eq-zoomed_delta1}), fractional anisotropy FA  (\ref{fig:FA_delta1}) for $q_h$ as in \eqref{eq_q_h} with $\delta =1$. Subfigure \ref{fig:q_pipe2_delta1}: mesoscopic tissue density $q_h$ for direction $\xi =\pi/2$.}	
	\label{fig:hyp-delta1-tissue}
\end{figure}

\begin{figure}[!htbp]
	\subfigure[][]{\includegraphics[width=0.23\textwidth]{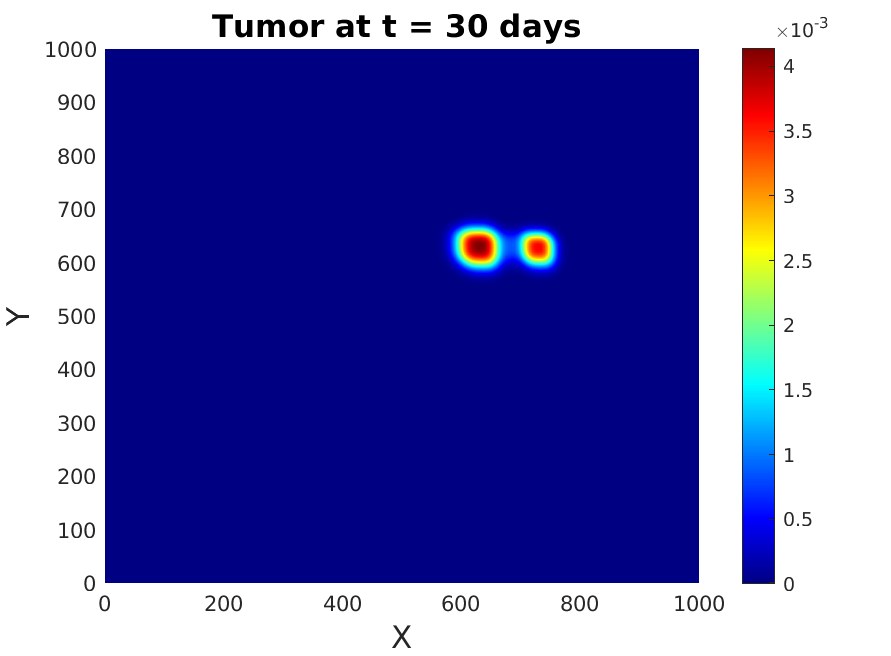}\label{fig:tumor-30days-delta1-hyp}}\quad \subfigure[][]{\includegraphics[width=0.23\textwidth]{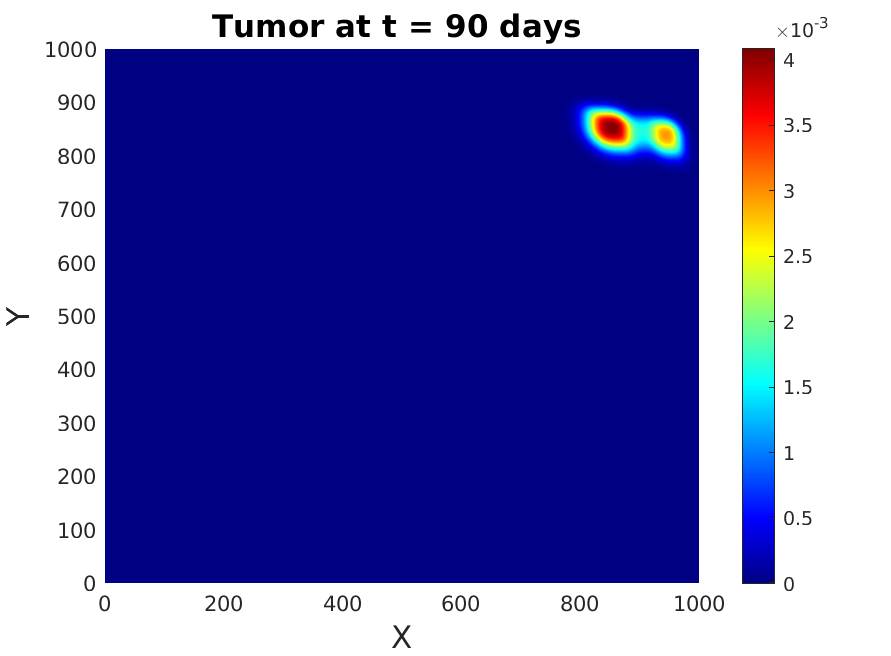}\label{fig:tumor-90days-delta1-hyp}}\quad \subfigure[][]{\includegraphics[width=0.23\textwidth]{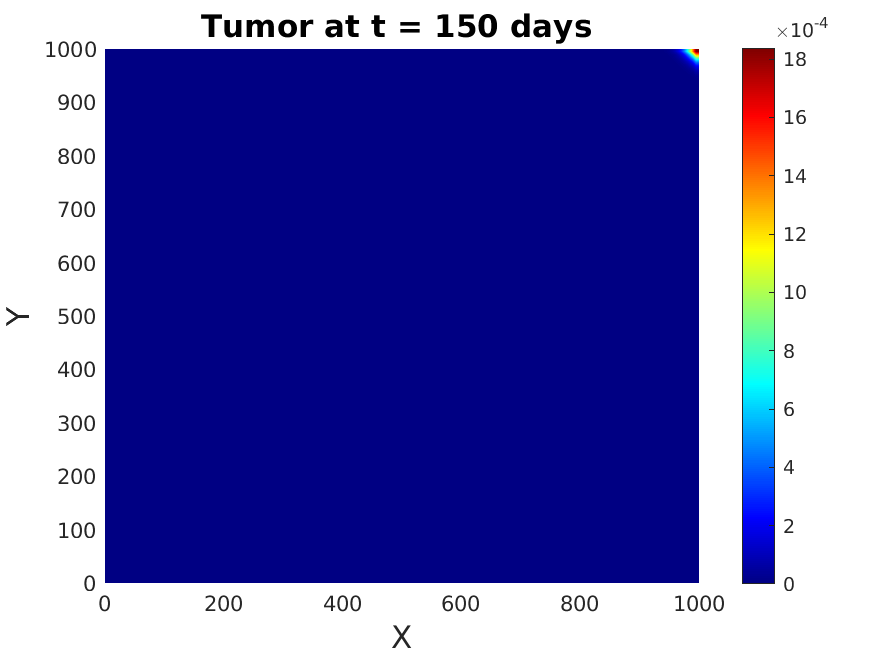}\label{fig:tumor-150days-delta1-hyp}}\quad \subfigure[][]{\includegraphics[width=0.23\textwidth]{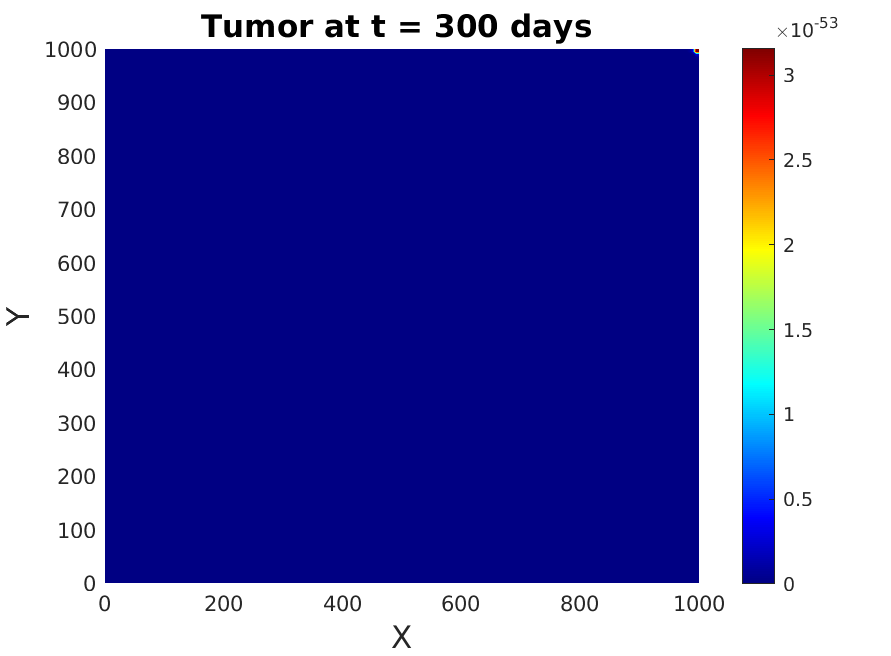}\label{fig:tumor-300days-delta1-hyp}}\\
	\subfigure[][]{\includegraphics[width=0.23\textwidth]{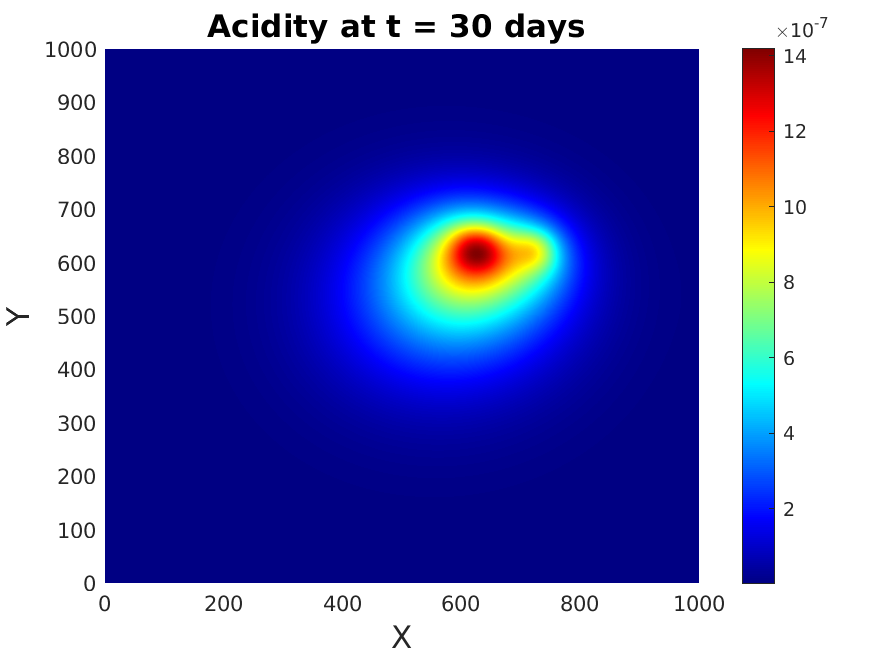}\label{fig:acid-30days-delta1-hyp}}\quad \subfigure[][]{\includegraphics[width=0.23\textwidth]{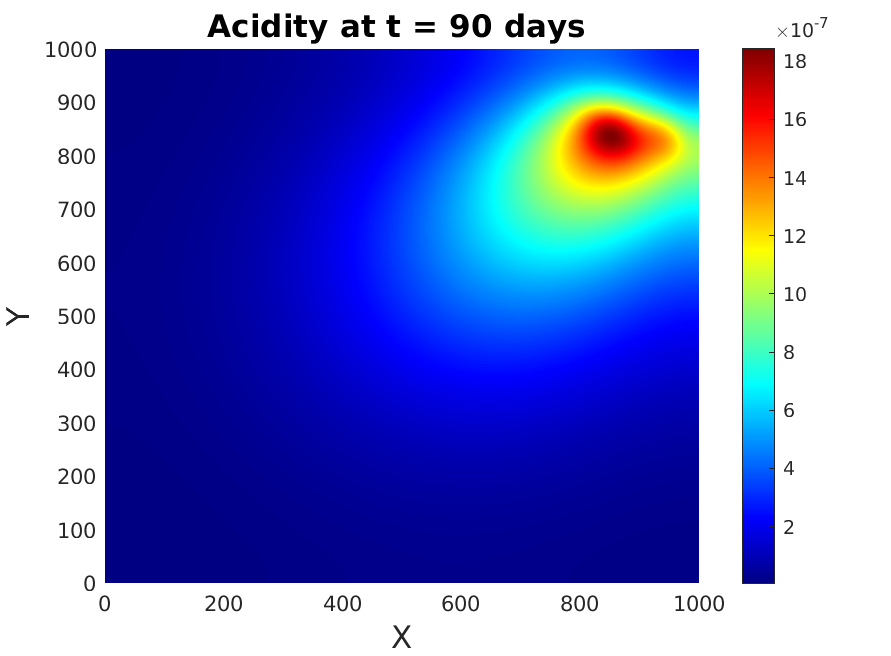}\label{fig:acid-90days-delta1-hyp}}\quad \subfigure[][]{\includegraphics[width=0.23\textwidth]{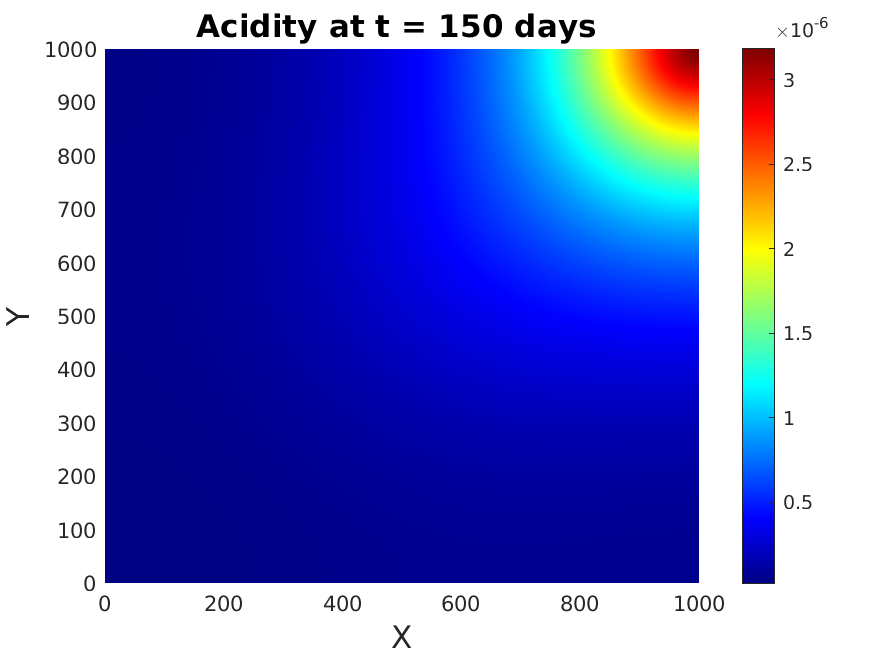}\label{fig:acid-150days-delta1-hyp}}\quad \subfigure[][]{\includegraphics[width=0.23\textwidth]{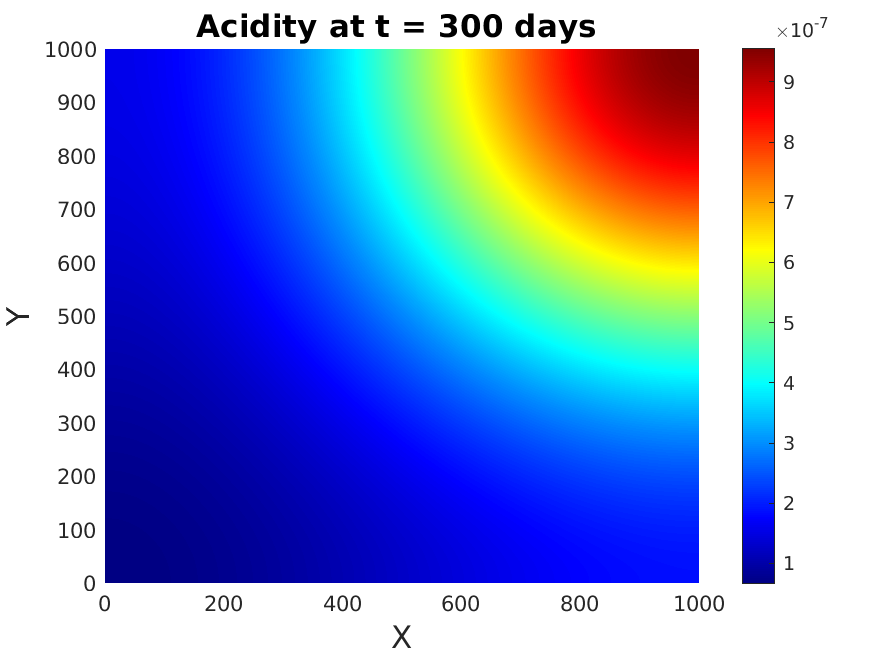}\label{fig:acid-300days-delta1-hyp}}\\	
	\caption[]{Tumor (upper row) and acidity (lower row) at several times for $q_h$ as in \eqref{eq_q_h} with $\delta =1$ and initial conditions \eqref{eq:ICs-neu}. Solutions of system \eqref{eq:macro-hyp}, \eqref{eq_S} obtained by hyperbolic scaling.}
	\label{fig:tumor-acid-delta1-hyp}
\end{figure}

\noindent
Similar observations apply when the first von Mises distribution in \eqref{eq_q_h} exerts full influence (for $\delta =1$). Figure \ref{fig:hyp-delta1-tissue} shows tissue characteristics for this case: fractional anisotropy FA, zoomed $\mathbb E_q$, and $q_h$ for $\xi =\pi/2$. The very low FA values indicate a highly isotropic tissue. Figure \ref{fig:tumor-acid-delta1-hyp} illustrates the behavior of tumor cell density and acidity for this case, in which the glioma cells are migrating very fast along the diagonal $\gamma$, accompanied by acidity they produce. When reaching the right uppermost corner of the domain they remain there (due to the no-flux boundary conditions) and further express acidity, eventually both solution components getting depleted. This is again in striking contrast to the solution behavior obtained by parabolic scaling for isotropic and undirected tissue (compare with Figure \ref{fig:tumor-acid-exp_isotropic-neu}) and, since such evolution is not seen in histologic patterns, it endorses the suspicion of the underlying tissue being undirected, at the same time speaking against hyperbolic scaling. As a casual observation there can be blow-up also in this case, however for a much (three orders of magnitude) stronger proton buffering than in the parabolic case. \\[-2ex]

\noindent
We also solved the system \eqref{eq:macro-hyp}, \eqref{eq_S} upon using several other initial conditions and parameter sets, none of which led to the formation of pseudopalisades. The observed behavior does not significantly change for any choice of the scaling parameter $\varepsilon\in [10^{-6},10^{-2}]$. Thus, since such patterns are actually  observed in histologic samples of glioblastoma, the simulations suggest that the fibers of brain tissue do not seem to be directed. This endorses the parabolic upscaling approach and goes along with a diffusion dominated motion, correspondingly biased by acidity gradients. With these, the cells are primarily driven by acidity, but also influenced by the underlying,  undirected tissue. The interplay between these actors leads to various types of patterns, depending on the parameter range and the relationship between the parameters.

\section{Qualitative analysis of the macroscopic reaction-diffusion-taxis system }\label{sec:analysis}

\subsection{Main results}
We consider system \eqref{macro_eq_non}, \eqref{macro_eq_non_S} with a slight modification of the source term in \eqref{macro_eq_non_S}:
{\setlength\abovedisplayskip{4pt}
	\setlength\belowdisplayskip{4pt}
	\begin{align}
	\left\{
	\begin{array}{ll}
	M_t=\nabla\cdot(\mathbb{D}_T(\xbf )\nabla M)+\nabla\cdot\left((g(S)\mathbb D_T(\xbf )\nabla S+\ubf (\xbf ))M\right)+f(M,S),\qquad (t,\xbf)\in (0,+\infty)\times\Omega ,
	\\
	S_t=\Delta S+\frac{\delta M }{1+M}-\alpha S,\qquad(t,\xbf)\in (0,+\infty)\times\Omega ,
	\\
	\Big (\mathbb{D}_T(\xbf )\nabla M+\ubf (\xbf )M\Big )\cdot \nubf=\nabla S\cdot \nubf =0,\qquad(t,\xbf)\in (0,+\infty)\times\partial\Omega ,
	\\
	M(0,\xbf )=M_0(\xbf ),\quad S(0,\xbf )=S_0(\xbf ),\qquad \xbf \in\Omega,
	\end{array}\right.\label{1.1}
	\end{align}
}
with $g(S):=\frac{\Lambda}{(S+K)^2(S+K+B)}$ and $f(M,S)=\mu_0M(1-M)(1-S)$, where we use $\Lambda =\lambda_1k_D$, $K=k_D$, and $B=\lambda_0$ to denote the corresponding constants occurring in the expression of $g(S)$ as given in Subsection \ref{subsec:params}. $\Om \subset \R^N$ is considered to be a bounded domain with sufficiently smooth boundary $\pom$, all involved constants are positive, $M_0\in L^{\infty }(\Om )$, $M_0\not \equiv 0$, $M_0, S_0\ge 0$, and $S_0\in W^{1,\infty}(\Omega)$.
The no-flux boundary conditions are obtained through the upscaling procedure (as done e.g., in \cite{CEKNSSW} for a related problem). 

\noindent
For the tumor diffusion tensor $\mathbb{D}_T$ we require
\begin{assumpt}\label{assumpt}\ \\[-3ex]
	
\begin{enumerate}
	\item[(A)]\label{assmp:A} $\mathbb{D}_T(\xbf )\in \Big (C^{2,\gamma}(\Omega)\cap C(\overline\Omega)\Big )^{N\times N}$, $\gamma \in (0,1)$, $\ubf=\nabla \cdot \mathbb{D}_T$ is uniformly bounded in $\Omega $, and $\ubf(\xbf)=0$ for $\xbf\in\pom $;
	\item[(B)]\label{assmp:B} there exists $\vartheta>0$ such that for any $\xibf\in\mathbb R^N$ and $\xbf \in\Omega$, $$ \xibf^{\top}\cdot\mathbb{D}_T(\xbf )\cdot \xibf\geq\vartheta |\xibf|^2.$$
\end{enumerate}
\end{assumpt}
\begin{theorem}\label{Theorem1.1}
	Let $N\geq1$. Suppose that Assumption \ref{assumpt} holds. Then if $\zeta<\alpha$ and $\|S_0\|_{L^\infty(\Omega)}<1$, system \eqref{1.1} admits a unique global bounded classical solution.
\end{theorem}

\begin{theorem}\label{Theorem1.2}
	Under the assumptions of Theorem \ref{Theorem1.1}, suppose moreover that $\nabla \cdot \ubf (\xbf )=0$ for all $\xbf \in \Om$. Then there exists $\mu^*>0$, such that if $\mu_0>\mu^*$, for any $\xbf \in\Omega$ we have
	\begin{align*}
	\lim_{t\to\infty}M(t,\xbf)=1,\quad\lim_{t\to\infty}S(t,\xbf)=\frac{\zeta}{2\alpha}.
	\end{align*}
Moreover, there exists $C>0$ and $D>0$ such that for all $t\in[0,+\infty)$,
\begin{align*}
\left \|M(t,\cdot)-1\right \|_{L^\infty(\Omega)}\leq Ce^{-\frac{Dt}{N+2}},
\\
\left \|S(t,\cdot)-\frac{\zeta}{2\alpha }\right \|_{L^\infty(\Omega)}\leq Ce^{-\frac{Dt}{N+2}}.
\end{align*}
\end{theorem}

\subsection{Global existence, uniqueness, and boundedness of solutions}

\noindent
Firstly, we state a result concerning local existence of classical solutions, which can be proved  by well-established methods involving standard parabolic regularity theory and an appropriate fixed point framework. Moreover, one can thereby derive a sufficient condition for extensibility of a given local-in-time solution(see \cite{winkler3} or \cite{cao} for example).

\begin{lemma}\label{Lemma1.1} Let $\Omega\subset\mathbb R^N$ ($N\geq 1$) be  a bounded domain with smooth boundary. Suppose that the nonnegative functions $M_0, S_0$ are in $W^{1,\infty}(\Omega)$. Then there exist $T_{max}\in(0,\infty]$   and a unique pair of non-negative functions
	$(M,S)$ satisfying
	$$M\in C^0([0,T_{max}); C^0(\overline{\Omega}))\cap C^{2,1}((0,T_{max})\times \overline{\Omega}),$$
	$$S\in C^0([0,T_{max}); C^0(\overline{\Omega}))\cap L^\infty_{loc}([0,T_{max}); W^{1,\infty}(\Omega))
	\cap C^{2,1}((0,T_{max})\times \overline{\Omega}),$$
	and solving \eqref{1.1} classically in $\Omega\times(0,T_{max})$. Moreover, if
	$T_{max}<\infty$, then
	$$\limsup_{t\to T_{max}}\left(\|M(t,\cdot)\|_{L^\infty(\Omega)}+
	\|S(t,\cdot)\|_{W^{1,\infty}(\Omega)}\right)=\infty .$$
\end{lemma}

\noindent
Next we prove results relating to the global boundedness of solutions to \eqref{1.1}.

\begin{lemma}\label{Lemma2.1}
	There exists $C_S>0$ such that
	\begin{align*}
	\|S\|_{L^\infty([0,T_{max})\times\Omega)}\leq&\max\left\{\frac\zeta\alpha,\|S_0\|_{L^\infty(\Omega)}\right\},
	\\
	\|\nabla S\|_{L^\infty([0,T_{max})\times\Omega)}\leq& C_S\left(\|\nabla S_0\|_{L^\infty(\Omega)}+1\right).
	\end{align*}
\end{lemma}

\noindent
\proof Taking $pS^{p-1}$ ($p>1$) as a test function for the $S$-equation of \eqref{1.1}, for any $\varepsilon\in(0,1)$, we obtain
\begin{align}
\frac{d}{dt}\int_{\Omega}S^p=&-\frac{4(p-1)}{p}\int_\Omega|\nabla S^{\frac p2}|^2+p\zeta \int_{\Omega}\frac{MS^{p-1}}{1+M}-\alpha p\int_{\Omega}S^p\label{1.11}
\\
&\leq-\frac{4(p-1)}{p}\int_\Omega|\nabla S^{\frac p2}|^2+\frac{p\zeta^p|\Omega|}{\alpha^{p-1}(1-\varepsilon)^{p-1}}-\varepsilon\alpha p\int_{\Omega}S^p,\nonumber
\end{align}
from which we obtain
\begin{align*}
\frac{d}{dt}\int_{\Omega}S^p\leq \frac{p\zeta^p|\Omega|}{\alpha^{p-1}(1-\varepsilon)^{p-1}}-\varepsilon\alpha p\int_{\Omega}S^p
\end{align*}
and then by Gronwall's inequality
\begin{align}
\int_{\Omega}S^p\leq \int_{\Omega}S_0^p+\frac{\zeta^p|\Omega|}{\varepsilon  \alpha^{p}(1-\varepsilon)^{p-1}},
\end{align}
from which we obtain that for any $t\in[0,T_{max})$,
\begin{align*}
\|S(t,\cdot)\|_{L^\infty(\Omega)}=
&\lim_{p\to\infty}\left(\int_{\Omega}S^p\right)^{\frac1p}
\\
\leq&\lim_{p\to\infty}\left(\int_{\Omega}S_0^p
+\frac{\zeta^p|\Omega|}{\varepsilon\alpha^p(1-\varepsilon)^{p-1}}\right)^{\frac1p}
\\
=&\max\left\{\frac{\zeta}{\alpha(1-\varepsilon)},\|S_0\|_{L^\infty(\Omega)}\right\}.
\end{align*}
From the arbitrariness of $\varepsilon\in(0,1)$ we therefore obtain $$\|S\|_{L^\infty([0,T_{max})\times \Omega)}\leq\max\left\{\frac{\zeta}{\alpha},\|S_0\|_{L^\infty(\Omega)}\right\}.$$

\noindent
On the other hand, from the $L^p$-$L^q$ estimates for the Neumann heat semigroup on a bounded domain and the fact that
\begin{align*}
S=e^{t\Delta}S_0+\int_0^te^{(t-s)\Delta}\left(\frac{\zeta M}{1+M}-\alpha S\right),
\end{align*}
we obtain for all $t\in(0,T_{max})$,
\begin{align*}
\|\nabla S(t,\cdot)\|_{L^\infty(\Omega)}&=\|\nabla e^{t\Delta}S_0\|_{L^\infty(\Omega)}+\int_0^t\|\nabla e^{(t-s)\Delta}\left(\frac{\zeta M}{1+M}-\alpha S\right)\|_{L^\infty(\Omega)}
\\
&\leq C_1e^{-\lambda_1 t}\|\nabla S_0\|_{L^\infty(\Omega)}+C_2\int_0^te^{-\lambda_1 (t-s)}(1+(t-s)^{-\frac12})\|\zeta+\alpha S\|_{L^\infty(\Omega)}
\\
&\leq C_S\left(\|\nabla S_0\|_{L^\infty(\Omega)}+1\right),
\end{align*}
where $\lambda_1>0$ denotes the first nonzero eigenvalue of $-\Delta$ in $\Omega\subset\mathbb R^N$ under the Neumann boundary condition.
\begin{lemma}\label{lem:Lemma2.3}
	Under the assumptions of Theorem \ref{Theorem1.1}, for any $p>1$, there exists $C(p)>0$ such that for $t\in(0, T_{max})$, we have
	$$\|M(t,\cdot)\|_{L^p(\Omega)}\leq C(p).$$
\end{lemma}

\noindent
\proof Taking $pM^{p-1}$ as a test function for the $M$-equation of \eqref{1.1} and denoting $D_0:=\|\mathbb{D}_T(\cdot)\|_{L^\infty(\Omega)}$, from the no-flux boundary, we obtain
\begin{align}
\frac{d}{dt}\int_{\Omega}M^p=&-\frac{4(p-1)}{p}\int_\Omega(\nabla M^{\frac p2})^{\top}\cdot\mathbb{D}_T(\xbf )\cdot\nabla M^{\frac p2}
-(p-1)\int_\Omega \ubf (\xbf )\cdot\nabla M^p\label{1.5}
\\
&-(p-1)\int_\Omega (\nabla M^p)^{\top}g(S)\mathbb{D}_T(\xbf )\nabla S+\mu_0 p\int_{\Omega}M^p (1-M)(1-S)\nonumber
\\
\leq&-\frac{4(p-1)\vartheta}{p}\int_\Omega|\nabla M^{\frac p2}|^2+\frac{2(p-1)\vartheta}{p}\int_\Omega|\nabla M^{\frac p2}|^2+\frac{p(p-1)}{2\vartheta}D_0^2\int_\Omega M^p\nonumber
\\
&+\frac{2(p-1)\vartheta}{p}\int_\Omega|\nabla M^{\frac p2}|^2+C_3\int_{\Omega}M^p
+\mu_0 p\int_{\Omega}M^p-C_4\int_{\Omega}M^{p+1}\nonumber
\\
=&(C_3+\mu _0p +\frac{p(p-1)}{2\vartheta}D_0^2)\int_{\Omega}M^p-C_4\int_{\Omega}M^{p+1}\nonumber
\\
\leq&C_5-\mu_0 p\int_{\Omega}M^{p},\nonumber
\end{align}
where 
  $$C_3:=\frac{(p-1)p}{2\vartheta}D_0^2C_S^2\left(\|\nabla S_0\|_{L^\infty(\Omega)}+1\right)^2\frac{\Lambda^2}{K^4(K+B)^2},$$
  $$C_4=\mu_0 p\left(1-\max\left\{\frac{\zeta}{\alpha},\|S_0\|_{L^\infty(\Omega)}\right\}\right),\quad
C_5:=(C_3+2\mu_0 p+\frac{p(p-1)}{2\vartheta}D_0^2)^{p+1}|\Omega|C_4^{-p}.$$

\noindent
Thus we obtain that for any $t\in(0,T_{max})$,
$$\|M(t,\cdot)\|_{L^p(\Omega)}\leq \left(\int_{\Omega}M_0^p+\frac{C_5}{\mu_0 p}\right)^{\frac1p}:=C(p).$$

\noindent
{\bf Proof of Theorem \ref{Theorem1.1}.} From Lemma \ref{lem:Lemma2.3} and the standard Moser iteration process,
there exists $C>0$ such that $\|M(t,\cdot)\|_{L^\infty(\Omega)}\leq C$ for all $t\in(0,T_{max})$. Then in view of Lemma \ref{Lemma1.1},  Theorem \ref{Theorem1.1} is a direct consequence of  Lemma \ref{Lemma2.1}. 

\subsection{Long time behavior}
\begin{lemma}\label{Lemma2.2}
	Under the assumptions of Theorem \ref{Theorem1.2}, there exists $\mu^*>0$ defined in \eqref{1.9} such that for $\mu_0>\mu^*$, for all $t>0$, the function
	$$F(t)=\int_{\Omega}(M-1-\ln M)+\frac{C_M}{2}\int_{\Omega}\left (S-\frac{\zeta}{2\alpha}\right )^2$$
	satisfies
	$$F'(t)\leq-D(t),$$
	where
	$$D(t)=D\left\{\int_{\Omega}(M-1)^2+\frac{C_M}{2}\int_{\Omega}\left (S-\frac{\zeta}{2\alpha}\right )^2\right\}$$
	with $D$ a constant defined in \eqref{1.10}.
\end{lemma}

\noindent
\proof According to the strong maximum principle and the assumption $M_0\not\equiv0$, $M$ is positive in $(0,\infty)\times\Omega$. Testing the $M$-equation of \eqref{1.1} by $1-\frac1M$, by Young's inequality and the fact of $\nabla\cdot \ubf (\xbf )=0$ for $\xbf \in\Omega$, $\ubf(\xbf)=0$ for $\xbf\in\partial\Omega$, using the no-flux boundary condition, we obtain that there exists $C_M>0$ such that
\begin{align}
\frac{d}{dt}\int_{\Omega}(M-1-\ln M)=&-\int_{\Omega}\left(\frac{\nabla M}{M}\right)^{\top}\mathbb{D}_T(\xbf )\frac{\nabla M}{M}-\int_{\Omega}\left(\frac{\nabla M}{M}\right)^{\top}g(S)\mathbb{D}_T(\xbf )\nabla S\nonumber
\\
&+\int_{\Omega}\nabla\cdot \ubf \ln M-\mu_0
\int_\Omega(M-1)^2\left(1-S\right)\label{1.7}
\\
\leq&-\vartheta\int_{\Omega}\left|\frac{\nabla M}{M}\right|^2+\vartheta\int_{\Omega}\left|\frac{\nabla M}{M}\right|^2+C_M\int_{\Omega}|\nabla S|^2\nonumber
\\
&-\mu_0\left(1-\max\left\{\frac{\zeta}\alpha,\|S_0\|_{L^\infty(\Omega)}\right\}\right)
\int_\Omega(M-1)^2\nonumber
\\
=& C_M\int_{\Omega}|\nabla S|^2-\mu_0\left(1-\max\left\{\frac{\zeta}\alpha,\|S_0\|_{L^\infty(\Omega)}\right\}\right)
\int_\Omega(M-1)^2\nonumber
\end{align}
with 
$C_M:=\frac{D_0^2\Lambda^2}{4\vartheta K^4(K+B)^2}.$ Testing the $S$-equation of \eqref{1.1} by $S-\frac{\zeta}{2\alpha}$, we obtain
\begin{align}
 \frac12\frac{d}{dt}\int_{\Omega}\left (S-\frac{\zeta}{2\alpha}\right )^2=&-\int_{\Omega}|\nabla S|^2+\frac{\zeta}{2}\int_{\Omega}\frac{M-1}{(1+M)}(S-\frac{\zeta}{2\alpha})
-\alpha\int_{\Omega}\left (S-\frac{\zeta}{2\alpha}\right )^2\label{1.8}
\\
\leq&-\int_{\Omega}|\nabla S|^2-\frac\alpha2\int_{\Omega}\left (S-\frac{\zeta}{2\alpha}\right )^2+\frac{\zeta^2}{8\alpha}\int_{\Omega}(M-1)^2.\nonumber
\end{align}
 Combining \eqref{1.7} and \eqref{1.8}, we obtain
 \begin{align*}
&\frac{d}{dt}\int_{\Omega}\left[M-1-\ln M+\frac{C_M}{2}\left (S-\frac{\zeta}{2\alpha}\right )^2\right]
\\
\leq &\left(\frac{\zeta^2C_M}{8\alpha}-\mu_0\left(1-\max\left\{\frac{\zeta}\alpha,\|S_0\|_{L^\infty(\Omega)}\right\}\right)\right)\int_\Omega(M-1)^2-\frac{\alpha C_M}2\int_{\Omega}\left (S-\frac{\zeta}{2\alpha}\right )^2.
\end{align*}
By choosing
\begin{align}
\mu^*:&=\frac{\zeta^2 C_M}{4\alpha\left(1-\max\{\frac\zeta\alpha,\|S_0\|_{L^\infty(\Omega)}\}\right)},\label{1.9}
\\
D:&=\min\left \{\frac{\zeta^2C_M}{8\alpha},\alpha\right \},\label{1.10}
\end{align}
we obtain that $\mu_0>\mu^*$ leads to $F'(t)\leq-D(t)$.\\

\noindent
{\bf Proof of Theorem \ref{Theorem1.2}.} The proof of Theorem \ref{Theorem1.2} is very standard. We include the proof here for completeness.
Denote $h(s):=s-1-\ln s$. Noticing that $h'(s)=1-\frac1s$ and $h''(s)=\frac1{s^2}>0$ for all $s>0$, we obtain that $h(s)\geq h(1)=0$ and $F(t)$ is nonnegative. From Lemma \ref{Lemma2.2}, we have $F'(t)\leq-D(t)$ and then
$$\int_0^tD(\tau)d\tau\leq F(0)$$
for all $t>0$, from which we have
$$\int_0^t \left\{\int_{\Omega}(M-1)^2+\frac{C_M}{2}\int_{\Omega}\left (S-\frac{\zeta}{2\alpha}\right )^2\right\}<\infty.$$
Using a similar argument as in Lemma 3.10 of \cite{Tao16}, we can obtain the uniform convergence of solutions, namely
$$\|M(t,\cdot)-1\|_{L^\infty(\Omega)}\to0,\quad \left \|S(t,\cdot)-\frac{\zeta}{2\alpha }\right \|_{L^\infty(\Omega)}\to0$$
as $t\to\infty$. Then there exists $t_0>0$ such that for all $t>t_0$, $\|M-1\|_{L^\infty(\Omega)}\leq\frac{1}2$, which together with the fact that
$$\frac13(s-1)^2\leq h(s)\leq (s-1)^2\qquad \text{for all }s>\frac12$$
implies that
\begin{align}
\frac1{3}\int_\Omega(M-1)^2+\frac{C_M}{2}\int_{\Omega}\left (S-\frac{\zeta}{2\alpha}\right )^2\leq F(t)\leq \frac1D D(t)\label{1.43}
\end{align}
for all $t>t_0$.
Hence $$F'(t)\leq-D(t)\leq-DF(t),$$
from which we obtain
\begin{align}
F(t)\leq F(t_0)e^{-D(t-t_0)}.\label{1.44}
\end{align}
Substituting \eqref{1.44} into \eqref{1.43},  we obtain
$$\frac1{3}\int_\Omega(M-1)^2+\frac{C_M}{2}\int_{\Omega}\left (S-\frac{\zeta}{2\alpha }\right )^2\leq F(t_0)e^{-D(t-t_0)},$$
which implies that there exists $C>0$ such that for all $t>t_0$,
\begin{align*}
\|M(t,\cdot)-1\|_{L^2(\Omega)}\leq Ce^{-Dt/2},
\\
\left \|S(t,\cdot)-\frac{\zeta}{2\alpha}\right \|_{L^2(\Omega)}\leq Ce^{-Dt/2}.
\end{align*}
Furthermore, notice that there exists a constant $C_1>0$ such that
$$\|M(t,\cdot)-1\|_{W^{1,\infty}(\Omega)}\leq C_1,\quad \left \|S(t,\cdot)-\frac{\zeta}{2\alpha}\right \|_{W^{1,\infty}(\Omega)}\leq C_1\quad
\hbox{for all}\,t>0.$$
Thus the Gagliardo--Nirenberg inequality yields
\begin{align*}
\|M(t,\cdot)-1\|_{L^\infty(\Omega)}&\leq C\left(\|M(t,\cdot)-1\|_{W^{1,\infty}(\Omega)}^{\frac{N}{N+2}}
\|M(t,\cdot)-1\|_{L^2(\Omega)}^{\frac{2}{N+2}}+\|M(t,\cdot)-1\|_{L^2(\Omega)}\right)
\\
&\leq C\|M(t,\cdot)-1\|_{L^2(\Omega)}^{\frac{2}{N+2}}\leq Ce^{-\frac{Dt}{N+2}}
\end{align*}
for all $t>0$. Similarly, we can obtain
$$\left \|S(t,\cdot)-\frac{\zeta }{2\alpha }\right \|_{L^\infty(\Omega)}\leq Ce^{-\frac{Dt}{N+2}}.$$
This concludes the proof of Theorem \ref{Theorem1.2}.\\[-2ex]

\begin{remark}\textup{For the above rigorous results to hold we required among others that $\zeta <\alpha$, which means that the acidity buffering by the tumor environment is stronger than the production of protons by the cancer cells. While this is true for lower grade tumors, it no longer holds for more advanced neoplasms like GBM. Numerical simulations show that no pseudopalisades are forming, unless $\zeta $ substantially exceeds $\alpha $.} \\[-2ex]

\noindent	
\textup{In fact, in Subsection \ref{subsec:simulations} we already observed that $\alpha $ (which controls proton buffering) was the decisive parameter for the fate of the patterns and even for singularity formation. The weakening of proton production considered in this Section enhances the influence of acidity depletion, which due to the form of $f(M,S)$ contributes to keeping the glioma density bounded by its carrying capacity. A similar result can be obtained by replacing in $f(M,S)$ the factor $1-S$ with $1/(1+S)$. In that case there is no smallness requirement for $\zeta $; moreover, the results hold even if \eqref{macro_eq_non_S} is considered instead of the $S$-equation in \eqref{1.1}. No pseudopalisades are forming in this case either (recall Figure \ref{fig:source-modif}). }	
\end{remark}

\section{Discussion}\label{sec:discussion}

\noindent
The multiscale approach employed in this work allows to obtain a macroscopic description for the evolution of glioma cell density featuring repellent pH-taxis and providing the concrete forms of involved diffusion, transport, and taxis coefficients, upon starting from modeling on the microscopic level of cell-acidity interactions. This fully continuous setting is quite different from previous models \cite{Alfonso,martinez2012hypoxic} of pseudopalisade formation and spread, which are rather  accounting for vascularization and necrosis than for direct effects of acidity. Nevertheless, the system of two PDEs of reaction-(myopic) diffusion-advection type obtained by parabolic upscaling from lower levels of description is able to reproduce biologically observed patterns, whereby repellent pH-taxis does not seem to effectively trigger, but merely to enlarge such structures; depending on the acidity buffering potential of the tumor cells and their environment in relationship to their ability to proliferate, the resulting patterns can be assigned to lower or higher tumor grades, with pseudopalisades corresponding to the latter. This endorses the idea that proton buffering might be beneficial for decelerating progression towards GBM, see e.g. \cite{Boyd2017,Koltai2020} and references therein. For instance, genetic targeting of carbonic anhydrase 9 (a common hypoxia marker catalyzing the conversion of carbon dioxide to bicarbonate and protons) provided evidence of delayed tumor growth in the GBM cell line U87MG \cite{McIntyre2012}.  \\[-2ex]

\noindent
In our deduction of the macroscopic system from the KTAP framework we used for the turning rate $\lambda (z)=\lambda_0-\lambda _1z>0$. This  could be made more general, e.g. upon considering any regular enough function $\lambda $, expanding it around the steady-state $y^*$, and keeping the first two terms of the expansion: $\lambda (z)\simeq \lambda (y^*)-\lambda'(y^*)z:=\lambda _0(S)-\lambda _1(S)z$. The higher order terms will get anyway lost during the scaling process, due to ignoring the higher order moments w.r.t. $z$. The new coefficients $\lambda _0, \lambda _1$ are no longer constants, but depend on the macroscopic variable $S$ by way of $y^*$.\footnote{An essential requirement on $\lambda$ is thereby to ensure that $\lambda (y^*)>0$ and $\lambda '(y^*)>0$.} Consequently, the obtained macroscopic PDE for the glioma population density will have diffusion and taxis coefficients depending on $S$, thus leading to a more intricate coupling of the PDE system for $M$ and $S$.\\[-2ex]

\noindent
Beside including subcellular level information via a transport term w.r.t. the activity variable(s) and a turning rate depending therewith, we also considered an alternative way to account for cell reorientations in response to acidity levels. Trying to recover the same macroscopic limit led to a well-determined choice of the acidity-dependent function $h$ involved in the turning rate $\lambda (\vbf ,S)$ from \eqref{eq:alt-turning_rate}.
\\[-2ex]

\noindent
For the sake of simplicity we considered in \eqref{eq_S} a genuinely macroscopic PDE of reaction-diffusion type for the evolution of acidity. More detailed models involving intra- and extracellular proton dynamics with randomness have been introduced in \cite{Hiremath2015,AthniHiremath2016,AthniHiremath2018,Kloeden2016}, some of them connecting it to the dynamics of tumor cells. The latter inferred, however, a rather heuristic, mainly macroscopic description, with coefficients possibly depending on such microscopic quantities like concentration of intracellular protons. Connecting multiscale formulations of proton and cell dynamics and identifying an appropriate way of upscaling to deduce the corresponding macroscopic equations would be a first step towards accounting for subcellular processes in a manner which is detailed enough to capture such low-scale events, but also eventually simplified enough to still enable efficient computations. \\[-2ex]

\noindent
The observation that no pseudopalisades seem to emerge for a transport-dominated system as obtained by hyperbolic scaling of the micro-meso setting suggests that the microscopic brain tissue is undirected, at least w.r.t. glioma migration along its constituent fibers. This is a relevant information for the existing models of glioma invasion built upon ideas commonly employed within the KTAP framework and which take into account the underlying brain structure and its properties in trying to predict the tumor extent and its aggressiveness; we refer to \cite{HillenM5} and \cite{CEKNSSW} for two works where such issue is explicitly addressed. On the other hand, this  could also be relevant from a biological viewpoint; indeed, to our knowledge such information is not available in the biological literature. We are far from claiming to have a watertight evidence; it is rather a cue to motivate such speculation which should of course be properly verified by appropriately designed biological experiments.\\[-2ex]

\noindent
The linear stability analysis performed in Appendix \ref{sec:append} for constant diffusion coefficients suggests that pseudopalisades are a rather nonstandard type of patterns -at least as far as this model is concerned. The pH-chemorepellence is enhancing the diffusive effect, driving the tumor cells away from the strongly hypoxic site(s). Thereby, the form of the space-dependent tumor diffusion coefficient seems to play a decisive role for the shape of the tumor pattern, as simulations show. The formation of garland-like structures can be observed during the first half of the simulation time, after which there is no 'ring-like closure' of the cell aggregates, although these  seem to develop on each side of a hypocellular, acidic region. A rigorous analysis has still to be done, even in the case $D_T(x)>0$ for all $x$.  \\[-2ex]

\noindent
To acquire more qualitative information about the solutions of the macroscopic system deduced via parabolic scaling, we also performed a well-posedness analysis. As the global behavior of solutions to \eqref{macro_eq_non}, \eqref{macro_eq_non_S} seems out of reach, we assumed the production of protons by tumor cells to infer saturation and proved that the corresponding system has a unique global bounded nonnegative solution in the classical sense - for which certain assumptions on the tumor diffusion tensor were needed. In the case of solenoidal drift velocity and sufficiently large tumor growth, we proved that the solution approaches asymptotically the steady-state in which the tumor is at its carrying capacity, with a corresponding acidity concentration. The patterning behavior for the system with saturated, but sufficiently high net proton production is the same as for system \eqref{macro_eq_non}, \eqref{macro_eq_non_S} and numerical simulations show, too,  the same qualitative behavior of solutions. The rigorous qualitative study of system \eqref{macro_eq_non}, \eqref{macro_eq_non_S} (without the modifications and assumptions made in Section \ref{sec:analysis}) in terms of global well-posedness and singularity formation remains open. \\[-2ex]

\noindent
The model could be extended to include effects of vascularization and necrosis. Indeed, it is largely accepted \cite{Brat2002,Brat2004,wippold2006neuropathology} that the hypoxic glioma cells induced to migrate away from sites with very low pH express, among others, proteases and vascular endothelial growth factors (VEGF) initiating and sustaining angiogenesis. Endothelial cells (ECs) are attracted chemotactically towards the garland-like structures of high glioma density surrounding the hypoxic area, which leads to further invasion and overall tumor expansion. A corresponding model should contain an adequate description of macroscopic EC dynamics, which could be obtained as well from an originally multiscale setting, similarly to that for glioma cells but taking into account the features specific to EC migration.

\section*{Acknowledgement}

PK acknowledges funding by DAAD in form of a PhD scholarship. CS was funded by BMBF in the project \textit{GlioMaTh} 05M2016.

\appendix \label{sec:appendix}
\renewcommand{\theequation}{\Alph{section}.\arabic{equation}}
\setcounter{equation}{0}
\section{Linear stability analysis for a version of \eqref{macro_eq_non}, \eqref{macro_eq_non_S} with constant tumor diffusion coefficient}\label{sec:append}

For simplicity we perform a 1D analysis; the extension to a higher dimensional case involving a constant tumor diffusion tensor in diagonal form is straightforward. \\[-2ex]

\noindent
The uniform steady-states are $P_1=(0,0)$, $P_2=(1,\frac{1}{\alpha})$, and $P_3=(\alpha,1)$. In the absence of diffusion and taxis, $P_1$ and $P_2$ are saddles, while $P_3$ is stable for $0<\alpha <1$. Thus, we investigate the possibility of Turing-like patterns only around $P_3$ for such biologically relevant $\alpha$. \\[-2ex]

\noindent
Let $P_*:=(M_*,S_*)$ be a steady-state and consider the perturbations $u:=M-M_*$, $\sigma:=S-S_*$. Linearizing  \eqref{macro_eq_non}, \eqref{macro_eq_non_S} (with $D_T=D$ constant) around $P_*$ leads to
\begin{equation}\label{eq:vectorial}
\begin{pmatrix}u\\\sigma\end{pmatrix}_t=\mathbb A \begin{pmatrix}u\\\sigma\end{pmatrix}_{xx}+\mathbb B\begin{pmatrix}u\\\sigma\end{pmatrix},
\end{equation}
where $\mathbb A=\begin{pmatrix} D&D\alpha g(S_*)S_*\\0&1\end{pmatrix}$ and
$\mathbb B=\begin{pmatrix}\mu_0(1-2\alpha S_*)(1-S_*)&-\alpha \mu_0S_*(1-\alpha S_*)\\
1&-\alpha \end{pmatrix}$. \\[-1ex]

\noindent
We look for solutions of the form $\sum _{k\neq 0}T_k(t)\Xbf _k(x)$, with $\Delta \Xbf _k(x)+k^2\Xbf _k(x)={\bf 0}$ in $\Omega $ and $\nabla \Xbf_k\cdot \nubf ={\bf 0}$ on $\partial \Omega $, thus $\Xbf _k$ are eigenfunctions of $-\Delta $, each corresponding to the wavenumber $k$.\\[-2ex] 

\noindent
The terms making up the solution $(u,\sigma)$ will involve exponents of the eigenvalues $\lambda _k$ of the matrix $\mathbb B-k^2\mathbb A$. It holds that
\begin{align}
&\lambda_{1,k}+\lambda_{2,k}=\text {trace} (\mathbb B-k^2\mathbb A)<0 \\
&\lambda_{1,k}\lambda_{2,k}=\text {det}(\mathbb B-k^2\mathbb A)\label{trace-det}.
\end{align}

\begin{figure}[!htbp]
	\subfigure[][]{\includegraphics[width=0.32\textwidth]{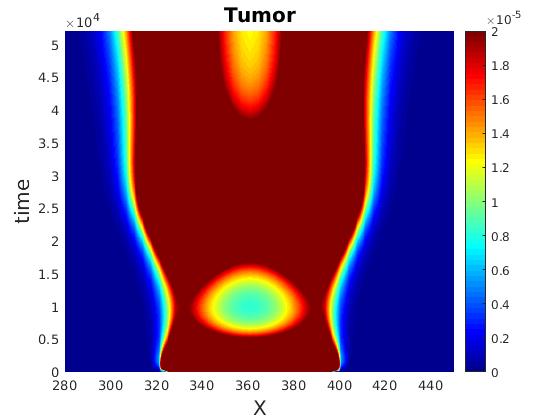}\label{fig:pattern-T-D_const}}\quad 	\subfigure[][]{\includegraphics[width=0.32\textwidth]{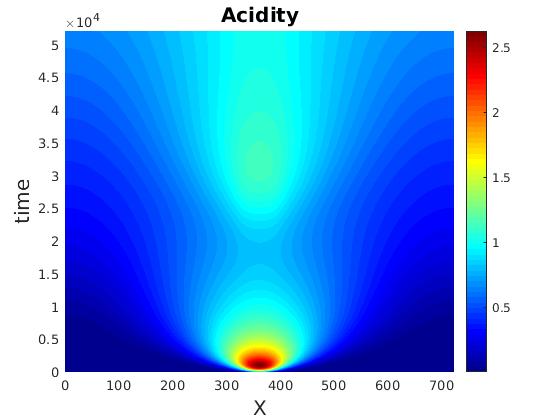}\label{fig:pattern-A-D_const}}\quad
	\subfigure[][]{\includegraphics[width=0.32\textwidth]{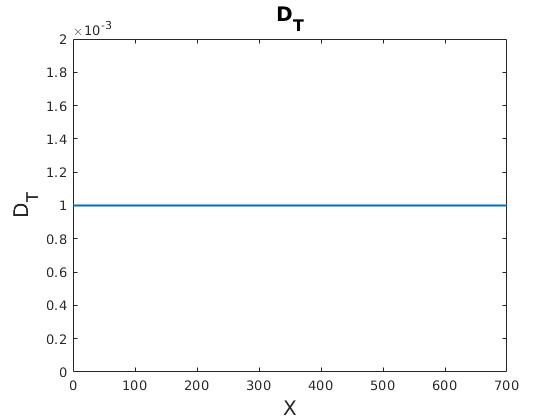}\label{fig:pattern-D-D_const}}\\
	\subfigure[][]{\includegraphics[width=0.32\textwidth]{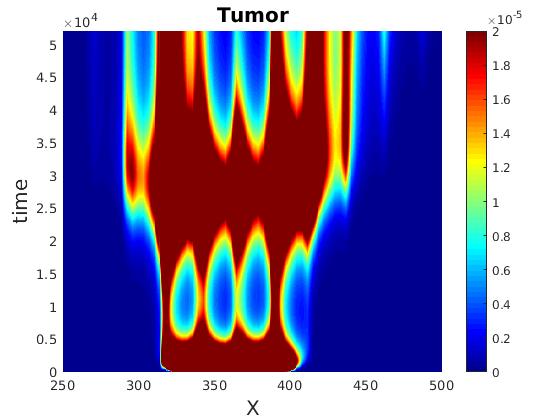}\label{fig:pattern-T-D_const}}\quad 	\subfigure[][]{\includegraphics[width=0.32\textwidth]{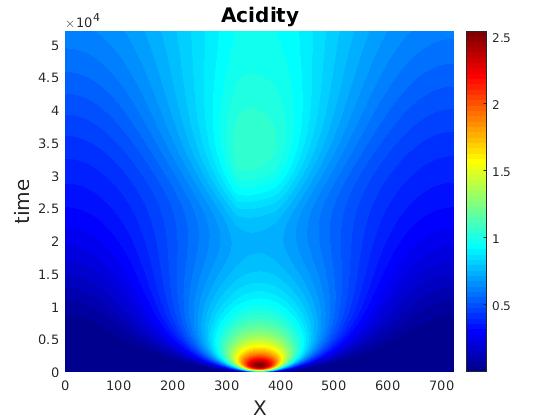}\label{fig:pattern-A-D_const}}\quad
	\subfigure[][]{\includegraphics[width=0.32\textwidth]{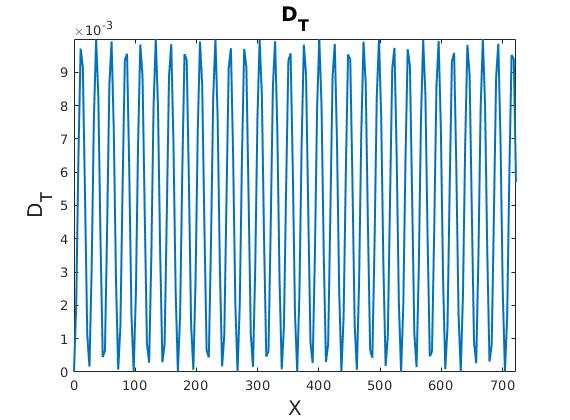}\label{fig:pattern-D-D_const}}\\
	\subfigure[][]{\includegraphics[width=0.32\textwidth]{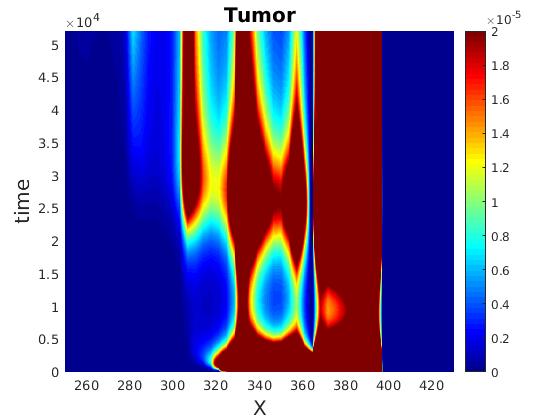}\label{fig:pattern-T-D_const}}\quad 	\subfigure[][]{\includegraphics[width=0.32\textwidth]{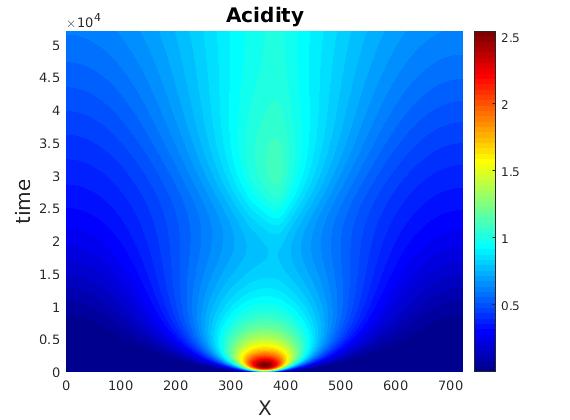}\label{fig:pattern-A-D_const}}\quad
	\subfigure[][]{\includegraphics[width=0.32\textwidth]{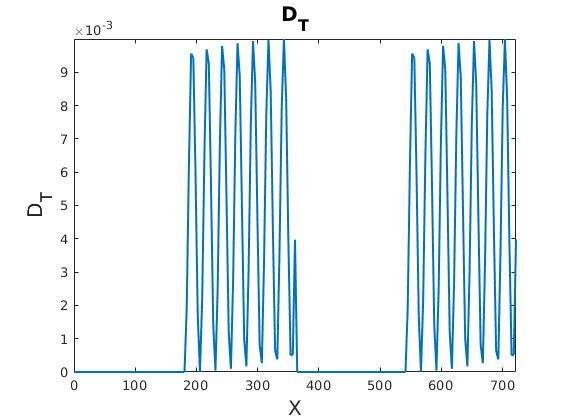}\label{fig:pattern-D-D_const}}
	\caption[]{Patterns (in 1D) for tumor (left column) and acidity (middle column) for several choices of the tumor diffusion coefficient $D_T(x)$ (plots of the latter shown in the right column). Simulations are done for the dimension-free system \eqref{macro_eq_non}, \eqref{macro_eq_non_S}, the maximum simulation time corresponds to 10 months in dimensional framework.}
	\label{fig:patterns-1D}
\end{figure}
\noindent
For Turing-like patterns around $P_*$ we need $\text {det}(\mathbb B-k^2\mathbb A)<0$, which means that there is a positive eigenvalue. This condition takes for $P_*=P_3$ the concrete form
\begin{equation}\label{eq:A4}
k^4D+\alpha Dk^2(1+g(1))+\mu_0\alpha (1-\alpha)<0.
\end{equation}
This cannot be satisfied for $\alpha < 1$, which is typical for higher grade tumors (especially for GBM). In fact, in view of the nondimensionalization done in Subsection \ref{subsec:params}, it is very improbable to have $\alpha > 1$ for this problem.  \\[-1ex]


\noindent
To see the effect of pH-taxis (repellence by acidity) we set $g(S)=0$, which only enhances the chances of \eqref{eq:A4} to hold (still for $\alpha <1$), thus the presence of pH-taxis does not dramatically change the patterning behavior; this is due to the tactic bias being repellent; an attractive pH-taxis (as proposed in \cite{Bartel} for melanoma cells or in \cite{paradise} for endothelial cells\footnote{The latter could serve, however, for a model extension including angiogenesis.}) could render \eqref{eq:A4} valid even for $\alpha <1$, but does not seem to be appropriate for describing pseudopalisade formation.\\[-1ex]

\noindent
The above suggests that pseudopalisades are not Turing-like patterns - at least as far as our model with constant diffusion coefficient is used for their description. To see the effect of the diffusion coefficient $D_T(x)$ we plot in Figure \ref{fig:patterns-1D} the patterns obtained in 1D for the same parameter combination and several choices of $D_T(x)$, including various kinds of degeneracy. Thus, the second row in Figure \ref{fig:patterns-1D} shows the case where $D_T$ degenerates on countable set, while the last row illustrates the situation with a strong degeneracy, i.e. on whole subintervals of the space domain ($D_T$ is of the type considered in \cite{WS17} for a closely related problem, however with haptotaxis).

\newcommand{\noopsort}[1]{}
\addcontentsline{toc}{section}{References}
\bibliographystyle{plain}
\bibliography{pp_bibtex_v14}

\end{document}